\documentclass[aps,prx,reprint,superscriptaddress,floatfix,nofootinbib,longbibliography]{revtex4-2}
\usepackage{amsmath,amssymb,amstext,dsfont,tikz,graphicx,physics,mathtools,bm,
simpler-wick,color,ifthen}
\usepackage[colorlinks=true, allcolors=purple]{hyperref}
\usepackage{circuitikz}

\usepackage{multirow}  
\definecolor{myBlue}{HTML}{0072B2}
\usetikzlibrary{arrows}
\graphicspath{{./}{figures/thresholds/}}
\newcommand{\mathbbm}{\mathds}
\renewcommand{\Pr}{\mathop{\mathrm{Pr}}\nolimits}

\makeatletter
\def\l@subsubsection#1#2{}
\DeclareRobustCommand{\AppendixTOCOnly}{%
  \let\l@subsection\@gobbletwo
}
\makeatother

\def\bra#1{\mathinner{\langle{#1}|}}
\def\ket#1{\mathinner{|{#1}\rangle}}

\newcommand{\cH}{\mathcal H}
\newcommand{\Irr}{\operatorname{Irr}}

\definecolor{myBlue}{HTML}{0072B2}
\definecolor{fluxGreen}{HTML}{009E73}

\begin{document}

\title{Fusion-assisted decoding of non-Abelian topological order}

\author{Rohith Sajith}
\affiliation{Department of Physics, Harvard University, Cambridge, MA 02138, USA}

\author{Yabo Li}
\affiliation{Department of Physics and Astronomy, Purdue University, West Lafayette, IN, 47907}
\affiliation{Center for Quantum Phenomena, Department of Physics, New York University, 726 Broadway, New York, NY 10003, USA}

\author{Zijian Song}
\affiliation{C. N. Yang Institute for Theoretical Physics, State University of New York at Stony Brook, Stony Brook, NY 11794, USA}

\author{Tsung-Cheng Lu}
\affiliation{Department of Physics and Astronomy, Center for Quantum Research and Technology, University of Oklahoma, Norman, OK 73069, USA}
\affiliation{Joint Center for Quantum Information and Computer Science, NIST/University of Maryland, College Park, MD, 20742, USA}

\author{Carolyn Zhang}
\affiliation{Department of Physics, Harvard University, Cambridge, MA 02138, USA}
\affiliation{Department of Physics and Astronomy, and Quantum Matter Institute,
University of British Columbia, Vancouver, BC, Canada V6T 1Z1}

\date{\today}

\begin{abstract}
Maximum likelihood (ML) decoders perform quantum error correction by measuring a set of syndromes, classically computing an optimal recovery based on these syndromes and the noise model, and then performing that recovery. ML decoders are well understood for the toric code and generalize straightforwardly to other surface codes with Abelian anyons, where they can saturate information-theoretic decoding thresholds. On the other hand, codes with non-Abelian anyons have become increasingly relevant because they can support transversal or constant-depth non-Clifford gates and admit efficient preparation protocols in modern quantum devices. However, they present new challenges for decoding: non-Abelian anyons cannot in general be deterministically annihilated, and the logical operators cannot be classified by Abelian homology. In this work, we show that refined syndrome sets, including measurements that resolve non-Abelian anyon fusion outcomes, systematically improve the performance of ML decoders. We demonstrate this concretely by constructing explicit decoders for the $D_4$ quantum double model under charge noise, and obtaining numerical results for thresholds via mappings to classical statistical-mechanics models. We show that there is a nonzero noise threshold below which the entire logical qudit can be recovered with a single round of measurements.
\end{abstract}

\maketitle

\tableofcontents

\section{Introduction}
\label{sec:introduction}

Topological quantum memories encode information in global degrees of freedom that cannot be changed by local operators. This allows them to have a nonzero threshold against local noise: when exposed to a local noise channel for a sufficiently short time, the resulting decohered state can be decoded back to the original code state with probability approaching one in the thermodynamic limit~\cite{wen1990topological,wen1990ground,kitaev_fault_2003,dennis2002topological}. 2+1d surface codes, such as the toric code, are known to have such a threshold, and have been realized on various experimental platforms, making the design of optimal decoders not only a theoretical question but also a problem relevant to fault-tolerant quantum computers~\cite{bravyi1998quantum,fowler2012surface,Bravyi2014efficient,satzinger2021realizing,krinner2022realizing,Bluvstein_2022_atom_arrays,foss2023experimental,henrik_2024_measurement_TO,google_below_threshold_2025}.

In recent years, non-Abelian topological orders have become increasingly relevant to practical implementation of quantum computation because they admit non-Clifford logical gates~\cite{huang2025D4, davydova_universal_2025, sajith_nonclifford_2026, bauer2025planarfaulttolerantcircuitsnonclifford,huang2026hybrid,kobayashi_clifford_2026, hsin2025automorphismgaugetheorieshigher,warman_constatnt_2026, manjunath_universal_2026,li2026engineering,li_anyon_2026} and efficient preparation protocols on modern devices~\cite{bravyi2022adaptive,tantivasadakarn2024long,lu2022measurement,Shortest_non_abelian_ruben_2023,li2023symmetry,Hierarchy,iqbal_non-Abelian_2024,lo2026universal}. The conventional approach to topological quantum computation encodes logical information in the collective fusion space of non-Abelian anyons and manipulates it through braiding and fusion measurements~\cite{mochon_anyons_2003,mochon_anyon_2004,bonderson2008interferometry,bonderson_measurement_2008,cui2015universal,cui2015universal2,chen2025universal,lo2026universal}. Another approach, extensively studied recently, encodes logical information in the ground state codespace and manipulates it to implement logical gates, including gates at higher levels of the Clifford hierarchy~\cite{kobayashi_clifford_2026,hsin2025automorphismgaugetheorieshigher,manjunath_universal_2026,warman_constatnt_2026,li2026engineering}, or magic state preparation~\cite{huang2025D4,huang2026hybrid}. Here, we focus on decoding quantum memories that use this ground-state encoding.

A useful starting point is maximum likelihood (ML) decoding, which has been studied in depth for Abelian topological orders~\cite{Bravyi2014efficient,ferris2014tensor,chubb2021general,higgott2023improved,Sundaresan_2023,googleAI2023,Berezutskii_2025}. The goal of a decoder is to remove the local noise-induced errors without introducing a logical error. A maximum likelihood (ML) decoder does this by measuring a syndrome record, which, in Abelian topological orders, records the locations of the anyon excitations. These mark the endpoints of the invisible error chains, or anyon strings. The ML decoder then uses its knowledge of the noise model to calculate the probability of each of the finite number of homology equivalence classes of the errors, and implements a correction appropriately~\cite{dennis2002topological}. Schematically, decoding succeeds when 
\begin{equation}\label{recoveryintro}
    \mathcal{R}[\mathcal{N}[\rho_0]]=\sum_{\mathcal{S}}\mathcal{R}_{\mathcal{S}}[P_{\mathcal{S}}\mathcal{N}[\rho_0]P_{\mathcal{S}}]\to \rho_0
\end{equation}
where $P_{\mathcal{S}}$ is the projector onto a particular syndrome $\mathcal{S}$ and $\mathcal{R}_{\mathcal{S}}$ is a unitary channel obtained from the ML decoder's lookup table mapping syndromes to the most likely error homology class. For the surface code, the ML decoder just needs to compare two different homology classes, which correspond to the $\mathbb{Z}_2$ winding around non-contractible cycles.

Several approaches to decoding non-Abelian topological orders have already been developed~\cite{wootton_error_2014,Hutter_2015,wootton_active_2016,schotte_quantum_2022,davydova_universal_2025,jing_intrinsic_2025,jing2026integerlinearprogrammingdecoder}. Extending the Abelian ML framework to non-Abelian topological orders, however, presents two linked challenges. First, the positions of local anyon excitations do not exhaust the error
information; decohered non-Abelian topological orders cannot be diagonalized by a simple local anyon excitation basis as in the toric code with incoherent Pauli noise. Multiple non-Abelian anyons also carry a collective fusion
state, which cannot be reconstructed from independent local syndrome measurements.
Relatedly, anyons in general cannot be deterministically annihilated: non-Abelian anyons fuse non-invertibly, as
\begin{equation}
    a\times \bar{a} = 1+\sum_{c\neq 1}N_{a\bar{a}}^cc
\end{equation}
so bringing together an anyon $a$ and its conjugate charge/``antiparticle" $\bar{a}$ may produce other anyons $c$ other than the vacuum. Naively, one might think that this would preclude single-shot decoding, which is decoding based on a single round of syndrome measurements followed by a correction $\mathcal{R}_\mathcal{S}$, because local anyon syndromes do not determine fusion outcomes. Indeed, previous decoding protocols use multiple rounds of measurement and correction~\cite{Brell_2014,wootton_error_2014,Hutter_2015,wootton_active_2016,Burton_2017,dauphinais_fault_2017,schotte_quantum_2022,jing_intrinsic_2025,sala_stability_2025,jing2026integerlinearprogrammingdecoder}. 

Second, Abelian ML decoders decide on $\mathcal{R}_{\mathcal{S}}$ by comparing the classical probabilities of different $\mathbb{Z}_N$ (where $N=2$ for the toric code) homology equivalence classes of errors. But for non-Abelian topological orders, errors do not form simple homology equivalence classes; a logical operator corresponding to a winding $a$ string combined with that of a winding $\bar{a}$ string may not give a trivial logical operator; in general this would give coherences between other logical operators. A related problem, as will be discussed in Sec.~\ref{sec:fusion-logical-backaction}, is that the same anyon fusion measurements that help us identify likely errors can actually accidentally measure logical information.

In this work, we take a first step toward developing a general theory of ML-type decoders for non-Abelian topological orders. To be explicit, we use microscopic models of quantum doubles based on non-Abelian groups $G$~\cite{kitaev_fault_2003}, and we focus on charge noise on a cylinder geometry (where the number of logical states is the number of irreducible representations (irreps) of $G$). We provide a general framework for designing ML-type decoders for these models, and we derive the corresponding statistical-mechanics models (analogs of the random-bond Ising model that appears in the ML decoder for the $\mathbb{Z}_2$ toric code~\cite{dennis2002topological}) and thresholds in detail for a minimal example of a non-Abelian topological order.

\begin{figure}[t]
 \centering
 \includegraphics[width=0.9\linewidth]{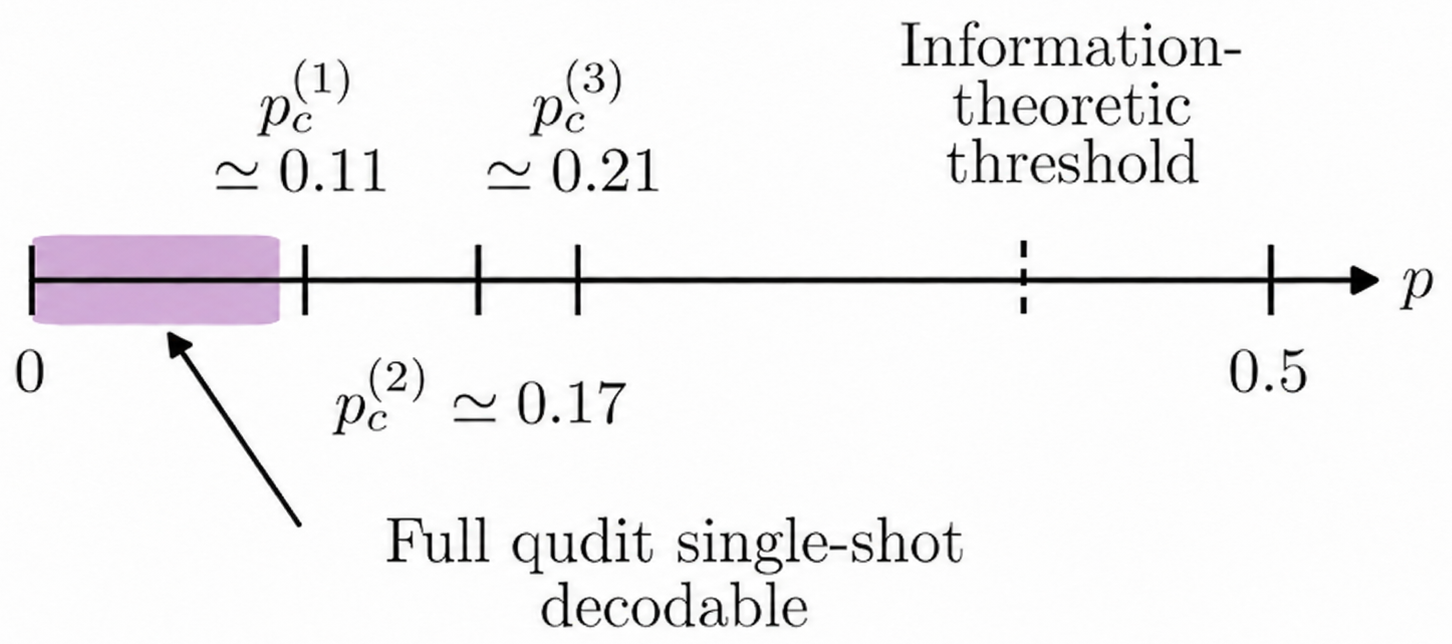}
 \caption{The phase diagram of the $D_4$ quantum double under the non-Abelian $E$ charge decoherence, with a noise strength parameterized by $p$. Refining the syndrome sets from $\mathcal{S}^{(1)}$ to $\mathcal{S}^{(2)}$ to $\mathcal{S}^{(3)}$ improves the decoding thresholds from $p_c^{(1)}$ to $p_c^{(2)}$ to $p_c^{(3)}$. In particular, $\mathcal{S}^{(1)}$ records the absence/presence of the two-dimensional charge $E$ on each vertex, $\mathcal{S}^{(2)}$ records the complete onsite anyon syndrome, and $\mathcal{S}^{(3)}$ further records pair fusion measurements of the $E$ anyons. The numerical threshold values for these three syndromes are obtained from finite-size numerics and are shown in Table~\ref{tab:num-conditional-thresholds}. We also show that the full logical qudit is single-shot decodable below the minimal-weight perfect matching threshold in Sec.~\ref{sec:full-D4-qudit-decoding}. The intrinsic information-theoretic threshold is shown schematically; R\'{e}nyi-2 coherent information calculations in Sec.~\ref{sec:renyi2-intrinsic-proxy} indicate a R\'{e}nyi-2 transition at $p_c\simeq 0.46$.}
 \label{fig:phasediag}
\end{figure}

\subsection{Main results}\label{sec:main_results}
In quantum double models based on non-Abelian groups decohered by charge noise, we find that
\begin{itemize}
    \item For each non-Abelian topological order, there are different ML decoders one can construct based on different syndrome sets. In general, one can improve thresholds by measuring anyon fusion outcomes in addition to local anyon excitations, at the cost of introducing more non-local measurements.
    \item The entire logical space of quantum doubles of nilpotent groups of class $2$ such as $D_4$ can be completely single-shot recovered up to a nonzero threshold, if we use anyon fusion measurements.
    \item Logical information stored in the ``central qudit", which is a subspace of the full logical qudit for groups that have a nontrivial center, can be recovered using $\mathbb{Z}_N$ homology similar to that used in Abelian topological orders.
    \item Fusion measurements of incorrectly grouped anyons, even when the measurements do not span the system, can in fact disturb logical information outside of the central qudit. This becomes dangerous above a certain noise threshold after which we are unlikely to correctly group all anyons. Nevertheless, if we choose to sacrifice this logical subspace, fusion measurement significantly improves the threshold of the central qudit. 
\end{itemize}

In the $D_4$ quantum double, we find three natural syndrome sets for decoding the central qudit. The decodability as a function of the noise strength $p$ for incoherent $E$ noise is illustrated in Fig.~\ref{fig:phasediag}. On a cylinder with smooth/flux condensed boundary conditions, there are 5 logical states labeled by the five conjugacy classes of $D_4$, and charge noise can induce logical charge operators that wind around the periodic direction of the cylinder. The charge operators are labeled by irreps $A_{00},A_{10},A_{01},A_{11},$ and $E$. The first four irreps are 1d, and correspond to Abelian anyons, while the $E$ irrep is 2d and labels a non-Abelian anyon. 

Preservation of the central qubit amounts to ensuring that we don't act with a logical $E$ operator; the Abelian anyons can be cleaned up arbitrarily if we just care about the central qubit. On this subspace an $E$ line winding around the cylinder is a logical Pauli $Z$, while a boundary-to-boundary shift by the central element is a logical Pauli $X$. Note that $E$ fuses with itself to give only the other irreps:
\begin{equation}\label{Efusion}
    E\times E=A_{00}+A_{10}+A_{01}+A_{11}
\end{equation}
This property implies that decoding the central qubit is similar to Abelian ML decoding in the sense that it is sufficient to compare the classical probabilities of two homology classes. We will explore when this simplification of decoding occurs more generally in Sec.~\ref{sec:general-central-decoding}, in particular showing that it occurs beyond cases where the quantum double group $G$ is nilpotent of class $2$, i.e. when the right hand side of the non-Abelian fusion consists only of Abelian anyons as in Eq.~\eqref{Efusion}. 

The three classes of syndrome measurements are:
\begin{itemize}
    \item $\mathcal{S}^{(1)}:$ on each site, we measure whether or not there is an $E$ anyon (i.e. whether there is a 1d irrep or the $E$ irrep). 
    \item $\mathcal{S}^{(2)}:$ on each site, we measure the on-site local charge, which may be any of the 5 irreps. The other irreps may occur from fusion of adjacent $E$ anyons.
    \item $\mathcal{S}^{(3)}:$ in addition to the on-site local charges, we also measure fusion outcomes of disjoint $E$ anyon pairs. Since the result is always an Abelian anyon, the rest of the fusion results are completely deterministic.
\end{itemize}
These three kinds of measurements give ML decoders with thresholds $p_c^{(1)},p_c^{(2)},$ and $p_c^{(3)}$ respectively. Since each one measures more information than the last, we have
\begin{equation}
    p_c^{(1)}\leq p_c^{(2)}\leq p_c^{(3)}
\end{equation}
For noise that only proliferates $E$ anyons, we find that the improvements are substantial as shown in Fig.~\ref{fig:phasediag}:
\begin{equation}
 p_c^{(1)}\simeq0.111
 <p_c^{(2)}\simeq0.165
 <p_c^{(3)}\simeq0.214.
 \label{eq:intro-d4-hierarchy}
\end{equation}

The improvement from $\mathcal S^{(3)}$ comes with a sharp limitation.  A fusion projector is defined relative to a pairing path. Two paths joining
the same pair can differ by a noncontractible loop; we will show that in some cases this can lead to fusion measurements measuring logical information outside of the central qubit. Therefore, the remarkably high threshold we obtain for the central qubit comes at the cost of inevitably losing the rest of the logical qudit. This property is unique to non-Abelian topological order, and is related to the fact that for non-Abelian topological orders, sequences of logical operators on a torus can result in a single leftover anyon~\cite{bombin2008family,iqbal_non-Abelian_2024}.

At a high level, for Abelian topological orders, it is easy to find a basis that diagonalizes the decohered density matrix $\mathcal{N}[\rho_0]$ even if the eigenvalues are difficult to compute: this basis is simply the local anyon excitation basis. $P_{\mathcal{S}}\mathcal{N}[\rho_0]P_{\mathcal{S}}$ in Eq.~\eqref{recoveryintro} then is of minimal rank (with one state in each logical sector) and does not destroy any information from coherences between different syndrome states. Syndromes constructed from this basis naturally give an ML decoder that saturates the information-theoretic threshold for decodability of the topological order. For non-Abelian topological orders, not only are the eigenvalues difficult to compute but also the basis that diagonalizes the decohered density matrix is not practical to obtain. Even though we can develop improved syndrome sets by resolving fusion outcomes, these syndromes still do not in general correspond to a basis that fully diagonalizes the density matrix. As a result, the decoder loses some information from $\mathcal{N}[\rho_0]$ by removing off-diagonal terms via the projectors onto $P_{\mathcal{S}}$. An ML decoder does not see the true $\mathcal{N}[\rho_0]$ but rather sees the effective density matrix with only the information it extracted from the syndrome measurements. While we can get improved decoders by making finer syndrome measurements, we do not expect to be able to reach the intrinsic information-theoretic threshold analogous to those discussed in Refs.~\cite{fan_diagnostics_2024,sala_stability_2025,sang_stability_2025} using a reasonable set of syndrome measurements. In Fig.~\ref{fig:phasediag}, we indicate where the true information-theoretic threshold might be based on the R\'{e}nyi-2 calculations and numerics in Sec.~\ref{sec:renyi2-intrinsic-proxy}. The R\'{e}nyi-2 numerics show a transition at $p_c\simeq 0.46$.

\subsection{Outline}
The rest of the manuscript is organized as follows: Section~\ref{QDReview} defines the general quantum double model based on a group $G$ on a cylinder, its ground-state subspace and logical operators, and a noise channel that creates irrep $\Gamma$ anyons. We will describe in particular the noise that creates 2d $E$-charges in the $D_4$ quantum double model. Section~\ref{sec:syndrome-decoding-main} presents syndrome projectors (including fusion outcome projectors) and computes syndrome probabilities for general quantum double models with irrep $\Gamma$ decoherence. For the particular case of $E$ noise in the $D_4$ quantum double, we derive simplifications of the syndrome probabilities that take the form of classical partition functions. We also describe in detail how fusion measurements can reveal logical information outside in general non-Abelian topological orders. Section~\ref{sec:decoding} focuses on the $D_4$ quantum double and formulates its ML recovery explicitly, together with numerical results on thresholds. Finally, in Section~\ref{sec:discussion} we outline several open questions for more general kinds of topological orders and decoherence models.

Appendices~\ref{app:onsite-probability}--\ref{app:flux-ribbon-proof} provide the detailed derivations of syndrome probabilities, path-dependent fusion measurements, and microscopic flux-ribbon operators. Appendix~\ref{app-current-rep} presents an auxiliary positive current representation of the charge decoherence channel. Appendix~\ref{app:central-numerical-checks} describes the details of our numerical calculations: physical-record sampling, exact partition-function contractions, and finite-size analysis. Finally, Appendix~\ref{app:s3-syndrome-probabilities} illustrates the framework for the $S_3$ quantum double, and Appendix~\ref{app:renyi2-purity-derivation} presents the derivation of the R\'enyi-2 coherent-information diagnostic and its numerical calculation for probing the intrinsic information-theoretic threshold of the decohered state.

\section{Quantum double memory and charge decoherence}
\label{QDReview}
In this section, we introduce the conventions we use for quantum-double models~\cite{kitaev_fault_2003}, identify their ground-state subspace and logical operators defined on a cylinder, and present the charge noise channels studied in this work. Specializing to the case of the group $D_4$, we identify a subspace of the logical space that we call the central qubit, and we specify the charge decoherence channel that proliferates non-Abelian anyons associated with the 2d $D_4$ irrep.

\subsection{Quantum double models on a cylinder}\label{sec:quantumdouble}

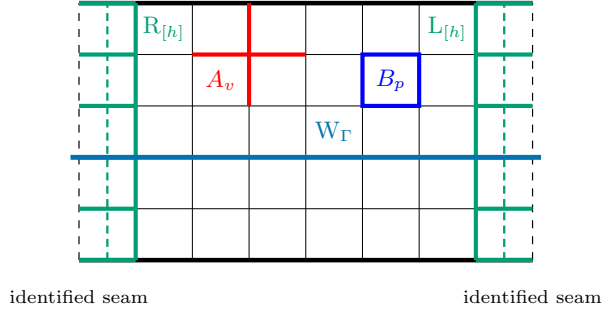
\begin{figure}[t]
\centering
\begin{tikzpicture}[x=0.75cm,y=0.68cm]

  \foreach \x in {1,2,3,4,5,6,7}{
    \draw[line width=0.35pt] (\x,0)--(\x,5);
  }
  \foreach \y in {1,2,3,4}{
    \draw[line width=0.35pt] (0,\y)--(8,\y);
  }

  \draw[line width=0.6mm] (0,0)--(8,0);
  \draw[line width=0.6mm] (0,5)--(8,5);

  \draw[
    dashed,
    line width=0.45pt
  ] (0,0)--(0,5);

  \draw[
    dashed,
    line width=0.45pt
  ] (8,0)--(8,5);


  \draw[
    fluxGreen,
    densely dashed,
    line width=0.3mm
  ] (0.5,0)--(0.5,5);

  \draw[
    fluxGreen,
    line width=0.55mm
  ] (1,0)--(1,5);

  \foreach \y in {0,1,2,3,4,5}{
    \draw[
      fluxGreen,
      line width=0.55mm
    ] (0,\y)--(1,\y);
  }

  \node[fluxGreen] at (1.5,4.5)
    {$\mathrm R_{[h]}$};


  \draw[
    fluxGreen,
    densely dashed,
    line width=0.3mm
  ] (7.5,0)--(7.5,5);

  \draw[
    fluxGreen,
    line width=0.55mm
  ] (7,0)--(7,5);

  \foreach \y in {0,1,2,3,4,5}{
    \draw[
      fluxGreen,
      line width=0.55mm
    ] (7,\y)--(8,\y);
  }

  \node[fluxGreen] at (6.5,4.5)
    {$\mathrm L_{[h]}$};


  \draw[
    red,
    line width=0.55mm
  ] (2,4)--(4,4);

  \draw[
    red,
    line width=0.55mm
  ] (3,3)--(3,5);

  \node[red] at (2.5,3.5)
    {$A_v$};

  \draw[
    blue,
    line width=0.55mm
  ] (5,3) rectangle (6,4);

  \node[blue] at (5.5,3.5)
    {$B_p$};


  \draw[
    myBlue,
    line width=0.65mm
  ] (-0.15,2)--(8.15,2);

  \node[myBlue] at (4.5,2.5)
    {$\mathrm W_{\Gamma}$};

  \node[below=8pt] at (0,0)
    {\scriptsize identified seam};

  \node[below=8pt] at (8,0)
    {\scriptsize identified seam};

\end{tikzpicture}

\caption{
Stabilizers and logical operators of the quantum double on a cylinder. The vertex stabilizer $A_v$ acts on the four edges incident on $v$, while the plaquette stabilizer $B_p$ acts on the edges surrounding $p$. The dashed vertical sides are identified. The charge line $\mathrm W_{\Gamma}$, labeled by an irreducible representation $\Gamma\in\operatorname{Irr}(G)$, winds around the periodic direction. The conjugacy-class-labelled flux ribbons $\mathrm R_{[h]}$ and $\mathrm L_{[h]}$ connect the two open boundaries along dual-lattice paths. In the displayed convention, $\mathrm R_{[h]}$ lies to right of the seam and $\mathrm L_{[h]}$ lies to the left of the seam. The dashed green centerlines mark the dual paths, while the solid green segments indicate the corresponding ribbon supports.
}
\label{fig:stabilizers-logicals-cylinder}
\end{figure}
The quantum double model is defined on a square lattice, and is based on a finite group $G$. We use an $L\times L$ lattice on a cylinder, periodic along the horizontal direction and open with a smooth boundary along the top and bottom edges (see Fig.~\ref{fig:stabilizers-logicals-cylinder}). Each oriented edge of the square lattice carries
the group algebra $\mathbb C[G]$ with basis $\{\ket g:g\in G\}$, giving a $|G|$ dimensional local Hilbert space. Throughout,
horizontal edges are oriented to the right and vertical edges upward. On the local edge
Hilbert space, $L^h$ and $R^h$ implement left and right group multiplication
respectively: 
\begin{equation}
 L^h_e|g\rangle_e=|hg\rangle_e,
 \qquad
 R^h_e|g\rangle_e=|gh^{-1}\rangle_e,
 \label{eq:cyl-LR-edge}
\end{equation}
For $G=\mathbb{Z}_2$, $L^g_e=R^g_e=X_e$, the Pauli $X$ operator, for the nontrivial group element. Note that the $h^{-1}$ in the action of $R^h$ is necessary for recovering the group law:
\begin{equation}
 L^g_eL^h_e=L^{gh}_e,
 \qquad R^g_eR^h_e=R^{gh}_e,
 \qquad [L^g_e,R^h_{e'}]=0.
 \label{eq:cyl-LR-algebra}
\end{equation}

For an irreducible representation (irrep) $\Gamma$ of dimension $d_\Gamma$, define the matrix-valued diagonal operator
\begin{equation}
 \begin{aligned}
 Z_e^\Gamma
   &\equiv
   \sum_{g\in G}\Gamma(g)\otimes\ket{g}_e\bra{g}_e
   =\vcenter{\hbox{
   \begin{tikzpicture}[scale=0.58]
     \draw[line width=0.35mm,red] (-0.45,0.35)--(0,0.35);
     \draw[line width=0.35mm,red] (1.30,0.35)--(1.75,0.35);

     \draw[fill=white] (0,0) rectangle (1.30,0.70);
     \node[scale=0.78] at (0.65,0.35) {$Z_e^\Gamma$};

     \draw (0.65,0.70)--(0.65,1.10);
     \draw (0.65,0)--(0.65,-0.40);
   \end{tikzpicture}}},\\
 Z_{ab,e}^{\Gamma}
   &\equiv
   {}_{\rm virt}\!\bra{a}Z_e^\Gamma\ket{b}_{\rm virt}
   =\sum_{g\in G}
   [\Gamma(g)]_{ab}\ket{g}_e\bra{g}_e .
 \end{aligned}
 \label{eq:cyl-Z-edge}
\end{equation}
Here the horizontal legs carry the virtual representation space
$V_\Gamma$, while the vertical legs act on the physical edge.
The indices $a,b=1,\ldots,d_\Gamma$ label the virtual space. $Z^{\Gamma}_{ab,e}$ is an operator that acts on the original edge Hilbert space. For $G=\mathbb{Z}_2$, $Z^\Gamma_e=Z_e$, the Pauli $Z$ operator, for the nontrivial $\mathbb{Z}_2$ irrep.

The Hamiltonian is a commuting projector Hamiltonian of the form $H = -\sum_{v} A_v - \sum_{p} B_{p},$ where $A_v$ are vertex projectors and $B_{p}$ are plaquette projectors. We define $A_v = \frac{1}{|G|} \sum_{g \in G} A_v^g,$ where
\begin{equation}
\raisebox{-0.5\height} {
    \begin{tikzpicture}

    \node at (4.5, 1) {$A_v^g = $};

    \draw (5,1) to node[currarrow, rotate=0] {} (6,1);
    \draw (6, 1) to node[currarrow, rotate=90] {} (6, 2);
    \draw (6, 1) to node[currarrow, rotate=0] {} (7, 1);
    \draw (6, 1) to node[currarrow, rotate=90] {} (6, 0);
    
    \node[scale=0.9] at (5.5, 1+0.3) {$R^g$};
    \node[scale=0.9] at (6+0.3, 1.5) {$L^g$};
    \node[scale=0.9] at (6.5, 1-0.3) {$L^g$};
    \node[scale=0.9] at (6-0.3, 0.5) {$R^g$};

\end{tikzpicture}}
\end{equation}
For a general non-Abelian group the set of operators $\{ A_v^{g} \mid g \in G \}$ do not commute and follow a $G$-group algebra.

$B_p^{h}$ is defined as a projector onto the subspace of the Hilbert space that has group-valued flux equal to $h$ on the plaquette $p$. We can write such an operator as
\begin{equation}
\raisebox{-0.5\height} {
    \begin{tikzpicture}

    \node at (4.5, 1) {$B_p^h = $};

    \node at (6.5, 1) {$\sum_{\{ g_i \}} \delta_{h, g_1g_2g_3^{-1}g_4^{-1}}$};

    \draw (8, 0.5) -- (8, 1.5);
    \draw (9.3, 1.5) -- (9.5, 1);
    \draw (9.5, 1) -- (9.3, 0.5);

    \draw (8.4, 1.3) to node[currarrow, rotate=0] {} (9, 1.3);
    \draw (8.4, 1.3) to node[currarrow, rotate=90] {} (8.4, 0.7);
    \draw (8.4, 0.7) to node[currarrow, rotate=0] {} (9, 0.7);
    \draw (9, 0.7) to node[currarrow, rotate=90] {} (9, 1.3);

    \node[scale=0.8] at (8.2, 1) {$g_1$};
    \node[scale=0.8] at (8.7, 1.5) {$g_2$};
    \node[scale=0.8] at (9.25, 1) {$g_3$};
    \node[scale=0.8] at (8.7, 0.5) {$g_4$};

    \draw (8.4+2-0.2, 1.3) to node[currarrow, rotate=0] {} (9+2-0.2, 1.3);
    \draw (8.4+2-0.2, 1.3) to node[currarrow, rotate=90] {} (8.4+2-0.2, 0.7);
    \draw (8.4+2-0.2, 0.7) to node[currarrow, rotate=0] {} (9+2-0.2, 0.7);
    \draw (9+2-0.2, 0.7) to node[currarrow, rotate=90] {} (9+2-0.2, 1.3);

    \node[scale=0.8] at (8.2+2-0.2, 1) {$g_1$};
    \node[scale=0.8] at (8.7+2-0.2, 1.5) {$g_2$};
    \node[scale=0.8] at (9.25+2-0.2, 1) {$g_3$};
    \node[scale=0.8] at (8.7+2-0.2, 0.5) {$g_4$};

    \draw (9.3+0.5, 1.5) -- (9.1+0.5, 1);
    \draw (9.1+0.5, 1) -- (9.3+0.5, 0.5);
    
    \draw (8+3.2+0.2-0.2, 0.5) -- (8+3.2+0.2-0.2, 1.5);

\end{tikzpicture}}
\label{S3BpOperator}
\end{equation}
The operator in the Hamiltonian is defined as $B_p \equiv B_p^1,$ which projects onto states with no flux. At the smooth upper and lower boundaries, we simply truncate the $A_v^h$ terms to the parts supported on the lattice (i.e. we get terms on three edges rather than four).

\subsection{Logical qudit on a cylinder}
\label{sec:qd-cylinder-groundspace}

We now identify the ground space of the quantum double on the cylinder, which forms the logical qudit. We will show that the number of states in the logical qudit is
\begin{equation}\label{quditdim}
    \mathrm{dim}(\mathcal{H}_L)=\#\operatorname{Conj}(G)=\#\Irr(G),
\end{equation}
where $\text{Conj}(G)$ is the set of conjugacy classes of $G$ and $\text{Irr}(G)$ is the set of irreps of $G$.\footnote{We could have also used a torus geometry, but use a cylinder geometry because it has a smaller $\mathrm{dim}(\mathcal{H}_L)$, slightly simplifying the problem.}

The $B_p=1$ $\forall p$ constraint define the flux-free subspace
\begin{equation}
 \mathcal H_{\mathrm{ff}}
 =\operatorname{span}\{\ket\psi\in \mathbb C[G]^{\otimes N_e}:B_p\ket\psi=\ket\psi
   \ \text{for every }p\}.
\end{equation}
Within this subspace, the ordered group product around every contractible plaquette is the identity.  We refer to this zero-flux condition on each plaquette, and hence
on every contractible path, as flatness.

Trivial holonomy around contractible paths does not imply trivial holonomy around the cylinder; the holonomy around the cylinder will label the $|\text{Conj}(G)|$ degenerate ground states. Specifically, for a state $|\{g_e\}\rangle$ and a path $\gamma$ that winds once around the cylinder, 
\begin{equation}
 g_\gamma
 =
 g_{e_1}^{\epsilon_1}
 g_{e_2}^{\epsilon_2}
 \cdots
 g_{e_\ell}^{\epsilon_\ell}.
\end{equation}
The multiplication order follows the traversal of $\gamma$
and is essential when $G$ is non-Abelian. For each $g_{e_k}$, $\epsilon_k=1$ if the direction the edge is traversed matches its orientation and $\epsilon_k=-1$ otherwise. The basepoint $v_0$ is the vertex between $e_1$ and $e_{\ell}$, and for non-Abelian groups, $g_{\gamma}$ can actually change under the action of $A_{v_0}^h$:
\[
A_{v_0}^h:\qquad g_\gamma\longmapsto h\,g_\gamma h^{-1}.
\]
It follows that projecting onto $A_v=1$ subspace produces one ground state - the equal-weight superposition state - for each such conjugacy class of $G$, as we mentioned earlier. We can span the ground state space by
\begin{equation}
 \begin{aligned}
 \cH_0=\operatorname{span}\{\ket{[g]}:[g]\in\operatorname{Conj}(G)\},
 \end{aligned}
 \label{eq:cyl-GSD}
\end{equation}
Later, we will often refer to a conjugacy class state as $|C\rangle$ for $C\in\mathrm{Conj}(G)$ to simplify notation. The second equality in (\ref{quditdim}) follows from character theory.  It is
made explicit by the orthonormal character basis
\begin{equation}
 \ket{\Gamma}_{\chi}
 =\frac1{\sqrt{|G|}}
 \sum_{[g]}\sqrt{|[g]|}\,\chi_\Gamma(g)\ket{[g]},
 \quad \Gamma\in\Irr(G).
 \label{eq:cyl-character-basis}
\end{equation}
where $\chi_{\Gamma}(g)=\mathrm{Tr}(\Gamma(g))$ is the character of the irrep $\Gamma$ evaluated on $g$. Character orthogonality gives
$\prescript{}{\chi}{\bra\Gamma}\Gamma'\rangle_\chi=\delta_{\Gamma\Gamma'}$,
so Eq.~\eqref{eq:cyl-character-basis} is a unitary change of basis between conjugacy classes and irreps.

We will study the example of $G=D_4$ in depth. There are eight elements: a $\mathbb{Z}_4$ rotation $r$ and a $\mathbb{Z}_2$ reflection $s$: $D_4=\langle r,s|r^4=s^2=1, srs=r^{-1}\rangle$. In this case, the conjugacy classes are
\begin{equation}
\begin{aligned}
 [1]   &= \{1\},
 & [r^2] &= \{r^2\},\\
 [r]   &= \{r,r^3\},
 & [s]   &= \{s,r^2s\},\\
 [rs]  &= \{rs,r^3s\}.
\end{aligned}
\label{eq:D4-conjugacy-classes}
\end{equation}

There are four one-dimensional representations \(A_{xz}\), labeled by \(x,z\in\mathbb Z_2\), and one two-dimensional representation \(E\) of $D_4$. A convenient choice is
\begin{equation}
A_{xz}(r)=(-1)^x,\qquad A_{xz}(s)=(-1)^z,
\label{abelianirreps}
\end{equation}
and
\begin{equation}
E(r)=
\begin{pmatrix}
0&-1\\
1&0
\end{pmatrix},
\qquad
E(s)=
\begin{pmatrix}
1&0\\
0&-1
\end{pmatrix}.
\label{eq:D4-real-E-representation}
\end{equation}

Since $D_4$ has five conjugacy classes and five irreps, 
\begin{equation}
 \operatorname{GSD}_{D_4}=5.
\end{equation}
on the cylinder.

A normalized conjugacy-class basis for the ground space is
\begin{equation}
\begin{aligned}
 \lvert[1]\rangle &= \lvert1\rangle_{\rm hol},
 &\lvert[r^2]\rangle &= \lvert r^2\rangle_{\rm hol},\\
 \lvert[r]\rangle
 &=\frac{\lvert r\rangle_{\rm hol}+\lvert r^3\rangle_{\rm hol}}{\sqrt2},
 &\lvert[s]\rangle
 &=\frac{\lvert s\rangle_{\rm hol}+\lvert r^2s\rangle_{\rm hol}}{\sqrt2},\\
 \lvert[rs]\rangle
 &=\frac{\lvert rs\rangle_{\rm hol}+\lvert r^3s\rangle_{\rm hol}}{\sqrt2}.
\end{aligned}
\label{eq:cyl-D4-class-basis}
\end{equation}
Here \(\lvert g\rangle_{\mathrm{hol}}\) denotes a superposition of all flux-free states
with holonomy \(g\) at a fixed basepoint (before averaging over the conjugation
action at the basepoint).

Since all of the $A_v$ and $B_p$ operators commute, the projector onto one of the five ground states is given by
\begin{equation}
    \rho_{L}=\prod_vA_v\prod_pB_p\cdot P_{L}
\end{equation}
where $L\in\mathrm{Conj}(G)$ and the projector $P_L$ is a global projector specifying that state.

\subsection{Logical operators}
\label{sec:cyl-charge-lines}

In this section, we define generalized $Z$ and $X$ logical operators for the quantum double on the cylinder. The $Z$ operators wrap around the periodic direction of the cylinder and are analogs of the $\prod_eZ_e$ operator in the Toric code. The basis in Eq.~(\ref{eq:cyl-GSD}) are eigenstates of these operators. On the other hand, the $X$ logical operators are analogs of the $\prod_eX_e$ operator between the top and bottom boundaries in the Toric code.

To define the logical $Z$ operators, we again choose a path $\gamma$ that winds once around the cylinder traversing edges $e_1,e_2,\dots e_\ell$. We place one copy of $Z_e^\Gamma$ on every edge of $\gamma$ and contract their virtual indices in the same order.\footnote{Instead, keeping the two virtual indices at the basepoint $v_0$ open defines
\begin{equation}
 \begin{aligned}
 \mathrm Z_{ab}^{\Gamma}(\gamma)
 &\equiv
 {}_{\rm virt}\!\bra{a}
 Z_{e_1}^\Gamma Z_{e_2}^\Gamma
 \cdots Z_{e_\ell}^\Gamma
 \ket{b}_{\rm virt}
 \\[-2pt]
 &=\vcenter{\hbox{
 \begin{tikzpicture}[scale=0.55]
   \draw[line width=0.35mm,red]
     (-0.45,0.35)--(4.8,0.35);

   \foreach \x/\lab in
     {0/{e_1},1.25/{e_2},3.45/{e_\ell}}{
     \draw[fill=white]
       (\x,0) rectangle +(0.90,0.70);
     \node[scale=0.68] at (\x+0.45,0.35)
       {$Z^\Gamma_{\lab}$};

     \draw (\x+0.45,0.70)--(\x+0.45,1.08);
     \draw (\x+0.45,0)--(\x+0.45,-0.38);
   }

   \node[fill=white,inner sep=1pt]
     at (2.72,0.35) {$\cdots$};
   \node[scale=0.75] at (-0.67,0.35) {$a$};
   \node[scale=0.75] at (5,0.35) {$b$};
 \end{tikzpicture}}}.
 \end{aligned}
 \label{eq:cyl-framed-Z}
\end{equation}

On an edge-basis state, this operator reads the holonomy directly:
\begin{equation}
 \mathrm Z_{ab}^{\Gamma}(\gamma)
 \ket{\{g_e\}}
 =
 [\Gamma(g_\gamma)]_{ab}\ket{\{g_e\}}.
 \label{eq:cyl-open-charge-action}
\end{equation}
In particular, on a state with based holonomy $g$, $
 \mathrm Z_{ab}^{\Gamma}(\gamma)
 \ket{g}_{\mathrm{hol}}
 =
 [\Gamma(g)]_{ab}\ket{g}_{\mathrm{hol}}$. A vertex action $A_{v_0}^h$ at the basepoint as mentioned before sends $
g\longmapsto hgh^{-1}$
and therefore conjugates the virtual matrix $
\Gamma(g)\longmapsto
\Gamma(h)\Gamma(g)\Gamma(h)^{-1}$. An individual matrix element is consequently not well defined on a conjugacy-class state.} The logical $Z^\Gamma$ operator is obtained by taking
the trace of this product:
\begin{equation}
 \begin{aligned}
 \mathrm W_\Gamma(\gamma)
 &\equiv
 \operatorname{Tr}_{\text{virt}}
 \left(
 Z_{e_1}^\Gamma Z_{e_2}^\Gamma
 \cdots Z_{e_\ell}^\Gamma
 \right)
 =
 \sum_a\mathrm Z_{aa}^{\Gamma}(\gamma)
 \\[-2pt]
 &=\vcenter{\hbox{
 \begin{tikzpicture}[scale=0.55]
   \draw[line width=0.35mm,red]
     (-0.45,0.35)--(4.35,0.35);

   \draw[red,line width=0.35mm,rounded corners=4pt]
     (-0.45,0.35)
     --(-0.72,0.35)
     --(-0.72,-0.82)
     --(5.07,-0.82)
     --(5.07,0.35)
     --(4.80,0.35);
    \draw[red,line width=0.35mm] (4.35,0.35)--(4.80,0.35);
   \foreach \x/\lab in
     {0/{e_1},1.25/{e_2},3.45/{e_\ell}}{
     \draw[fill=white]
       (\x,0) rectangle +(0.90,0.70);
     \node[scale=0.68] at (\x+0.45,0.35)
       {$Z^\Gamma_{\lab}$};

     \draw (\x+0.45,0.70)--(\x+0.45,1.08);
     \draw (\x+0.45,0)--(\x+0.45,-0.38);
   }

   \node[fill=white,inner sep=1pt]
     at (2.72,0.35) {$\cdots$};
 \end{tikzpicture}}}.
 \end{aligned}
 \label{eq:cyl-W-definition}
\end{equation}

These charge lines $W_{\Gamma}$ act diagonally on the states $|[g]\rangle$: 
\begin{equation}
 \begin{aligned}
 \mathrm W_\Gamma\ket{[g]}
 &=\chi_\Gamma(g)\ket{[g]},\\
 \frac{\mathrm W_\Gamma}{d_\Gamma}\ket{[g]}
 &=\frac{\chi_\Gamma(g)}{d_\Gamma}\ket{[g]}.
 \end{aligned}
 \label{eq:cyl-W-class-action}
\end{equation}
where again, $\chi_{\Gamma}(g)$ is the character of the irrep $\Gamma$, which is invariant within a conjugacy class (since $\mathrm{Tr}(\Gamma(g))=\mathrm{Tr}(\Gamma(h)\Gamma(g)\Gamma(h)^{-1})$. 

In the irrep basis, the same charge lines act by irrep fusion.
Character multiplication gives
\begin{equation}
 \mathrm W_\Gamma\ket{\Lambda}_\chi
 =
 \sum_{\Omega\in\Irr(G)}
 N_{\Gamma\Lambda}^{\Omega}
 \ket{\Omega}_\chi,
 \label{eq:cyl-W-fusion-action}
\end{equation}
where $N_{\Gamma\Lambda}^{\Omega}$ are integer fusion multiplicities. For the case of $D_4$, this becomes
\begin{align}
 \mathrm W_{A_{xz}}\ket{A_{x'z'}}_\chi
 &=\ket{A_{x+x',z+z'}}_\chi,
 \nonumber\\
 \mathrm W_{A_{xz}}\ket E_\chi
 &=\ket E_\chi,
 \label{eq:cyl-D4-A-fusion}\\
 \mathrm W_E\ket{A_{xz}}_\chi
 &=\ket E_\chi,
 \nonumber\\
 \mathrm W_E\ket E_\chi
 &=\sum_{x,z\in\mathbb Z_2}
   \ket{A_{xz}}_\chi,
 \label{eq:cyl-D4-E-fusion}
\end{align}
where addition in the first line is modulo two.

The logical $X$ operators consist of flux ribbon operators that go from the top to the bottom boundary on the dual lattice. We denote the ribbons drawn on the two sides of the identified seam by $\mathrm L_{[h]}$ and $\mathrm R_{[h]}$ (see Fig.~\ref{fig:stabilizers-logicals-cylinder}) (but they can be along any path from the top boundary to the bottom boundary). In order to define these conjugation invariant operators, we choose a basepoint and define the operators $\mathcal M_{\rho_L}^{h}$ and $\mathcal M_{\rho_R}^{h}$ as 
\begin{equation}
 \mathcal M_{\rho_L}^{h}\ket{g}_{\mathrm{hol}}
 =\ket{hg}_{\mathrm{hol}},
 \qquad
  \mathcal M_{\rho_R}^{h}\ket{g}_{\mathrm{hol}}
 =\ket{gh^{-1}}_{\mathrm{hol}}.
 \label{eq:cyl-framed-shift-action}
\end{equation}

For a non-Abelian group, the ribbon is not generally obtained by applying
the same $h$ shift to every crossed edge. Instead, its flux label is parallel
transported along the ribbon so that every bulk plaquette remains flux free.
The explicit construction and a direct proof that all stabilizers are
preserved are given in Appendix~\ref{app:flux-ribbon-proof}.

As with before, $\mathcal M_{\rho_L}^{h}$ and $\mathcal M_{\rho_R}^{h}$ still depend on the choice of basepoint.  A vertex action $A_{v_0}^{q}$ at the basepoint conjugates the ribbon label,
\begin{equation}
 A_{v_0}^{q}\mathcal M_{\rho}^{h}
 \bigl(A_{v_0}^{q}\bigr)^{-1}
 =
 \mathcal M_{\rho}^{qhq^{-1}}.
 \label{eq:cyl-ribbon-conjugation}
\end{equation}
The physical operator is therefore obtained by averaging over the
conjugacy class of $h$:
\begin{align}
 \mathrm L_{[h]}
 &\equiv
 \frac{1}{|[h]|}\sum_{k\in[h]}\mathcal M_{\rho_L}^{k},
 \label{eq:cyl-physical-L}\\
 \mathrm R_{[h]}
 &\equiv
 \frac{1}{|[h]|}\sum_{k\in[h]}\mathcal M_{\rho_R}^{k}.
 \label{eq:cyl-physical-R}
\end{align}
For $h$ in the center of $G$ (i.e. commuting with all elements of $G$), these operators do not include any sum, but otherwise they need not be unitary. These operators act on the ground space, and since in this space $\mathrm R_{[h]}
    =\mathrm L_{[h^{-1}]}
    =\mathrm L_{[h]}^\dagger$, it is sufficient to retain the single family
$\{\mathrm L_{[h]}\}$. 
In the irrep basis of the ground state space, the operators act diagonally:
\begin{equation}
 \mathrm L_{[h]}\ket{\Lambda}_{\chi}
 =
 \frac{\chi_\Lambda(h^{-1})}{d_\Lambda}
 \ket{\Lambda}_{\chi}
 \label{eq:cyl-shift-eigenvalues}
\end{equation}

\begin{table}[t]
\caption{Eigenvalue
$\ell_\Lambda([h])=\chi_\Lambda(h)/d_\Lambda$ of the normalized physical
$D_4$ shift ribbon on the different ground states.  All $D_4$ characters are real, so the $\mathrm L$ and
$\mathrm R$ eigenvalues agree.}
\label{tab:cyl-D4-shifts}
\begin{ruledtabular}
\begin{tabular}{c|ccccc}
$\Lambda$ & $[1]$ & $[r^2]$ & $[r]$ & $[s]$ & $[rs]$\\
\hline
$A_{00}$ & $1$ & $1$ & $1$ & $1$ & $1$\\
$A_{10}$ & $1$ & $1$ & $-1$ & $1$ & $-1$\\
$A_{01}$ & $1$ & $1$ & $1$ & $-1$ & $-1$\\
$A_{11}$ & $1$ & $1$ & $-1$ & $-1$ & $1$\\
$E$      & $1$ & $-1$ & $0$ & $0$ & $0$
\end{tabular}
\end{ruledtabular}
\end{table}

\subsection{The central logical qudit}
\label{sec:cyl-central-subspace}

In much of the paper, we will focus on the decoding of a subspace of the full ground state manifold on a cylinder. This subspace is associated with the center of the group. For any finite group $G$, the center $Z(G)$ (an abelian group consisting of all elements that commute with all other group elements) naturally form a set of $|Z(g)|$ singleton conjugacy classes, giving a subspace of size $|Z(G)|$. On this subspace, consider the case when $|Z(G)| = \mathbb{Z}_N$, the logical $X$ operators act simply as $Z_N$ $X$ qudit operators:
\begin{equation}
 \mathrm L_{[h]}|[g]\rangle = |[hg]\rangle
\end{equation}
where $\mathrm L_{[h]}$ is a single operator without any conjugacy class averaging. Similarly for any irrep $\Gamma$, Schur's lemma says $\Gamma(g)=[\chi_\Gamma(g)/d_\Gamma]\mathbf{1}_{d_{\Gamma}}$ for $g\in Z(G)$, with $\chi_{\Gamma}(g)/d_{\Gamma}\in U(1)$. So  $\mathrm W_{\Gamma}/d_{\Gamma}$ acts simply as a $\mathbb{Z}_N$ $Z$ qudit operator by (\ref{eq:cyl-W-class-action}). The central qudit therefore has a structure similar to abelian topological orders, and therefore admits an ML recovery of the familiar type, where we just compare weights of $\mathbb{Z}_N$ homology classes. The difference is that the noise may correspond to non-Abelian anyons, so clean-up of the central qudit gives a clear example of how more familiar ML recovery can be enhanced by fusion measurements. Recovery of the more general logical space may use full optimal recovery logical channels with coherent operations in the logical space, which is beyond ML decoders of this type. 

For $D_4$, the ground state space is $\mathbb{C}^2\oplus \mathbb{C}^3$ and we focus on decoding the $\mathbb{C}^2$ central qubit. Specifically, the central elements $1$ and $r^2$ of $D_4$ each form a singleton conjugacy class,
$[1]=\{1\}$ and $[r^2]=\{r^2\}$. We therefore have
\begin{equation}
 \cH_c=\operatorname{span}\{\ket{[1]},\ket{[r^2]}\}
 \subset\cH_0.
 \label{eq:cyl-central-qubit}
\end{equation}
The normalized $E$ Wilson loop acts as
\begin{equation}
 \frac{\mathrm W_E}{2}\ket{[1]}=\ket{[1]},
 \qquad
 \frac{\mathrm W_E}{2}\ket{[r^2]}=-\ket{[r^2]},
 \label{eq:cyl-central-Z}
\end{equation}
and is therefore the Pauli $Z$ operator on $\cH_c$. $\mathrm L_{[r^2]}=\mathrm R_{[r^2]}$ exchanges $\ket{[1]}$ with $\ket{[r^2]}$, so is the Pauli $X$ on
$\cH_c$.

\subsection{Charge decoherence channel}
Having fixed the code space and its logical operators, we now introduce
noise that creates electric charge of irrep $\Gamma$. The
operator $Z_{ab,e}^{\Gamma}$ is a single-edge segment of an electric
$\Gamma$-ribbon: it commutes with every plaquette projector, but can violate
the vertex constraints at the two endpoints of $e$. An incoherent mixture
of these ribbon segments therefore proliferates electric charges without
creating magnetic flux, which is why we call the resulting channel charge
decoherence~\cite{ellison2025toward,sun2025anomalous,schafernameki2025symtft,song2026strong,sakura}. Therefore, the $\prod_pB_p$ part of the density matrix remains unchanged.

For an irreducible representation
$\Gamma\in\operatorname{Irr}(G)$ of dimension $d_\Gamma$, the local charge
decoherence channel we consider on edge $e$ is
\begin{equation}
 \mathcal N_{e,\Gamma}(\rho)
 =(1-p)\rho+\frac{p}{d_\Gamma}\sum_{a,b=1}^{d_\Gamma}
 Z^\Gamma_{ab,e}\rho Z^{\bar{\Gamma}}_{ab,e}.
 \label{ChargeDecoherenceChannel}
\end{equation}
Here $Z^{\bar\Gamma}_{ab,e}=(Z^{\Gamma}_{ab,e})^\dagger$, where $\bar\Gamma$ is the complex-conjugate representation. The factor
$1/d_\Gamma$ ensures trace preservation because
\begin{equation}
 \sum_{a,b}Z^{\bar\Gamma}_{ab,e}Z^\Gamma_{ab,e}
 =d_\Gamma\,\mathbbm 1_e,
\end{equation}
which follows immediately from unitarity of $\Gamma(g)$ for every $g\in G$.

The most useful way to view Eq.~\eqref{ChargeDecoherenceChannel} is as a Schur multiplier in the group basis.  Since
\begin{equation}
 Z^\Gamma_{ab}\ket g=\Gamma(g)_{ab}\ket g,
\end{equation}
we have exactly
\begin{align}
\begin{split}
 \mathcal N_{e,\Gamma}(\ket g\!\bra h)
 &=K_{\Gamma,p}(gh^{-1})\ket g\!\bra h,\\
 K_{\Gamma,p}(x)
 &=(1-p)+\frac{p}{d_\Gamma}\chi_\Gamma(x).
 \end{split}\label{eq:single-charge-kernel}
\end{align}
Thus all dependence on the representation enters through its character.
Notice that $K_{\Gamma,p}(1)=1$, as required for preservation of the
diagonal matrix elements.

The full noise process
applies $\mathcal N_{e,\Gamma}$ independently on every edge. Its occupied edges
form $\Gamma$-charge ribbons: open endpoints carry $\Gamma$, while closed components and junctions may carry fusion outcomes of adjacent $\Gamma$ anyons.  In Appendix~\ref{app-current-rep}, we show a result that may be of independent interest: this channel for irreps of certain non-Abelian groups admits a rewriting in terms of a mixture of unitaries acting on the state with some probability. This occurs even when the $Z^\Gamma_{ab,e}$ operators are not unitary. 

\section{Three physical measurement records}
\label{sec:syndrome-decoding-main}

In this section, we consider the charge-decoherence channel defined in Eq.~\eqref{ChargeDecoherenceChannel} and develop ML decoders based on three possible sets of syndromes, which provide progressively refined syndrome information. Starting from the ground-state density matrix
$\rho_L$ in logical state $L$, we apply the channel
independently on every edge:
\begin{equation}
 \rho_{L,p}
 =\left(\prod_{e\in\mathcal E}\mathcal N_{e,\Gamma}\right)
  \!\left[\rho_L\right]=\mathcal{N}_{\Gamma,p}[\rho_L].
 \label{eq:sectionIII-noisy-state}
\end{equation}
An ML decoder has access to the noise model and computes a Born probability 
\begin{equation}
 \Pr_p(\mathcal{S})=\Tr\!\left[P_\mathcal{S}\rho_{p,L}\right].
 \label{eq:sectionIII-Born-rule}
\end{equation}
for any syndrome record $\mathcal{S}$, where $P_{\mathcal{S}}$ is the syndrome projector. It then also needs to be able to divide the full $\mathrm{Pr}_p(\mathcal{S})$ into equivalence classes based on logical operators, to determine the correct recovery. 

We will begin with giving some intuition for how the refined syndrome sets build on top of each other and help with error correction, using the concrete example of decoherence in the $D_4$ quantum double. Then we will provide technical tools for translating the intuition to a concrete decoder, by computing the syndrome probabilities for the different syndrome sets for the quantum double of any group $G$ with decoherence under any irrep $\Gamma$. We derive the syndrome probabilities by constructing the syndrome projectors and evaluating the trace explicitly. While the general syndrome probabilities are complicated expressions in terms of irrep dimensions and characters,  they simplify in the case of $G=D_4$ and $\Gamma=E$ to classical partition functions with clear physical interpretations. We will then use these results in Sec.~\ref{sec:decoding} for numerically evaluating thresholds.

The explicit expressions for the syndrome projectors also reveal an interesting physical phenomenon that only occurs in non-Abelian topological orders: fusion measurements, which are not global (i.e. don't wind around the cylinder), can nevertheless disturb logical information. The last part of this section studies this phenomenon in detail and shows that the central qudit always remains intact under these fusion measurements, so we can still safely apply ML decoding on the central qudit. 

\subsection{Motivation for refined syndromes}

We begin by studying the structure of refined syndrome sets in a simple example of the $D_4$ quantum double, giving physical intuition for how these measurements increase the decoding threshold through concrete examples where a decoder with coarser information cannot correctly guess a correct recovery but a decoder with finer information succeeds. After applying the noise channel in Eq.~(\ref{ChargeDecoherenceChannel}), we consider three syndrome sets $\mathcal{S}^{(1)},\mathcal{S}^{(2)},$ and $\mathcal{S}^{(3)}$. Each syndrome set has progressively more information:
\begin{equation}
 \mathcal R(\rho)=\sum_\mathcal{S}\mathcal R_\mathcal{S}\!\left(P_\mathcal{S}\rho P_\mathcal{S}\right),
 \label{eq:syndrome-conditioned-recovery}
\end{equation}
where $P_\mathcal{S}$ is the projector onto the syndrome $\mathcal{S}$ and
$\mathcal R_\mathcal{S}$ depends only on that classical outcome. We compare three nested records for decoding the central qubit:
\begin{enumerate}
 \item $\mathcal S^{(1)}=\gamma$ records whether each vertex carries the
 two-dimensional charge $E$.
 \item $\mathcal S^{(2)}=(\gamma,\mathbf a)$ records the complete onsite
 irrep, including the four Abelian charges at vertices outside $\gamma$.
 \item $\mathcal S^{(3)}=(\gamma,\mathbf a,\mathbf q)$ additionally records
 the fusion outcomes of disjoint $E$ anyon pairs under a deterministic, i.e. minimal-weight perfect matching (MWPM) pairing of the $E$ anyons. 
\end{enumerate}

Note that in this case, $\mathcal{S}^{(1)}$ is sufficient to recover the central qubit because the remaining errors after $E$ anyon cleanup are all abelian, and cannot fuse to new $E$ anyons. Note also that $\mathcal{S}^{(3)}$ above takes advantage of the fact that in $D_4$, the entire fusion tree is determined by pairwise $E$ anyon fusion when there are only charge anyons. This is because the fusion outcome of two $E$ anyons is always abelian and the rest of the fusion tree is completely deterministic. 

We will now give intuition for how decoders based on the above three syndrome sets should have increasingly higher thresholds. In particular, each syndrome set should get increasingly better at determining a recovery that, composed with the actual error, does not give a logical $\mathrm{W}_E$ operator that winds around the cylinder. 

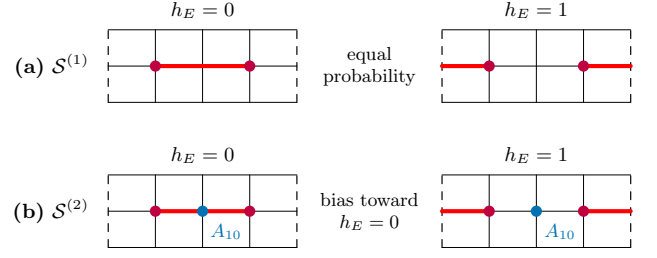
\begin{figure}[t]
\centering
\begin{tikzpicture}[
    x=0.48cm,y=0.48cm,
    line cap=round,line join=round
]
  \tikzset{
    lattice/.style={black,line width=0.35pt},
    seam/.style={black,dashed,line width=0.45pt},
    history/.style={red,line width=1.55pt}
  }

  \node[anchor=east,scale=0.86] at (-0.32,5)
    {\textbf{(a)} $\mathcal S^{(1)}$};
  \begin{scope}[shift={(0,4)}]
    \draw[lattice] (0,0)--(5.2,0) (0,1)--(5.2,1) (0,2)--(5.2,2);
    \foreach \x in {1.3,2.6,3.9}{\draw[lattice] (\x,0)--(\x,2);}
    \draw[seam] (0,0)--(0,2) (5.2,0)--(5.2,2);
    \draw[history] (1.3,1)--(3.9,1);
    \foreach \x in {1.3,3.9}{\fill[purple] (\x,1) circle (2.1pt);}
    \node[scale=0.84] at (2.6,2.55) {$h_E=0$};
  \end{scope}
  \node[align=center,scale=0.79] at (7.2,5)
    {equal\\ probability};
  \begin{scope}[shift={(9.2,4)}]
    \draw[lattice] (0,0)--(5.2,0) (0,1)--(5.2,1) (0,2)--(5.2,2);
    \foreach \x in {1.3,2.6,3.9}{\draw[lattice] (\x,0)--(\x,2);}
    \draw[seam] (0,0)--(0,2) (5.2,0)--(5.2,2);
    \draw[history] (0,1)--(1.3,1) (3.9,1)--(5.2,1);
    \foreach \x in {1.3,3.9}{\fill[purple] (\x,1) circle (2.1pt);}
    \node[scale=0.84] at (2.6,2.55) {$h_E=1$};
  \end{scope}

  \node[anchor=east,scale=0.86] at (-0.32,1)
    {\textbf{(b)} $\mathcal S^{(2)}$};
  \begin{scope}[shift={(0,0)}]
    \draw[lattice] (0,0)--(5.2,0) (0,1)--(5.2,1) (0,2)--(5.2,2);
    \foreach \x in {1.3,2.6,3.9}{\draw[lattice] (\x,0)--(\x,2);}
    \draw[seam] (0,0)--(0,2) (5.2,0)--(5.2,2);
    \draw[history] (1.3,1)--(3.9,1);
    \foreach \x in {1.3,3.9}{\fill[purple] (\x,1) circle (2.1pt);}
    \fill[myBlue] (2.6,1) circle (2.1pt);
    \node[myBlue,scale=0.79] at (3.25,0.45) {$A_{10}$};
    \node[scale=0.84] at (2.6,2.55) {$h_E=0$};
  \end{scope}
  \node[align=center,scale=0.79] at (7.2,1)
    {bias toward\\$h_E=0$};
  \begin{scope}[shift={(9.2,0)}]
    \draw[lattice] (0,0)--(5.2,0) (0,1)--(5.2,1) (0,2)--(5.2,2);
    \foreach \x in {1.3,2.6,3.9}{\draw[lattice] (\x,0)--(\x,2);}
    \draw[seam] (0,0)--(0,2) (5.2,0)--(5.2,2);
    \draw[history] (0,1)--(1.3,1) (3.9,1)--(5.2,1);
    \foreach \x in {1.3,3.9}{\fill[purple] (\x,1) circle (2.1pt);}
    \fill[myBlue] (2.6,1) circle (2.1pt);
    \node[myBlue,scale=0.79] at (3.25,0.45) {$A_{10}$};
    \node[scale=0.84] at (2.6,2.55) {$h_E=1$};
  \end{scope}
\end{tikzpicture}
\caption{A schematic of how additional onsite charge information modifies the posterior distribution over homology classes and improves the threshold of maximum-likelihood decoders. The dashed sides are identified, the purple dots are $E$ anyons, the red lines
are representative error histories, and homology is defined relative to the
short reference closure.
\textbf{(a)} The coarse record $\gamma$
gives equal weight to the two displayed histories.  \textbf{(b)} Resolving the
intervening Abelian charge $A_{10}$ makes the direct history compatible with
an $E$-containing component, whereas the displayed winding history leaves an
isolated nontrivial charge in an $E$-free component.  This comparison concerns
representative low-weight histories; the full sector weights sum over all
compatible histories.}
\label{fig:onsite-syndrome-bias}
\end{figure}

The simplest example compares the coarse record $\mathcal S^{(1)}$ with the complete onsite charge record $\mathcal S^{(2)}$.  In Fig.~\ref{fig:onsite-syndrome-bias}, we show a toy example of a cylinder of circumference $4,$ and the two $E$ anyons are separated by half the circumference (of course, this is a highly unlikely syndrome in the thermodynamic limit, but it illustrates the advantage of $\mathcal{S}^{(2)}$ very clearly). The two error strings shown in Fig.~\ref{fig:onsite-syndrome-bias}(a), from the perspective of a decoder based on $\mathcal{S}^{(1)}$, are equally likely. This therefore gives a 50\% probability that the decoder chooses a recovery that incurs a logical $E$ error. In Fig.~\ref{fig:onsite-syndrome-bias}(b), we make the more refined onsite measurement that also resolves an intervening Abelian charge $A_{10}$. Given that the noise model only creates error strings of non-Abelian $E$ anyons, any Abelian anyon in the syndrome must arise at the junction of two non-Abelian anyon error strings. This allows the decoder to select the correct recovery, which passes through the $A_{10}$ anyon. This example shows how a decoder with access to $\mathcal{S}^{(2)}$ can outperform one with only access to $\mathcal{S}^{(1)}$. 

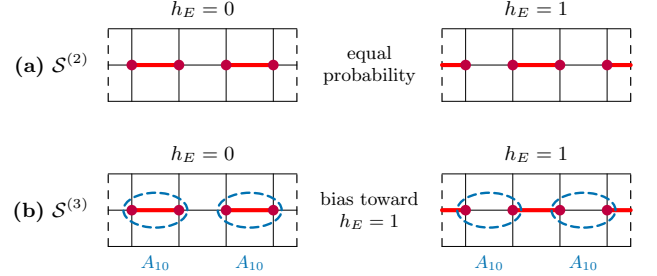
\begin{figure}[t]
\centering
\begin{tikzpicture}[
    x=0.48cm,y=0.48cm,
    line cap=round,line join=round
]
  \tikzset{
    lattice/.style={black,line width=0.35pt},
    seam/.style={black,dashed,line width=0.45pt},
    history/.style={red,line width=1.55pt},
    fusion/.style={myBlue,densely dashed,line width=0.90pt}
  }

  \node[anchor=east,scale=0.86] at (-0.32,5)
    {\textbf{(a)} $\mathcal S^{(2)}$};
  \begin{scope}[shift={(0,4)}]
    \draw[lattice] (0,0)--(5.2,0) (0,1)--(5.2,1) (0,2)--(5.2,2);
    \foreach \x in {0.65,1.95,3.25,4.55}{\draw[lattice] (\x,0)--(\x,2);}
    \draw[seam] (0,0)--(0,2) (5.2,0)--(5.2,2);
    \draw[history] (0.65,1)--(1.95,1) (3.25,1)--(4.55,1);
    \foreach \x in {0.65,1.95,3.25,4.55}{
      \fill[purple] (\x,1) circle (2.0pt);
    }
    \node[scale=0.84] at (2.6,2.55) {$h_E=0$};
  \end{scope}
  \node[align=center,scale=0.79] at (7.2,5)
    {equal\\ probability};
  \begin{scope}[shift={(9.2,4)}]
    \draw[lattice] (0,0)--(5.2,0) (0,1)--(5.2,1) (0,2)--(5.2,2);
    \foreach \x in {0.65,1.95,3.25,4.55}{\draw[lattice] (\x,0)--(\x,2);}
    \draw[seam] (0,0)--(0,2) (5.2,0)--(5.2,2);
    \draw[history] (0,1)--(0.65,1) (1.95,1)--(3.25,1)
                   (4.55,1)--(5.2,1);
    \foreach \x in {0.65,1.95,3.25,4.55}{
      \fill[purple] (\x,1) circle (2.0pt);
    }
    \node[scale=0.84] at (2.6,2.55) {$h_E=1$};
  \end{scope}

  \node[anchor=east,scale=0.86] at (-0.32,1)
    {\textbf{(b)} $\mathcal S^{(3)}$};
  \begin{scope}[shift={(0,0)}]
    \draw[lattice] (0,0)--(5.2,0) (0,1)--(5.2,1) (0,2)--(5.2,2);
    \foreach \x in {0.65,1.95,3.25,4.55}{\draw[lattice] (\x,0)--(\x,2);}
    \draw[seam] (0,0)--(0,2) (5.2,0)--(5.2,2);
    \draw[history] (0.65,1)--(1.95,1) (3.25,1)--(4.55,1);
    \foreach \x in {0.65,1.95,3.25,4.55}{
      \fill[purple] (\x,1) circle (2.0pt);
    }
    \draw[fusion] (1.30,1) ellipse (0.88 and 0.48);
    \draw[fusion] (3.90,1) ellipse (0.88 and 0.48);
    \node[myBlue,scale=0.75] at (1.30,-0.45) {$A_{10}$};
    \node[myBlue,scale=0.75] at (3.90,-0.45) {$A_{10}$};
    \node[scale=0.84] at (2.6,2.55) {$h_E=0$};
  \end{scope}
  \node[align=center,scale=0.79] at (7.2,1)
    {bias toward\\$h_E=1$};
  \begin{scope}[shift={(9.2,0)}]
    \draw[lattice] (0,0)--(5.2,0) (0,1)--(5.2,1) (0,2)--(5.2,2);
    \foreach \x in {0.65,1.95,3.25,4.55}{\draw[lattice] (\x,0)--(\x,2);}
    \draw[seam] (0,0)--(0,2) (5.2,0)--(5.2,2);
    \draw[history] (0,1)--(0.65,1) (1.95,1)--(3.25,1)
                   (4.55,1)--(5.2,1);
    \foreach \x in {0.65,1.95,3.25,4.55}{
      \fill[purple] (\x,1) circle (2.0pt);
    }
    \draw[fusion] (1.30,1) ellipse (0.88 and 0.48);
    \draw[fusion] (3.90,1) ellipse (0.88 and 0.48);
    \node[myBlue,scale=0.75] at (1.30,-0.45) {$A_{10}$};
    \node[myBlue,scale=0.75] at (3.90,-0.45) {$A_{10}$};
    \node[scale=0.84] at (2.6,2.55) {$h_E=1$};
  \end{scope}
\end{tikzpicture}
\caption{A schematic of how additional pair-fusion information modifies the posterior distribution over homology classes beyond the onsite charge record. The purple dots are $E$ anyons, and the blue dashed loops indicate the measured pairs $(1,2)$ and $(3,4)$, with fusion outcomes $q_{12}=q_{34}=A_{10}$. The two displayed error histories have the same onsite syndrome. For the $h_E=0$ history, adjoining the matching edges produces two components, each carrying a single nontrivial charge $A_{10}$, so this history is incompatible with the fusion record. For the $h_E=1$ history, the augmented graph forms one component whose total charge is $A_{10}\times A_{10}=A_{00}=1$, so this history remains compatible. Thus the fusion outcomes distinguish histories that the onsite record alone cannot distinguish. This comparison concerns representative low-weight histories; the full sector weights sum over all compatible histories.}
\label{fig:fusion-syndrome-bias}
\end{figure}

We will now show how a decoder based on $\mathcal{S}^{(3)}$ can further help pick a correct recovery. The four-$E$ example in Fig.~\ref{fig:fusion-syndrome-bias}(a) has two possible errors with
the same syndrome $\mathcal{S}^{(2)}$ but differing by an $E$ logical.  Now suppose that we pair the anyons as $(1,2)$ and $(3,4)$ and suppose that both pair-fusion measurements return
$A_{10}$.  Given that any closed component of error strings must have vacuum fusion, this gives evidence that the error strings actually pair the $E$ anyons as $(2,3)$ and $(4,1)$. The fusion record therefore distinguishes error histories that cannot be distinguished by onsite charge measurements.

These examples illustrate the general advantage of syndrome refinement. A decoder with finer syndrome data can always forget that information and decode according to coarser syndromes, so 
\begin{equation}
 p_c^{(1)}\le p_c^{(2)}\le p_c^{(3)}.
 \label{pcinequality}
\end{equation}
We cannot establish analytically that the inequalities are strict, i.e. that $p_c^{(1)}\neq p_c^{(2)}\neq p_c^{(3)}$, but the numerics in Sec.~\ref{sec:d4-central-numerics} show that the improvements are substantial.

\subsection{Onsite charge measurements}
\label{sec:charge-povm-corrected}

In this subsection we construct the two onsite records, which are analogs of $S^{(1)}$ and $S^{(2)}$ above for general quantum doubles with irrep $\Gamma$ decoherence. We derive the exact syndrome probabilities by defining charge projectors for a general quantum double and inserting them into the Born rule. We will then simplify the expression for $\Gamma=E$ in the $D_4$ quantum double to get a more explicit classical partition function representation that we can use for numerics.

We reserve $\Gamma$ for the irrep appearing in the decoherence channel
of Eq.~\eqref{ChargeDecoherenceChannel}. For a general irrep
$\lambda\in\operatorname{Irr}(G)$, the projector onto electric charge
$\lambda$ at vertex $v$ is
\begin{equation}
 P_v^\lambda=\frac{d_\lambda}{|G|}\sum_{h\in G}\chi_\lambda(h^{-1})A_v^h.
 \label{eq:central-charge-projector}
\end{equation}
We discuss the relationship between these projectors and fusion projectors in Sec.~\ref{pairfusionprojectors}, which are more involved.

These projectors are local, supported on the four edges around a vertex. They are also mutually orthogonal and complete:
\begin{equation}
 P_v^\lambda P_v^\mu=\delta_{\lambda\mu}P_v^\lambda,
 \qquad
 \sum_{\lambda\in\operatorname{Irr}(G)}P_v^\lambda=\mathbbm 1,
 \label{eq:charge-projector-completeness}
\end{equation}
by the character-convolution identity. They commute at distinct vertices.
For comparison, in the toric code these are simply the projectors onto the
$+1$ (trivial irrep) and $-1$ (nontrivial irrep) eigenvalues of the star stabilizer.

We now specialize the records to the self-dual $D_4$ charge $E$. The coarse
outcome $\mathcal S^{(1)}=\gamma$ records the set of vertices carrying $E$,
with projector
\begin{equation}
 P_\gamma
 =\prod_{v\in\gamma}P_v^E
  \prod_{v\notin\gamma}(\mathbbm 1-P_v^E).
 \label{eq:coarse-onsite-record-projector}
\end{equation}
The complete outcome $\mathcal S^{(2)}=(\gamma,\mathbf a)$ assigns an irrep
$\lambda_v$ to every vertex: $\lambda_v=E$ on $\gamma$ and
$\lambda_v=A_{x_vz_v}$ elsewhere. Its projector is
\begin{equation}
 P_{(\gamma,\mathbf a)}=\prod_vP_v^{\lambda_v}.
 \label{eq:complete-onsite-record-projector}
\end{equation}
As a sanity check, summing over the Abelian labels returns the coarse measurement associated with $\mathcal S^{(1)}$:
\begin{equation}
 \sum_{\mathbf a}P_{(\gamma,\mathbf a)}=P_\gamma.
 \label{eq:onsite-record-coarse-graining}
\end{equation}

\subsubsection{Onsite syndrome probabilities}
\label{sec:onsiteprob}

Below we will present the calculation for the onsite syndrome probabilities using the syndrome projectors in the previous section. We will first compute the syndrome probability for $\mathcal{S}^{(2)}$ using Eq.~(\ref{eq:central-charge-projector}), and then we will sum over $\mathbf{a}$ to get the syndrome probability for $\mathcal{S}^{(1)}=\gamma$. We will defer many details to Appendix~\ref{app:onsite-probability}. 

For the purpose of evaluating and simplifying the syndrome probabilities, it will be useful to expand out the quantum double ground state as a superposition of flux-free states, and compute the action of the channel (\ref{ChargeDecoherenceChannel}) using the Schur multiplier form in Eq.~(\ref{eq:single-charge-kernel}). Let $|C\rangle$ be the ground state associated
with conjugacy class $C$, and the set $\mathcal O_C$ of flat group-basis configurations
in this state, with $N_C=|\mathcal O_C|$. The encoded density matrix is
\begin{equation}
 \begin{aligned}
 \ket C&=\frac{1}{\sqrt{N_C}}
 \sum_{\mathbf g\in\mathcal O_C}\ket{\mathbf g},\\
 \rho_L&=\sum_{C,D\in\operatorname{Conj}(G)}
 \rho_{CD}\ket C\!\bra D.
 \end{aligned}
 \label{eq:general-logical-density}
\end{equation}
with $\sum_C\rho_{CC}=1$. The noisy state $\rho_{L,p}=\mathcal{N}_{\Gamma,p}[\rho_L]$ can be computed using 
\begin{equation}
 \begin{aligned}
 \mathcal N_{\Gamma,p}(\ket{\mathbf g}\!\bra{\mathbf k})
 &=\left[\prod_e K_{\Gamma,p}(g_ek_e^{-1})\right]
   \ket{\mathbf g}\!\bra{\mathbf k},
 \end{aligned}
 \label{eq:onsite-noisy-state-explicit}
\end{equation}
where $K_{\Gamma,p}(x)=(1-p)+\frac{p}{d_\Gamma}\chi_\Gamma(x)$ as in Eq.~\eqref{eq:single-charge-kernel}. Thus the channel changes coherences between states $|\mathbf{g}\rangle$ but does not change the
edge group labels themselves.

For the complete onsite record $\mathcal S^{(2)}=\{\lambda_v\}$,
the quantity to be evaluated is
\begin{equation}
\Pr_p(\mathcal S^{(2)}|\rho_{L}) =\Tr\!\left[P_{\mathcal S^{(2)}}\rho_{L,p}\right]
 \label{eq:onsite-probability-trace-main}
\end{equation}

Expanding the charge projectors in $P_{\mathcal{S}^{(2)}}$ in terms of $A_v^h$ operators as in Eq.~(\ref{eq:central-charge-projector}) introduces a group variable $h_v$ at every vertex, with weight $d_{\lambda_v}\chi_{\lambda_v}(h_v^{-1})/|G|$. The associated vertex actions at the two endpoints of an edge transform an oriented edge $e=\langle i,f\rangle$ as
\begin{equation}
    g_e\to g_e^{\mathbf{h}}=h_ig_eh_f^{-1}
\end{equation}
due to the $L^h_e$ and $R^h_e$ operators in $A_v^h$. The trace of $\rho_{L,p}$ with $P_{\mathcal{S}^{(2)}}$ therefore pairs $\mathbf{g}$ with $\mathbf{k}=\mathbf{g}^{\mathbf{h}}$. Because these vertex actions preserve the conjugacy class of the holonomy around the cylinder, the off diagonal matrix elements vanish: $\langle C|P_{\mathcal{S}^{(2)}}|D\rangle = 0$. Therefore, we only need to evaluate $\langle C|P_{\mathcal{S}^{(2)}}|C\rangle$.

To evaluate the contribution from a fixed sector $C$, we choose a representative group element $g_C$ in the conjugacy class $C\subseteq G$ and a dual cut joining the two boundaries of the cylinder. This cut can for example occur along the seam in Fig.~\ref{fig:stabilizers-logicals-cylinder}. Any flat configuration $\mathbf{g}\in \mathcal{O}_C$ can be parameterized by group elements $s_v\in G$ assigned to the vertices (intuitively, this comes from ungauging back to a theory with global $G$ symmetry) with edge labels 
\begin{equation}
    g_e=s_{i_e}g_C^{c_e}s_{f_e}^{-1}\quad\text{for every oriented edge }e=\langle i_ef_e\rangle
\end{equation}
where $c_e=0$ away from the cut and $c_e=\pm 1$ for an edge crossing it, with the sign fixed by orientation. So away from the cut, the edge labels simply take the form $s_{i_e}s_{f_e}^{-1}$ while the cut carries the group element $g_C^{c_e}$ that ensures the global holonomy is correct. To further simplify, we make a change of variables for the vertex charge projectors of $h_v\to r_v=s_v^{-1}h_vs_v$ (this is allowed since the $h_v$'s are summed over). Using $\mathbf{k}=\mathbf{g}^\mathbf{h}$, the argument of the noise kernel $K_{\Gamma,p}$ becomes 
\begin{equation}
    g_ek_e^{-1}=s_i(g_C^{c_e}r_fg_C^{-c_e}r_i^{-1})s_i^{-1}
\end{equation}
For simplicity, we will proceed with the calculation on the central qubit, so we can drop the $g_C^{c_e}$ terms by commuting them through $r_f$. We will therefore just write $\mathrm{Pr}_p(\mathcal{S}^{(2)})$ for shorthand, because there is no dependence on $L$ in the central qubit. We present the more general calculation in Appendix~\ref{app:onsite-probability}. Using the fact that both the noise kernel and the characters are invariant under conjugation, we get the simplified expression for the syndrome probability 
\begin{equation}
\begin{aligned}
 \Pr_p(\mathcal S^{(2)})
 &=\sum_C\rho_{CC}\,\mathcal Z_C,\\
 \mathcal Z_C
 &=\frac{1}{|G|^{N_v}}\sum_{\{r_v\in G\}}
   \prod_v d_{\lambda_v}\chi_{\lambda_v}(r_v^{-1})\\
 &\quad\times\prod_{e=\langle i_ef_e\rangle}
 K_{\Gamma,p}(r_{f_e}r_{i_e}^{-1}).
 \end{aligned}
 \label{eq:general-logical-projector-syndrome-probability}
\end{equation}

The vertex factor encodes the observed charge, and the edge factor encodes the noise. For general $G$ the individual character weights can be
complex due to the characters $\chi_{\lambda_v}$, although the complete Born probability is real and nonnegative.

Finally, the probability associated with the coarse record $\mathcal{S}^{(1)}$ is obtained by summing over all $\mathcal{S}^{(2)}$ with the same locations $\gamma$ of $\Gamma$ charges: 
\begin{equation}\label{ps1}
 \Pr_p(\mathcal S^{(1)}=\gamma)
 =\sum_{\mathbf a}\Pr_p(\mathcal S^{(2)}=(\gamma,\mathbf a)).
\end{equation}

\subsubsection{Statistical-mechanics representation}\label{statmechs1s2}

We now specialize to $G=D_4$ and $\Gamma=E$. Direct evaluation of the characters turns the syndrome probabilities for $\mathcal{S}^{(1)}$ and $\mathcal{S}^{(2)}$ into classical partition functions. Specifically, the syndrome probability associated with the coarse record $\mathcal{S}^{(1)}$ gives rise to an Ising-type model, while the complete onsite record $\mathcal{S}^{(2)}$ decorates the Ising model with additional labels coming from the 1d charges/abelian anyons.

We begin with Eq.~\eqref{ps1}, because this simplifies more easily. The four irreps other than $E$ are one-dimensional, and their characters obey
\begin{equation}
    \sum_{\lambda\neq E}\chi_\lambda(h)=4\mathbf{1}_{h=1,r^2}
\end{equation}
For the 2d irrep,
\begin{equation}
    \chi_E(h)=\begin{cases}2\qquad h=1\\ -2 \qquad h=r^2\\ 0\qquad\text{otherwise}\end{cases}
    \label{d4characters}
\end{equation}

Therefore, every $r_v$ giving a nonzero contribution to Eq.~\eqref{ps1} consists only of $\{1,r^2\}$. Writing $r_v=r^{2t_v}$ with $t_v\in\mathbb{Z}_2$, we can restrict $\sum_{\mathbf{r}}\to \sum_{t_v\in\mathbb{Z}_2}$. There is only a $-1$ factor when $\gamma_v=t_v=1$ (when there is an $E$ irrep and $r_v=r^2)$, so we get
\begin{align}
\begin{split}
    \Pr_p(\mathcal{S}^{(1)}=\gamma)&=\frac{1}{2^{N_v}}\sum_{\{t_v\}}(-1)^{\sum_v\gamma_vt_v}\\
    &\times \prod_{e=\langle i_ef_e\rangle}[(1-p)+p(-1)^{t_{i_e}+t_{f_e}}]
\end{split}
\end{align}
Let $F\subseteq\mathcal E$ be the set of edges for which we choose the second term. Then
\begin{align}
\begin{split}
    &\prod_{e\in\mathcal E}[(1-p)+p(-1)^{t_{i_e}+t_{f_e}}]\\
    &=\sum_{F\subseteq\mathcal E}(1-p)^{N_e-|F|}p^{|F|}(-1)^{\sum_{e\in F}(t_{i_e}+t_{f_e})}
\end{split}
\end{align}
Combining this with the $(-1)^{\sum_v\gamma_vt_v}$ term gives
\begin{align}
\begin{split}
    \Pr_p(\mathcal{S}^{(1)}=\gamma)&=\sum_{F\subseteq\mathcal E}(1-p)^{N_e-|F|}p^{|F|}\\
    &\times \frac{1}{2^{N_v}}\sum_{\{t_v\}}(-1)^{\sum_v\gamma_vt_v+\sum_{e\in F}(t_{i_e}+t_{f_e})}
\end{split}
\end{align}
Since every $t_v$ occurs once for every edge incident on $v$, $\sum_{e\in F}(t_{i_e}+t_{f_e})=\sum_{v\in\partial F}t_v$. The sum over $\{t_v\}$ then enforces $\partial F=\gamma$ so we get
\begin{equation}\label{sumF}
    \Pr_p\!\left(\mathcal{S}^{(1)}=\gamma\right)
=
\sum_{\substack{F\subseteq\mathcal E\\ \partial F=\gamma}}
(1-p)^{N_e-|F|}p^{|F|}.
\end{equation}

This syndrome probability can be mapped to the partition function of an Ising model with antiferromagnetic bonds that terminate on the syndrome~\cite{dennis2002topological}. Computing decoding quantities averaged over the syndrome distribution can be mapped to quenched disorder averages of the random bond Ising model along the Nishimori line.

We next evaluate the probability associated with the complete onsite record $\mathcal{S}^{(2)} = (\gamma,\bm a)$, where  $\gamma$ is again the set of vertices carrying $E$, and $\bm a$ records one of the four Abelian
irreps at every vertex outside $\gamma$. Eq.~\eqref{sumF} already gives the coarse probability for $\mathcal{S}^{(1)}$ as a sum over subgraphs $F$ satisfying $\partial F=\gamma$, with weights $(1-p)^{N_e-|F|}p^{|F|}$. Resolving $\mathbf{a}$ refines the same expansion:
\begin{equation}
    \mathrm{Pr}_p(\mathcal{S}^{(2)}=(\gamma,\mathbf{a}))=\sum_{\substack{F\subseteq\mathcal E\\ \partial F=\gamma}}(1-p)^{N_e-|F|}p^{|F|}\mathcal{A}_F(\mathbf{a})
\end{equation}
where $\mathcal{A}_F(\mathbf{a})$ is the probability of $\mathbf{a}$ given $F$. To evaluate this additional weight, we write the vertex variables as $r_v=r^{2t_v+u_v}s^{w_v}$ for $t_v,u_v,w_v\in\mathbb{Z}_2$. We again plug 
\begin{equation}
 K_{E,p}(r_fr_i^{-1})
 =
 (1-p)
 +
 p\,\frac{\chi_E(r_fr_i^{-1})}{2}.
 \label{eq:char-loop-edge-kernel}
\end{equation}
into Eq.~(\ref{eq:general-logical-projector-syndrome-probability}) and expand the product over edges by choosing either the
first or the second term in
Eq.~\eqref{eq:char-loop-edge-kernel} on every edge. For each edge in $F$,
\begin{equation}
    \frac{\chi_E(r_fr_i^{-1})}{2}=(-1)^{t_i+t_f}\delta_{u_i,u_f}\delta_{w_i,w_f}
\end{equation}
The sum over $t_v$ enforces the same condition $\partial F=\gamma$ as in the $\mathcal{S}^{(1)}$ calculation, and $\mathcal{A}_F(\mathbf{a})$ comes from the remaining delta functions enforcing $u_v,w_v$ to be constant on each connected component $B$ of $F$:
\begin{equation}
    u_v=u_B\quad w_v=w_B\quad\forall v\in B
\end{equation}
We will now divide $F$ into components $B$ and evaluate the contribution from each component to $\mathcal{A}_F(\mathbf{a})$.

$F$ has closed components, without any endpoints in $\gamma$, as well as open components, with endpoints in $\gamma$. For open components, any vertex in $B$ with $E$ charge has onsite weight
\begin{equation}
    d_E\chi_E(r_v^{-1})=4(-1)^{t_v}\delta_{u_v,0}\delta_{w_v,0}
\end{equation}
forcing $u_B=w_B=0$ for every $v\in B$. The 1d charge vertex contributes
\begin{equation}
\chi_{A_{x_v,z_v}}(r_v^{-1})=(-1)^{x_vu_v+z_vw_v}
\end{equation}
so on open components, abelian vertices always give $(-1)^{x_vu_B+z_vw_B}=1$ regardless of $x_v,z_v$. As a result, there is no constraint on Abelian labels on these components, reflecting the fusion rule $E\times A_{xz}=E$. On the other hand on closed components, $u_B,w_B$ remain free and the total abelian charge on such a component is
\begin{equation}
    X_B=\oplus_{v\in B}x_v\qquad Z_B=\oplus_{v\in B}z_v
\end{equation}
giving a remaining sum
\begin{equation}
    \sum_{u_B,w_B\in\mathbb{Z}_2}(-1)^{u_BX_B+w_BZ_B}=4\delta_{X_B,0}\delta_{Z_B,0}
\end{equation}
so these closed components differ by a factor of 4 and a total vacuum fusion constraint on the abelian anyons.

\begin{figure}[t]
 \centering
 \includegraphics[width=0.9\linewidth]{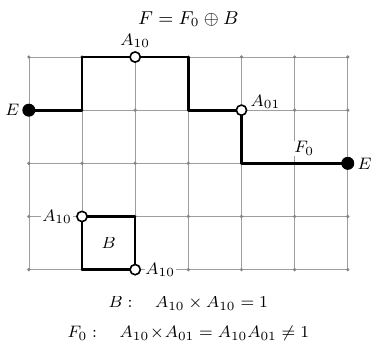}
 \caption{Edge configurations contributing to the complete onsite record $\mathcal{S}^{(2)}$. For a fixed set $\gamma$ of measured $E$ charges, we choose a reference configuration $F_0$ satisfying $\partial F_0=\gamma$. Every compatible configuration can then be written as $F=F_0\oplus B$, where $B$ consists of neutral components with only Abelian anyons. For example, on component $B$ shown, we indeed have $A_{10} \times A_{10} = 1$. In contrast, a component terminating on $E$ charges need not be Abelian-neutral because $E\times A\simeq E$ for every Abelian charge $A$; hence the open component $F_0$ may carry $A_{10}\times A_{01}=A_{11} \neq1$.}
 \label{fig:errorstring}
\end{figure}

Combining the above results for closed and open components gives the desired
loop-model representation:
\begin{align}
 \Pr_p\!\left(\mathcal{S}^{(2)}=(\gamma,\bm a)\right)
 &=
 4^{|\gamma|-N_v}
 \sum_{\substack{F\subseteq\mathcal E\\
                  \partial F=\gamma}}
 (1-p)^{N_e-|F|}
 p^{|F|}
 \nonumber\\
 &\quad\times
 4^{|\mathcal{B}_0(F)|}
 \prod_{B\in\mathcal B_0(F)}
 \delta_{X_B,0}
 \delta_{Z_B,0}.
 \label{eq:char-loop-final}
\end{align}
where $\mathcal{B}_0$ is the set of closed components. Note that the $\gamma=\emptyset$ syndrome gives a probability proportional to a partition function similar to the $\mathcal{O}(4)$ loop model. It is somewhat different because we are working on a square lattice, and closed components can have crossings where all four edges around a vertex are occupied. As we increase $|\gamma|$, there are more and more open components, leading to a reduced effect of the $4^{|\mathcal{B}_0(F)|}$ factor. 
As shown in Fig.~\ref{fig:errorstring}, we can summarize the physical picture of the syndrome probability as a sum over error configurations $F$ and Abelian decorations such that:
\begin{itemize}
    \item $F$ ends on $E$ anyons at vertices in $\gamma$: $\partial F=\gamma$
    \item All Abelian anyons lie on $F$ (vertices not in $F$ form isolated vertices which are elements of $\mathcal{C}_0(F)$ and thus can only carry the vacuum Abelian anyon type $A_{00}$)
    \item Abelian anyons on connected components without $E$ anyons must fuse together to the trivial irrep
    \item On connected components with $E$ anyons, the Abelian anyons can fuse to any 1d irrep
\end{itemize}

As a sanity check, we can explicitly sum Eq.~\eqref{eq:char-loop-final} over the
Abelian record $\bm a$. Indeed, the \(N_v-|\gamma|\) non-\(E\) vertices each admit four Abelian labels, while every \(E\)-free component imposes two independent \(\mathbb Z_2\) neutrality constraints, reducing the count by a factor of \(4\) per component and giving \(4^{N_v-|\gamma|-|\mathcal{C}_0(F)|}\). Thus for a fixed $F$, the number of
allowed Abelian records is
\begin{equation}
 4^{N_v-|\gamma|-|\mathcal{B}_0(F)|}.
 \label{eq:char-loop-record-count}
\end{equation}
This cancels the factor
$4^{|\gamma|-N_v+|\mathcal{B}_0(F)|}$ in
Eq.~\eqref{eq:char-loop-final}, leaving
\begin{equation}
 \Pr_p(\mathcal{S}^{(1)}=\gamma)
 =
 \sum_{\substack{F\subseteq\mathcal E\\
                  \partial F=\gamma}}
 (1-p)^{N_e-|F|}
 p^{|F|},
 \label{eq:char-loop-binary-check}
\end{equation}
which agrees with Eq.~(\ref{sumF}).

\begin{figure}[t]
 \centering
 \includegraphics[width=0.98\linewidth]{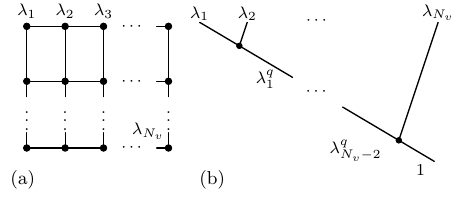}
 \caption{Information resolved by onsite and fusion measurements.  (a) The
 complete onsite record assigns an irrep $\lambda_v$ to every vertex.  (b) A
 fusion-tree measurement additionally resolves the intermediate total charges
 $\lambda_j^q$.  The final label is the vacuum because the channel creates
 trivial total electric charge.  The pair-fusion protocol used for $D_4$ measures a
 fixed, local subset of these intermediate labels.}
 \label{fig:fusiontree}
\end{figure}

\subsection{Anyon fusion measurements}
\label{sec:fusion-motivation-main}
The onsite record discussed in the previous subsection is complete as a set of
\emph{local} charge measurements, but it does not resolve the collective
fusion state of the non-Abelian anyons. In the $D_4$ quantum double, one has the fusion rule
\begin{equation}
 E\times E=A_{00}\oplus A_{01}\oplus A_{10}\oplus A_{11}.
 \label{eq:EE-fusion-main}
\end{equation}
This implies that an onsite record containing two $E$ charges does not determine which of
the four Abelian channels carries their total charge. A fusion measurement
supported on a region containing the pair reveals precisely this missing
collective information, as is shown schematically in Fig.~\ref{fig:fusiontree}. In the following, we first define the physical projectors associated with the fusion measurement in the $E$-charge decohered $D_4$ model, then derive the Born probability, and finally discuss its interpretation as a classical statistical-mechanics model.

Based on the fusion rule (\ref{eq:EE-fusion-main}) in the $D_4$ model, the outcome of every $E\times E$ fusion is Abelian, and the subsequent fusions of these resulting Abelian anyons are deterministic. Therefore, a deterministic pairing of all measured $E$ locations resolves a complete fusion-tree basis.
We choose the pairing by minimum-weight perfect matching, and we remark that this is a concrete measurement protocol, which may not be information-theoretically optimal.

With the extra information revealed by the fusion measurement, the complete syndrome is 
\begin{equation}
 \mathcal S^{(3)}=(\gamma,\mathbf a,\mathbf q).
 \label{eq:S3-record-main}
\end{equation}
where $\mathbf q=\{q_{ij}\}$ is the paired Abelian fusion outcomes.

\subsubsection{Pair-fusion charge projectors}
\label{pairfusionprojectors}

\begin{figure}[t]
\centering
\begingroup
\definecolor{pairFlux}{RGB}{126,58,153}
\definecolor{pairSpine}{RGB}{0,110,104}
\noindent\resizebox{\linewidth}{!}{%
\begin{tikzpicture}[
 x=2.15cm,y=1.90cm,
 lattice/.style={draw=black!65,line width=.45pt},
 shifted/.style={draw=pairFlux,line width=1.55pt},
 spine/.style={draw=pairSpine,line width=1.35pt},
 every node/.style={font=\normalsize,inner sep=1pt}
]
\foreach \y in {-1,0,1}{
 \foreach \x in {-1,0,1,2,3}{
  \draw[lattice] (\x,\y)--(\x+1,\y);
  \draw[lattice,->] (\x,\y)--(\x+.62,\y);
 }}
\foreach \x in {-1,0,1,2,3,4}{
 \foreach \y in {-1,0}{
  \draw[lattice] (\x,\y)--(\x,\y+1);
  \draw[lattice,->] (\x,\y)--(\x,\y+.62);
 }}
\foreach \x/\ell in {0/1,1/2,2/3}{
 \draw[spine] (\x,0)--(\x+1,0);
 \draw[spine,->] (\x,0)--(\x+.62,0);
 \node[text=pairSpine] at (\x+.5,.20)
   {$T_{s_{\ell},e_{\ell}}^{x_{\ell}}$};
}
\foreach \x in {0,1,2,3}{
 \draw[shifted] (\x,0)--(\x,1);
 \draw[shifted,->] (\x,0)--(\x,.62);
 \draw[shifted] (\x,-1)--(\x,0);
 \draw[shifted,->] (\x,-1)--(\x,-.38);
 \node[anchor=west,text=pairFlux] at (\x+.09,.76)
   {$L^{h_{\x}}$};
 \node[anchor=west,text=pairFlux] at (\x+.09,-.76)
   {$R^{h_{\x}}$};
}
\draw[shifted] (-1,0)--(0,0);
\draw[shifted,->] (-1,0)--(-.38,0);
\draw[shifted] (3,0)--(4,0);
\draw[shifted,->] (3,0)--(3.62,0);
\node[text=pairFlux] at (-.78,.20) {$R^{h_0}$};
\node[text=pairFlux] at (3.78,.20) {$L^{h_3}$};
\draw[pairFlux,line width=1pt,dash pattern=on 3.8pt off 2.6pt]
 (-.5,-.5) rectangle (3.5,.5);
\node[text=pairFlux] at (2.55,.63) {$\rho_p$};
\node[text=pairSpine] at (1.5,-.30) {$p_{i,j}$};
\foreach \x in {1,2}{
 \fill[black] (\x,0) circle (1.9pt);
}
\foreach \x in {0,3}{
 \filldraw[draw=black,fill=white,line width=.8pt] (\x,0) circle (4pt);
 \fill[black] (\x,0) circle (2.1pt);
}
\node[anchor=north west] at (.08,-.10) {$i=v_0$};
\node[anchor=north west] at (1.08,-.10) {$v_1$};
\node[anchor=north west] at (2.08,-.10) {$v_2$};
\node[anchor=north east] at (2.92,-.10) {$j=v_3$};
\node at (1.5,1.30)
 {$\mathcal F_{p_{i,j}}(h)$};
\end{tikzpicture}
}
\endgroup
\caption{Closed flux ribbon surrounding a pair of charges.
The transport path \(p_{i,j}\) joins the two circled
vertices $i$ and $j$.
The teal spine edges carry only diagonal
\(T_{s_{\ell},e_\ell}^{x_\ell}\) operators.
The product of the star operators gives the emanating left and right shift operators, with
\(h_\ell=X_\ell^{-1}hX_\ell\).
Edge subscripts on \(L\) and \(R\) are suppressed.
The figure shows one edge group label term; the full ribbon sums
over all $\{\mathbf{x}\}$.}
\label{fig:pair-ribbon-spine}
\end{figure}
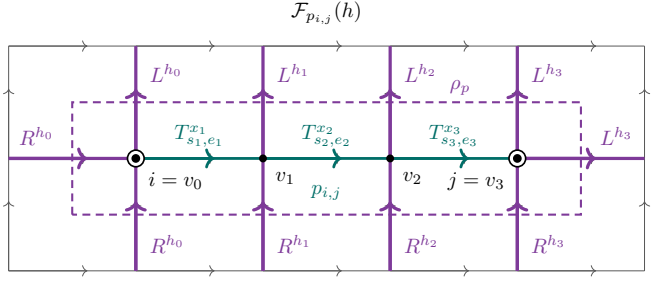

We now construct the observable that measures the total electric charge in
a narrow strip containing a selected pair.\footnote{We can measure the fusion outcome using an operator on any loop that encloses the pair, but we choose the minimal strip for simplicity and to avoid enclosing spurious anyons.} The fusion projectors follow a structure similar to the onsite charge projectors, but we must replace the operators $A_v^h$ with operators acting around the strip that encloses the pair we want to fuse. We will obtain
\begin{equation}
 P_{p_{i,j}}^\lambda
 =
 \frac{d_\lambda}{|G|}
 \sum_{h\in G}\chi_\lambda(h^{-1})\mathcal F_{p_{i,j}}(h).
 \label{app:pair-fusion-projector}
\end{equation}
where $\mathcal{F}_{p_{i,j}}(h)$ are ribbon operators enclosing the oriented pair $i,j\in \gamma$. When the strip reduces to a single site, we will recover $\mathcal{F}_{p_{i,j}}(h)=A_i^h$. One might naively take $\mathcal{F}_{p_{i,j}}(h)$ to be a product of $A_v^h$ inside the desired strip, but this does not quite work because there would be a nontrivial action on edges inside the strip (the green edges in Fig.~\ref{fig:pair-ribbon-spine}): they would get conjugated by $h$. Below, we will carefully define the ribbon operators $\mathcal{F}_{p_{i,j}}(h)$ that (1) leave the interior edges alone, and (2) multiply according to the group law like the $A_v^h$ operators. We do this by appropriately parallel-transporting the group element $h$ along the path $p_{i,j}$.

Specifically, we choose an oriented path $p_{i,j}$ from $i$ to $j$ and a narrow strip surrounding it, as shown in Fig.~\ref{fig:pair-ribbon-spine}. If an edge along the path has a value $x_e$, then an action $h$ at its initial vertex must be paired with $x_e^{-1}hx_e$ at its final vertex, so the total action leaves the edge unchanged:
\begin{equation}
    x_e\to hx_e(x_e^{-1}hx_e)^{-1}=x_e
\end{equation}
we repeat this prescription along the path to obtain operators that preserve the edges within the strip $p_{i,j}$ and shift the edges emanating from the strip.

More explicitly, we write the path as $v_0=i,\cdots v_m=j$ and we let $x_{\ell}$ be its $\ell$th edge read in the direction of the path. Then we define
\begin{equation}
    X_0=1\qquad X_{\ell}=x_1\cdots x_{\ell}
\end{equation}
The group element acting on vertex $v_{\ell}$ is then $h_{\ell}=X_{\ell}^{-1} h X_{\ell}$. The resulting ribbon operator $\mathcal{F}_{p_{i,j}}(h)$ has action controlled coherently by the path-edge labels. 

To make the above label-dependent action an explicit operator, we use the edge label projector $T^{x_{\ell}}_{s_{\ell},e_{\ell}}$ which projects into the edge label $x_{\ell}$ read in the direction of the path:
\begin{equation}
 T_{s_{\ell},e_\ell}^{x_\ell}=
 \begin{cases}
 T_{+,e_\ell}^{x_\ell},&\text{path follows the edge orientation},\\
 T_{-,e_\ell}^{x_\ell},&\text{path opposes the edge orientation}.
 \end{cases}
 \label{app:pair-path-projector}
\end{equation}
where 
\begin{equation}
 T_{+,e}^h|g\rangle_e=\delta_{h,g}|g\rangle_e,
 \qquad
 T_{-,e}^h|g\rangle_e=\delta_{h^{-1},g}|g\rangle_e.
 \label{eq:cyl-T-edge}
\end{equation}
The projector onto the full path configuration is 
\begin{equation}
 \begin{aligned}
 \mathsf T_{p_{i,j}}(\mathbf x)&=\prod_{\ell=1}^m T_{s_{\ell},e_\ell}^{x_\ell},
 \end{aligned}
 \label{app:pair-transported-group-elements}
\end{equation}

The full ribbon operator is then 
\begin{equation}
 \mathcal F_{p_{i,j}}(h)
 =
 \sum_{\mathbf x\in G^m}
 \left[\prod_{\ell=0}^m A_{v_\ell}^{X_\ell^{-1}hX_\ell}\right]
 \mathsf T_{p_{i,j}}(\mathbf x).
 \label{app:pair-regional-operator}
\end{equation}
Note that the actions of the vertex terms cancel along $p_{i,j}$ as desired; $\mathcal{F}_{p_{i,j}}(h)$ only changes labels of edges emanating from $p_{i,j}$ in Fig.~\ref{fig:pair-ribbon-spine}, and depends coherently on the unchanged path labels in $p_{i,j}$.

One can check that
\begin{equation}
\begin{aligned}
 \mathcal F_{p_{i,j}}(h)\mathcal F_{p_{i,j}}(k)
 =\mathcal F_{p_{i,j}}(hk).
 \end{aligned}
\end{equation}
Likewise,
\begin{equation}
 \mathcal F_{p_{i,j}}(h)^\dagger
 =\mathcal F_{p_{i,j}}(h^{-1}),\qquad
 \mathcal F_{p_{i,j}}(1)=\mathbf1.
 \label{app:pair-regional-representation}
\end{equation}
Hence \(h\mapsto\mathcal F_{p_{i,j}}(h)\) is a unitary \(G\)-representation. 

Plugging $\mathcal{F}_{p_{i,j}}(h)$ into (\ref{app:pair-fusion-projector}) gives a projector onto the total charge $\lambda$ of the vertices in the strip $p_{i,j}$. When the intermediate vertices are vacuum, the measured total charge is the fusion channel of the two endpoint $E$ irreps.

As a consistency check, when the path has zero length,
\begin{equation}
 \mathcal F_{p_{i,i}}(h)=A_i^h,
 \qquad
 P_{p_{i,i}}^\lambda=P_i^\lambda.
 \label{app:pair-onsite-limit}
\end{equation}
Thus the onsite and pair-fusion projectors are the same construction applied to representations supported on regions of different sizes.

We now specialize to the $D_4$ case. For two $E$ charges, the four possible pair-fusion outcomes are measured by
\begin{equation}
 P_{p_{i,j}}^{A_{xz}}
 =
 \frac{1}{8}
 \sum_{h\in D_4}
 \chi_{A_{xz}}(h^{-1})
 \mathcal F_{p_{i,j}}(h),
 \qquad
 x,z\in\mathbb Z_2.
 \label{app:D4-pair-fusion-projectors}
\end{equation}
On the two-$E$ sector, if there are no intermediate abelian charges along $p_{i,j}$,
\begin{equation}
 P_{p_{i,j}}^E=0,
 \qquad
 \sum_{x,z\in\mathbb Z_2}
 P_{p_{i,j}}^{A_{xz}}
 =
 \mathbbm 1.
\end{equation}
Since every $A_{xz}$ is one-dimensional and occurs with multiplicity one,
this measurement completely resolves the fusion outcome of the two
$E$ charges.

\subsubsection{Fusion-syndrome probabilities}
\label{sec:fusion-syndrome-probabilities}

We now evaluate the joint probability of the onsite charges and the measured fusion outcomes, again focusing on the central qubit. The result has the same edge weights as the onsite model, but one joint source factor replaces the onsite factors on each measured ribbon. Appendix~\ref{app:fusion-probability} gives the full trace expansion in direct parallel with Appendix~\ref{app:onsite-probability}.

We begin with the $\mathcal{S}^{(2)}$ syndrome probability in (\ref{eq:general-logical-projector-syndrome-probability}). For each observed onsite record $\mathcal S^{(2)}=(\gamma,\mathbf a)$, we fix the pairing $M=\operatorname{pair}(\gamma)$ before measuring their fusion outcomes. $m\in M$ are oriented paths $p_{i,j}$ for pairs $i,j\in \gamma$.  We assume disjoint microscopic ribbon supports, including their diagonal spine projectors, and write
\begin{equation}
 C_{m}=V(m),\qquad \mathcal P=\bigsqcup_{m\in M}C_{m}.
 \label{eq:S3-ribbon-vertex-sets}
\end{equation}
Here $C_{m}$ is a set of vertices: it includes both endpoints and every intermediate vertex in the oriented path $m$. Its measured total charge $q_m$ is not assumed to be the endpoint-pair charge: no vacuum assumption is made about the interior.

The refined record is $\mathcal S^{(3)}=(\gamma,\mathbf a,\mathbf q)$. Because the corresponding charge projectors commute, its Born probability is
\begin{equation}
 \begin{aligned}
 &\Pr_p(\mathcal S^{(3)})\\
 &\quad=\Tr\!\left[
 \rho_p(\rho_L^{\mathrm{cen}})
 \left(\prod_vP_v^{\lambda_v}\right)
 \left(\prod_{m\in M}P_{m}^{q_m}\right)\right],
  \end{aligned}
 \label{eq:S3-projector-trace}
\end{equation}
where $P^{q_m}_{m}$ is of the form (\ref{app:pair-fusion-projector}) with irrep $\lambda=q_m$ and path $p_{i,j}=m$. 

To compute the above trace, we use the group variable $c_v\in G$ on each vertex, with 
\begin{equation}
    c_v=\begin{cases} r_vk_m\quad v\in C_m\\ r_v\quad v\notin \mathcal{P}
    \end{cases}
\end{equation}
The multiplication by $k_m$ for $v\in C_m$ comes from the action of the ribbon operators $\mathcal{F}_{m}(k)$ in $P^{q_m}_m$. We show in Appendix~\ref{app:fusion-probability} that the noise term takes the same form, $K_{E,p}(c_fc_i^{-1})$. On the measured ribbon, the substitution $r_v=c_vk_m^{-1}$ changes the onsite character to $\chi_{\lambda_v}(c_v^{-1}k_m)$. Summing over the ribbon variable gives the factor
\begin{equation}
 \begin{aligned}
 \mathcal V_{q_m}^{C_m}(\{c_v\}_{v\in C_m})
 ={}&\frac{d_{q_m}}{|G|}\sum_{k\in G}\chi_{q_m}(k^{-1})\\
 &\times\prod_{v\in C_m}
 d_{\lambda_v}\chi_{\lambda_v}(c_v^{-1}k).
 \end{aligned}
 \label{eq:S3-general-regional-factor}
\end{equation}
The same $k$ appears at every vertex in the ribbon, so this factor accounts for the onsite charges and selects their measured total charge together. The resulting total syndrome probability is 
\begin{equation}
\begin{aligned}
 \Pr_p(\mathcal S^{(3)})
 ={}&\frac{1}{|G|^{N_v}}\\
 &\times\sum_{\{c_v\in G\}}
 \prod_eK_{\Gamma,p}(c_{f_e}c_{i_e}^{-1})\\
 &\times\prod_{v\notin\mathcal P}
 d_{\lambda_v}\chi_{\lambda_v}(c_v^{-1})\\
 &\times\prod_{m\in M}
 \mathcal V_{q_m}^{C_m}(\{c_v\}_{v\in C_m}).
 \end{aligned}
 \label{eq:S3-partition-main}
\end{equation}
Relative to the onsite model (\ref{eq:general-logical-projector-syndrome-probability}), the edge term encoding the noise is unchanged. For each measured pair ribbon, $\mathcal{V}_{q_m}^{C_m}$ replaces the separate onsite charge factors with a single factor that also selects the measured total charge $q_m$. Summing over $\mathbf{q}$ restores the product of onsite
factors on $C_m$, so
\begin{equation}
 \sum_{\mathbf q}\Pr_p(\gamma,\mathbf a,\mathbf q)
 =\Pr_p(\gamma,\mathbf a).
 \label{eq:fusion-syndrome-marginalization}
\end{equation}

\subsubsection{Statistical-mechanics representation including fusion-syndrome probabilities}
\label{sec:fusion-D4-components}
We again focus on $G=D_4$ and $\Gamma=E$ to rewrite the syndrome probability (\ref{eq:S3-partition-main}) as a classical partition function. Compared to the $\mathcal{S}^{(2)}$ partition function in (\ref{eq:char-loop-final}), the weights $(1-p)^{N_e-|F|}p^{|F|}$ and the constraint $\partial F=\gamma$ remain unchanged. The new ingredient is the measured fusion channel of each paired set of $E$ anyons. We will represent each such measurement by a new degree of freedom, living on an additional edge joining the measured pair. The stat mech model will therefore be on an augmented graph that consists of the original square lattice together with additional edges from fusion measurements. The measurements pair up all of the $E$ anyons so we have only closed components on this augmented graph, each with a vacuum fusion constraint on the abelian labels (including both on-site charges and fusion charges). Because we only perform pairwise fusion rather than measurement of an entire fusion tree, the extra lattice at most increases the number of degrees of freedom by 50\%. 

As in the on-site charge measurement case, we write
\begin{equation}
 c_v=r^{2t_v+u_v}s^{w_v},
 \qquad (t_v,u_v,w_v)\in\mathbb Z_2^3.
\end{equation}
now with $c_v$ instead of $r_v$. First, we distinguish the measured charge of a ribbon from the fusion channel of its endpoints. For a ribbon $m$ joining $i_m,j_m$, let $I_m$ be the intermediate vertices. Each carries a known Abelian charge $A_{x_v,z_v}$, recorded by the onsite measurements. If the measured total ribbon charge is $q_m=A_{x_mz_m}$, then the endpoint pair channel must be $q_m$ fused with the inverse of the onsite charges in $I_m$ (in this case, all charges are $\mathbb{Z}_2$, so we do not need to take the inverse). Call this $E$ pair fusion result $A_{\bar{x}_m\bar{z}_m}$ (note that $\bar{x}_m$ here does not mean $1-x_m$).\footnote{Specifically, $\bar{x}_m=x_m\oplus_{v\in I_m}x_v$, and $\bar{z}_m$ is defined similarly.} 

Because the intermediate charges in $I_m$ are 1d, their characters factor out:
\begin{equation}
    \mathcal{V}^{C_m}_{q_m}=\left[\prod_{v\in I_m}\chi_{A_{x_vz_v}}(c_v^{-1})\right]V_{\bar{x}_m\bar{z}_m}(c_{i_m},c_{j_m})
\end{equation}
where $c_{i_m},c_{j_m}$ are the group elements at the endpoints of $m$. Evaluating the remaining pair factor gives
\begin{equation}
    V_{\bar{x}_m\bar{z}_m}(c_i,c_j)=4(-1)^{t_i+t_j+\bar{x}_mu_j+\bar{z}_mw_j}\delta_{u_i,u_j}\delta_{w_i,w_j}
\end{equation}
The sum over $t_v$ again enforces $\partial F=\gamma$ just as in Sec.~\ref{statmechs1s2}. The new information lies in the remaining factors: the measured pair must share the same $u,w$ values and its fusion channel contributes the phase $(-1)^{\bar{x}_mu_j+\bar{z}_mw_j}$.

\begin{figure}[tbp]
 \centering
 \includegraphics[width=0.9\linewidth]{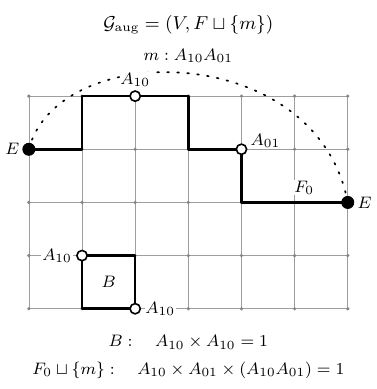}
 \caption{An augmented graph that contributes to the probability of the $\mathcal{S}^{(3)}$ syndrome. A subgraph $F$ of error strings has endpoints $\partial F=\gamma$.  Each ribbon measurement
 adds an abstract matching edge carrying the endpoint-corrected label
 $A_{\bar x_m\bar z_m}$, obtained from its measured total charge and the known internal charges.  A term contributes only when the total Abelian
 charge on every component of $F\sqcup M$ is trivial.}
 \label{fig:fusionclosed}
\end{figure}

We now add one abstract matching edge for every measured ribbon and form the
augmented graph (Fig.~\ref{fig:fusionclosed})
\begin{equation}
 \mathcal G_{\rm aug}(F,M)=(V,F\sqcup M),
 \label{eq:augmented-fusion-graph}
\end{equation}
where $\sqcup$ retains both edges if a physical edge and a matching
edge share the same vertices.  Let $\mathcal B_{\rm aug}(F,M)$ be its set of
connected components, including isolated vertices. Since the measurements pair all of the $E$ anyons, all of the components of $\mathcal{B}_{\mathrm{aug}}(F,M)$ are closed. The occupied physical edges already require $u_v,w_v$ at their endpoints. The matching edges impose the same condition between measured partners, so $u_v,w_v$ are now constant throughout each connected component $B$ of the augmented graph. For each component $B\in \mathcal B_{\rm aug}(F,M)$,
define
\begin{align}
 X_B^{\rm aug}
 &=\bigoplus_{v\in B\setminus\gamma}x_v
   \oplus\bigoplus_{m\in M_B}\bar x_m,
 \nonumber\\
 Z_B^{\rm aug}
 &=\bigoplus_{v\in B\setminus\gamma}z_v
   \oplus\bigoplus_{m\in M_B}\bar z_m,
 \label{eq:S3-D4-augmented-charge}
\end{align}
where $M_B$ denotes the matching edges in $B$. Summing its two remaining variables gives
\begin{equation}\label{eq:S3-D4-component-sum-main}
    \sum_{u_B,w_B\in\mathbb{Z}_2}(-1)^{u_bX_B^{\mathrm{aug}}+w_bZ_B^{\mathrm{aug}}}=4\delta_{X_B^{\mathrm{aug}},0}\delta_{Z_B^{\mathrm{aug}},0}
\end{equation}

Since $|M|=|\gamma|/2$, applying this identity to every component yields
\begin{equation}
 \begin{aligned}
 \Pr_p(\mathcal S^{(3)})
 ={}&4^{|\gamma|/2-N_v}
 \sum_{\substack{F\subseteq\mathcal E\\\partial F=\gamma}}
 (1-p)^{N_e-|F|}p^{|F|}
 4^{|\mathcal B_{\rm aug}(F,M)|}
 \\
 &\times
 \prod_{B\in\mathcal B_{\rm aug}(F,M)}
 \delta_{X_B^{\rm aug},0}\delta_{Z_B^{\rm aug},0}.
 \end{aligned}
 \label{eq:S3-D4-closed-polymer}
\end{equation}

The physical interpretation of the above partition function is that the fusion measurements close up all of the open components on the original graph, so all components on the augmented graph are closed and supply a factor of $4$. Furthermore, the fusion outcomes supply an Abelian label
$A_{\bar x_m\bar z_m}$ on each matching edge. These labels must fuse with the onsite abelian charges along the component to vacuum, because all of the components are closed, with trivial total charge. 

This means that even though the fusion measurements in general help recovery of the central qudit, it may preclude recovering the full logical qudit if it uses incorrect fusion tree measurements. This generally does not occur for sufficiently small noise, i.e., for the $D_4$ quantum double below the MWPM threshold for the $E$ anyons or its heralded version~\cite{jing_intrinsic_2025,jing2026integerlinearprogrammingdecoder}.

\subsection{Fusion measurements disturbing logical information}
\label{sec:fusion-logical-backaction}

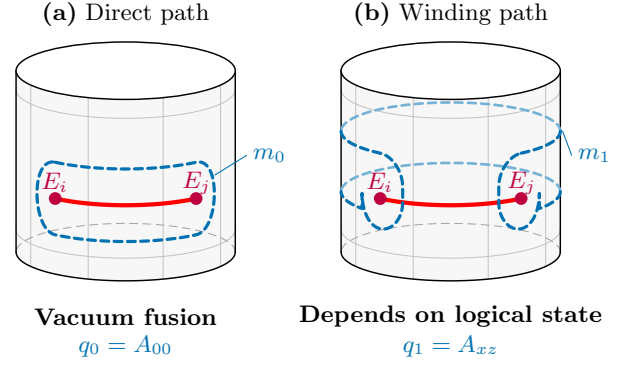
\begin{figure}[t]
\centering
\begin{tikzpicture}[x=1cm,y=1cm,
  line cap=round,line join=round,font=\small]
  \definecolor{fusionBlue}{HTML}{0072B2}
  \tikzset{
    rim/.style={black,line width=0.55pt},
    mesh/.style={black!24,line width=0.3pt},
    history/.style={red,line width=1.6pt},
    fusion/.style={fusionBlue,densely dashed,line width=1.2pt},
    hidden fusion/.style={fusionBlue!55,densely dashed,line width=1.0pt}
  }
  \foreach \dx in {0,4.3}{
    \begin{scope}[shift={(\dx,0)}]
      \fill[black!3] (-1.45,0)--(-1.45,2.4)
        arc[start angle=180,end angle=0,x radius=1.45,y radius=0.38]
        --(1.45,0)
        arc[start angle=0,end angle=-180,x radius=1.45,y radius=0.38]
        --cycle;
      \foreach \ang in {210,250,290,330}{
        \draw[mesh] ({1.45*cos(\ang)},{0.38*sin(\ang)})
          --({1.45*cos(\ang)},{2.4+0.38*sin(\ang)});
      }
      \foreach \zz in {0.3,2.15}{
        \draw[mesh] plot[domain=180:360,samples=45,variable=\t]
          ({1.45*cos(\t)},{\zz+0.38*sin(\t)});
      }
      \draw[black!35,densely dashed,line width=0.4pt]
        plot[domain=0:180,samples=45,variable=\t]
          ({1.45*cos(\t)},{0.38*sin(\t)});
      \draw[rim] (-1.45,0)--(-1.45,2.4) (1.45,0)--(1.45,2.4);
      \draw[rim] plot[domain=180:360,samples=45,variable=\t]
        ({1.45*cos(\t)},{0.38*sin(\t)});
      \filldraw[fill=white,draw=black,line width=0.55pt]
        (0,2.4) ellipse[x radius=1.45,y radius=0.38];
      \draw[history] plot[domain=230:310,samples=35,variable=\t]
        ({1.45*cos(\t)},{1+0.38*sin(\t)});
      \node[purple] at (-0.94,0.92) {$E_i$};
      \node[purple] at (0.94,0.92) {$E_j$};
    \end{scope}
  }

  \draw[fusion]
    plot[domain=230:310,samples=35,variable=\t]
      ({1.45*cos(\t)},{1.48+0.38*sin(\t)})
    --plot[domain=0:180,samples=35,variable=\t]
      ({1.45*cos(310+14*sin(\t))},
       {1+0.48*cos(\t)+0.38*sin(310+14*sin(\t))})
    --plot[domain=310:230,samples=35,variable=\t]
      ({1.45*cos(\t)},{0.52+0.38*sin(\t)})
    --plot[domain=0:180,samples=35,variable=\t]
      ({1.45*cos(230-14*sin(\t))},
       {1-0.48*cos(\t)+0.38*sin(230-14*sin(\t))})
    --cycle;
  \draw[fusionBlue,line width=0.4pt] (1.18,1.02)--(1.60,1.28);
  \node[fusionBlue,anchor=west,inner sep=0pt]
    at (1.68,1.28) {$m_0$};
  \node at (0,3.15) {\textbf{(a)} Direct path};
  \node at (0,-0.85) {\textbf{Vacuum fusion}};
  \node[fusionBlue] at (0,-1.24) {$q_0=A_{00}$};

  \begin{scope}[shift={(4.3,0)}]
    \foreach \zz in {0.80,1.60}{
      \draw[hidden fusion] plot[domain=0:180,samples=55,variable=\t]
        ({1.45*cos(\t)},{\zz+0.38*sin(\t)});
    }
    \draw[fusion] plot[domain=-50:0,samples=25,variable=\t]
      ({1.45*cos(\t)},{1.60+0.38*sin(\t)});
    \draw[fusion] plot[domain=180:230,samples=25,variable=\t]
      ({1.45*cos(\t)},{1.60+0.38*sin(\t)})
      --plot[domain=0:90,samples=25,variable=\t]
        ({1.45*cos(230+14*sin(\t))},
         {1.20+0.40*cos(\t)+0.38*sin(230+14*sin(\t))})
      --({1.45*cos(244)},{1+0.38*sin(244)})
      --plot[domain=0:180,samples=35,variable=\t]
        ({1.45*cos(230+14*cos(\t))},
         {1-0.40*sin(\t)+0.38*sin(230+14*cos(\t))})
      --({1.45*cos(216)},{0.80+0.38*sin(216)})
      --plot[domain=216:180,samples=25,variable=\t]
        ({1.45*cos(\t)},{0.80+0.38*sin(\t)});
    \draw[fusion] plot[domain=0:-36,samples=25,variable=\t]
      ({1.45*cos(\t)},{0.80+0.38*sin(\t)})
      --({1.45*cos(-36)},{1+0.38*sin(-36)})
      --plot[domain=0:180,samples=35,variable=\t]
        ({1.45*cos(-50+14*cos(\t))},
         {1-0.40*sin(\t)+0.38*sin(-50+14*cos(\t))})
      --({1.45*cos(-64)},{1.20+0.38*sin(-64)})
      --plot[domain=0:90,samples=25,variable=\t]
        ({1.45*cos(-50-14*cos(\t))},
         {1.20+0.40*sin(\t)+0.38*sin(-50-14*cos(\t))});
    \draw[fusionBlue,line width=0.4pt] (1.48,1.60)--(1.60,1.28);
    \node[fusionBlue,anchor=west,inner sep=0pt]
      at (1.68,1.28) {$m_1$};
    \node at (0,3.15) {\textbf{(b)} Winding path};
    \node at (0,-0.85) {\textbf{Depends on logical state}};
    \node[fusionBlue] at (0,-1.24) {$q_1=A_{xz}$};
  \end{scope}

  \foreach \dx in {0,4.3}{
    \begin{scope}[shift={(\dx,0)}]
      \foreach \ang in {230,310}{
        \fill[purple] ({1.45*cos(\ang)},{1+0.38*sin(\ang)})
          circle (2.4pt);
      }
    \end{scope}
  }
\end{tikzpicture}
\caption{The fusion outcomes measured along $m_0$ and
$m_1$ are denoted $q_0$ and $q_1$, respectively.
\textbf{(a)} The direct path $m_0$ closes contractibly with the red
error string, giving vacuum fusion $A_{00}$.
\textbf{(b)} On the other hand, the path $m_1$ closes with the red string to wind
once around the cylinder. In $D_4$, the outcome is $A_{01}$, $A_{10}$,
or $A_{11}$ for holonomy $[r]$, $[s]$, or $[rs]$, respectively.
Central holonomies $[1]$ and $[r^2]$ both give $A_{00}$.}
\label{fig:path-dependent-fusion}
\end{figure}

Unlike the complete onsite syndrome measurement, a fusion measurement is specified along a path of nonzero length. Interestingly, this implies that the fusion measurement along two homologically distinct paths may differ depending on the logical state. This is depicted schematically in Fig.~\ref{fig:path-dependent-fusion}. More precisely, every state outside of the central qubit can be detected by a suitable fusion measurement, whereas the central logical subspace associated with the central qubit is not disturbed by the projective fusion measurements. This may be counterintuitive, because the fusion measurements do not wrap around the cylinder. However, when combined with the physical error string, an incorrect homology fusion measurement would tell you the result of creating two anyons and bringing them around the cylinder before re-annihilating. The outcome may actually not be vacuum by the same mechanism that gives a single anyon on a torus after the application of anyon creation, transport, and fusion processes around the two different cycles~\cite{bombin2008family,iqbal_non-Abelian_2024,chan2024torus}. 

For our ML decoding procedure, we use for the most part MWPM to pair the $E$ anyons. Below the MWPM threshold, the fusion measurements are unlikely to disturb the logical information, but above the MWPM threshold (which we do indeed probe because the threshold for the central qubit is a lot higher), we are likely to disturb the logical information outside of the central qubit. Therefore, above the MWPM threshold, the rest of the qudit is sacrificed to improve decoding for the central qubit.
\subsubsection{General finite-group result}

To study the aforementioned difference between two homologically different fusion measurements, we will compare the two projectors $P_{m_0}^\lambda$ and
$P_{m_1}^\lambda$ for paths $m_0$ and $m_1$ oriented
from $i$ to $j$. We assume here that the state is flux-free and that, for simplicity, intermediate
vertices carry vacuum charge. The important point is that $m_0\overline m_1$ winds once around the cylinder (where $\overline m_1$ has opposite orientation, from $j$ to $i$). We will show that for a general conjugacy class logical state, these two measurements differ in a way that reveals logical information. 

For an edge group configuration $\mathbf g$, let $X_\mu(\mathbf g)$ be
the ordered product of edge labels along $m_\mu$, using inverse labels
when the path opposes the edge orientation. Their relative holonomy is
\begin{equation}
 H(\mathbf g)=X_1(\mathbf g)X_0(\mathbf g)^{-1}.
 \label{eq:path-relative-lattice-holonomy}
\end{equation}
In logical sector $[H]$, the element $H(\mathbf g)$ belongs
to that conjugacy class, but its representative can vary
between edge configurations. The microscopic ribbon operator
uses the actual edge labels in each configuration, rather
than a fixed representative of $[H]$.

To see how these labels enter, consider one configuration
$\mathbf g$. The diagonal group label projectors $T^{x_{\ell}}_{s_{\ell},e_{\ell}}$ in
$\mathcal F_{m_\mu}(h)$ select the corresponding path labels.
These labels determine the group elements used in the shifts
on the ribbon boundary: starting with the vertex action of $h$ at $i$, after each
successive path with edge configuration $x_\ell$, the action at the next vertex is the conjugate of the last action by $x_\ell^{-1}$ as described in Sec.~\ref{pairfusionprojectors}. Consequently, after following
the entire path, the endpoint action at $j$ is
$X_\mu^{-1}hX_\mu$.

Now compare the two ribbon paths while keeping the starting action
$h$ at $i$ fixed. Writing $X_\mu=X_\mu(\mathbf g)$ and
$H=H(\mathbf g)$ for compactness, the actions at $j$ for two paths are given by
\begin{equation}
 \begin{aligned}
 Y_0(h;\mathbf g)
 &=X_0^{-1}hX_0,\\
 Y_1(h;\mathbf g)
 &=(HX_0)^{-1}h(HX_0)\\
 &=X_0^{-1}H^{-1}hHX_0,
 \end{aligned}
 \label{eq:path-fusion-projector-conjugation}
\end{equation}
where we used the relation $X_1=HX_0$. Thus, for homologically distinct paths the actions at $j$ are different with a given action $h$ at $i$. Under the vacuum-interior
assumption, this relative endpoint action is the relevant
distinction between the two fusion measurements. A full derivation of the fusion projectors is given in Appendix~\ref{app:path-dependent-fusion}. 

The consequences are different depending on whether $H$ is in the center of $G$ or not. For central holonomy $H=z\in Z(G)$,
the extra conjugation is trivial, so
\begin{equation}
\left. P_{m_1}^{\lambda}\right|_{[z]}
 =\left.P_{m_0}^{\lambda}\right|_{[z]}.
 \label{eq:central-path-independence}
\end{equation}

On the other hand, any noncentral holonomy can be detected by a suitable charge pair. Suppose that the actual error string created $\Gamma,\bar{\Gamma}$ along $m_0$, so the state of interest is
\begin{equation}
 \mathcal Z_\Gamma(m_0)|[H]\rangle
 =\frac{1}{\sqrt{|\mathcal{O}_H|}}\sum_{\mathbf{g}\in \mathcal{O}_H}\chi_\Gamma(X_0(\mathbf g))|\mathbf g\rangle.
 \label{eq:path-reference-charge-line-main}
\end{equation}
The vacuum probabilities for $m_0$ and $m_1$, using $P^1_{m_0}$ and $P^1_{m_1}$, are
\begin{equation}
 \begin{aligned}
 \Pr_{[H]}(\mathbf1\mid m_0)&=1,\\
 \Pr_{[H]}(\mathbf1\mid m_1)
 &=\frac{|\chi_\Gamma(H)|^2}{d_\Gamma^2}.
 \end{aligned}
 \label{eq:path-dependent-vacuum-probability}
\end{equation}
The first line follows from the fact that preparing the two charges along $m_0$ means they fuse back to vacuum along $m_0$. Along $m_1$, the additional holonomy conjugation does not necessarily give vacuum, as shown by explicit calculation in Appendix~\ref{app:path-dependent-fusion}.

For central $H$, both probabilities are one. If $H$ is not central, its winding-path measurement has a nonzero nonvacuum probability for some $\Gamma$ and thus reveals logical information.

\subsubsection{Specialization to \(D_4\)}

For \(D_4\), a single pair of \(E\) charges detects all three noncentral
conjugacy classes. Indeed,
\begin{equation}
 \chi_E(H)=
 \begin{cases}
  2,  & H=1,\\
 -2,  & H=r^2,\\
  0,  & H\notin Z(D_4).
 \end{cases}
\end{equation}
Eq.~\eqref{eq:path-dependent-vacuum-probability} therefore gives
\begin{equation}
 \Pr_{[H]}(A_{00}\mid m_1)
 =
 \begin{cases}
  1, & H\in Z(D_4),\\
  0, & H\notin Z(D_4).
 \end{cases}
 \label{eq:D4-path-vacuum-probability}
\end{equation}
In other words, provided that the two \(E\) charges fuse to vacuum (\(A_{00}\)) along \(m_0\), for \(\ket{[1]}\) and \(\ket{[r^2]}\) they still fuse deterministically to vacuum along $m_1$. On the other hand, for \(\ket{[r]}\), \(\ket{[s]}\), and \(\ket{[rs]}\) they never fuse to vacuum.

In the following, we show that for \(\ket{[r]}\), \(\ket{[s]}\), and \(\ket{[rs]}\), the fusion outcomes along $m_1$ are also deterministic. Write
\begin{equation}
 H=r^\alpha s^\beta,\qquad g=r^\mu s^\nu,
\end{equation}
with exponents understood modulo \(4\) or \(2\), as appropriate.
Since \(D_4/Z(D_4)\) is Abelian,
\begin{equation}
 \begin{aligned}
 H^{-1}gH&=r^{2\epsilon_H(g)}g,\\
 \epsilon_H(g)&=\alpha\nu+\beta\mu\pmod 2.
 \end{aligned}
 \label{eq:D4-commutator-character}
\end{equation}
The sign
\begin{equation}
 \eta_H(g)=(-1)^{\epsilon_H(g)}
\end{equation}
is the one-dimensional character
\begin{equation}
 \begin{aligned}
 \eta_H&=A_{x_Hz_H},\\
 (x_H,z_H)&=(\beta,\alpha\bmod 2),
 \end{aligned}
 \label{eq:D4-holonomy-character}
\end{equation}
where \(A_{xz}(r)=(-1)^x\) and \(A_{xz}(s)=(-1)^z\).
Because \(r^2\) acts as \(-\mathbbm 1\) on an \(E\) charge, the extra
conjugation in Eq.~\eqref{eq:D4-commutator-character} multiplies each
character-projector term by \(\eta_H(g)\). On the two-\(E\) sector at
fixed holonomy \([H]\), character multiplication therefore gives
\begin{equation}
 P_{m_1}^{A_{xz}}
 =
 P_{m_0}^{A_{x\oplus x_H,\,z\oplus z_H}}.
 \label{eq:main-path-projector-permutation}
\end{equation}

Let \(\ket{\psi_H}\) denote the holonomy of a state on which two \(E\) charges fuse to vacuum along \(m_0\). Eq.~\eqref{eq:main-path-projector-permutation} gives
\begin{equation}
 \begin{aligned}
 P_{m_1}^{A_{xz}}\ket{\psi_H}
 &=\delta_{x,x_H}\delta_{z,z_H}\ket{\psi_H},\\
 P_{m_0}^{A_{00}}\ket{\psi_H}
 &=\ket{\psi_H},
 \end{aligned}
 \label{eq:main-path-dependent-fusion}
\end{equation}
with
\begin{equation}
 \renewcommand{\arraystretch}{1.15}
 \begin{array}{c|c}
 \text{Holonomy} & \text{Outcome along }m_1\\ \hline
 {[1],\,[r^2]} & A_{00}\\
 {[r]}         & A_{01}\\
 {[s]}         & A_{10}\\
 {[rs]}        & A_{11}
 \end{array}
 \label{tab:main-path-dependent-fusion}
\end{equation}

Thus, after preparing a pair of $E$ anyons along \(m_0\), the \(m_1\) fusion measurement
induces the following projective measurement on the five-dimensional
ground space:
\begin{equation}
 \begin{aligned}
 \mathsf E_{00}&=\Pi_{[1]}+\Pi_{[r^2]},\\
 \mathsf E_{01}&=\Pi_{[r]},\\
 \mathsf E_{10}&=\Pi_{[s]},\\
 \mathsf E_{11}&=\Pi_{[rs]},
 \end{aligned}
 \label{eq:D4-induced-logical-PVM}
\end{equation}
where \(\Pi_{[H]}=\ket{[H]}\!\bra{[H]}\).
The same two anyons therefore give a trivial outcome along \(m_0\), but
measure the noncentral logical sector when fused along \(m_1\).

Crucially, \(\mathsf E_{xz}\) does not distinguish \([1]\) from \([r^2]\),
so the central qubit is preserved. The central qubit is therefore invisible to these projective fusion outcomes, even when
the complementary three-dimensional part of the ground space is measured.

\subsection{Single-shot decoding of the full \(D_4\) logical qudit}
\label{sec:full-D4-qudit-decoding}

In this section, we will show that despite the dangers of measuring fusion outcomes explained in the previous section, the full $D_4$ logical qudit can still be recovered using a single round of measurements below a certain threshold. In particular, we will show that
\begin{equation}\label{fulldecode}
 \begin{aligned}
p_{\mathrm{full}}
 \geq p_{\mathrm{HMWPM}}\geq p_{\mathrm{MWPM}}\simeq 0.103.
 \end{aligned}
\end{equation}
where $p_{\mathrm{full}}$ is the threshold for decoding the full qudit with a single round of measurements. Here, $\mathrm{HMWPM}$ is the error correction in Ref.~\cite{jing_intrinsic_2025} (here, $H$ means ``heralded") that corrects the $E$ anyons in the $D_4$ quantum double by a MWPM algorithm that selects the minimal weight $E$ recovery conditioned on the fact that the $E$ strings must pass through all measured abelian on-site charges. Ref.~\cite{jing_intrinsic_2025} obtained a numerical estimate of the HMWPM threshold for their model, but because the lattice model is different from our $D_4$ quantum double, we cannot directly use their threshold. All we can say is that $p_{\text{HMWPM}}\geq p_{\text{MWPM}}\simeq 0.103$ on the square lattice. 

We will show Eq.~\eqref{fulldecode} by constructing a decoder that succeeds for $p<p_{\text{HMWPM}}$. The basic idea is that instead of pairing the $E$ anyons with MWPM, which we did in previous sections just for simplicity (while we didn't need to specify the precise pairing in the analytic calculations, we did need to specify the pairing for the numerics in Sec.~\ref{sec:decoding}), we instead pair with HMWPM to maximize the threshold for single shot decoding. Each syndrome for full decoding is, schematically,
\begin{equation}
P_{\tilde{\mathcal{S}}}=P_{f,\text{HMWPM}}P_{\mathcal{S}^{(2)}}
\end{equation}
where $P_{f,\text{HMWPM}}$ is a projector onto fusion outcomes with paths dictated by HMWPM of the onsite syndromes in $P_{\mathcal{S}^{(2)}}$. The full recovery then succeeds below the HMWPM threshold because
\begin{itemize}
    \item $P_{\tilde{S}}$ does not connect $E$ anyons with the wrong holonomy below the HMWPM threshold
    \item $P_{f,\text{HMWPM}}$ provides enough information to clean up the $E$ anyons and then the abelian ones
\end{itemize}
Roughly speaking, we are simply combining the two steps of Ref.~\cite{jing_intrinsic_2025} into one step of measurement. In Ref.~\cite{jing_intrinsic_2025}, they first measured the on-site charges, then annihilated pairs of $E$ anyons according to HMWPM, then measured the resulting abelian anyons to finish correcting the Abelian anyons. There are two rounds of measurement: the first to perform HMWPM, and the second to clean up the remaining abelian anyons. Our protocol differs a bit beyond combining the steps of Ref.~\cite{jing_intrinsic_2025} in that the recovery of the $E$ anyons needs to be done coherently, without creating new abelian anyons beyond the measured fusion outcomes.

Heralding uses the observed Abelian charges to constrain the non-Abelian correction paths. Below $p_{\mathrm{HMWPM}}$, we expect the union of the physical error strings and correction paths to decompose, with probability approaching one as the system grows, into contractible, vacuum-fusing components. Provided the subsequent Abelian cleanup remains confined to these components, it cannot distinguish logical states or implement a nontrivial logical operation. Under this confinement assumption, Abelian cleanup does not introduce a lower threshold, meaning if $E$ recovery succeeds then the Abelian anyon recovery also succeeds. This expectation is supported by Ref.~\cite{jing_intrinsic_2025}, which finds statistically consistent thresholds with and without subsequent Abelian correction in its $D_4$ realization. 

Our observation here is simply that instead of moving the $E$ anyons together to annihilate into an Abelian anyon, we can instead measure the fusion outcome at the beginning. This would tell us what we would get if we moved the $E$ anyons together to fuse into an Abelian anyon. Specifically, let $U$ denote the coherent transport used in the two-round protocol and $Q_{\boldsymbol\lambda}$ its post-transport fusion projector. The corresponding initial fusion projector is
\begin{equation}
 \begin{aligned}
 P_{\tilde{\mathcal{S}}}
 &=U^\dagger Q_{\boldsymbol\lambda}U,\\
 U P_{\tilde{\mathcal{S}}}
 &=Q_{\boldsymbol\lambda}U.
 \end{aligned}
\end{equation}
Thus measuring before transport reproduces both the outcome
probabilities and the conditional states of the original protocol.

Since there is only one round of (albeit, not necessarily finite range) measurements, this would constitute a single-shot recovery. In practice, to minimize the weight of measurements and avoid costly coherent non-Abelian anyon transport, one would probably prefer a multi-step procedure similar to that of Ref.~\cite{jing_intrinsic_2025} for full qudit recovery.

\subsection{Central decoding of general non-Abelian TO}
\label{sec:general-central-decoding}

The central-decoding problem admits a generalization whenever the group has a nontrivial center. When the center is a cyclic $\mathbb{Z}_N$ group, the irreps 
admit a $\mathbb{Z}_N$ grading since $\Gamma(g)=[\chi_{\Gamma}(g)/d_{\Gamma}]\mathbf{1}_{d_{\Gamma}}$ with $\chi_{\Gamma}(g)/d_{\Gamma}$ an $N$th root of unity for group elements in the center (see Sec.~\ref{sec:cyl-central-subspace}). Even if all non-Abelian anyons do not fuse to produce only Abelian anyons (as in the case of $D_4$),
there can still be an Abelian grading on the set of anyons. This allows an Abelian homology
decoder to protect a corresponding logical sector without resolving
all non-Abelian fusion channels. We will show in this section that, for example, there is a similar logical qubit in the $D_6$ quantum double that admits ML decoding even when the charges do not fuse only to abelian charges. This section presents a different perspective on the comments in Sec.~\ref{sec:cyl-central-subspace} that decoding the central qudit admits an ML algorithm similar to Abelian decoding.

\subsubsection{Abelian grading and the syndrome}
Suppose that the charge types carry labels
$q(a)\in\mathbb Z_N$ satisfying
\begin{equation}
 N_{ab}^{c}>0
 \quad\Longrightarrow\quad
 q(c)=q(a)+q(b)\pmod N,
 \label{eq:charge-grading-rule}
\end{equation}
where $N_{ab}^{c}$ is the multiplicity of $c$ in $a\times b$. In particular, the vacuum has grade zero and $q(\bar a)=-q(a)$. Eq.~\eqref{eq:charge-grading-rule} means that every allowed outcome of a fusion measurement has the same total grade, even when the measurement has several possible outcomes.

Consider incoherent local errors creating charge--anticharge pairs and their associated ribbon networks. Replace each oriented ribbon of type $a$ by its grade $q(a)$. The resulting network defines a $\mathbb Z_N$-valued chain $c$. Grade conservation at every fusion vertex gives
\begin{equation}
 \partial c=s,
 \qquad
 s_v=q(a_v),
 \label{eq:graded-syndrome-boundary}
\end{equation}
where $a_v$ is the charge remaining at vertex $v$. Thus the measured grades are precisely the boundary of an Abelian error chain. Reversing a ribbon reverses its grade; for $\mathbb Z_2$ this orientation distinction disappears.

For a measured syndrome $\mathcal{S}$ giving $q(a_v)$ on every site (not $a_v$), we can choose a reference recovery $c_{\mathcal{S}}$ with $\partial c_\mathcal{S}=s$. Assuming the relevant charges cannot disappear through the cylinder boundaries due to the smooth boundary condition, the difference $c-c_\mathcal{S}$ is closed, and its remaining ambiguity is
\begin{equation}
 h=[c-c_{\mathcal{S}}]
 \in H_1(\mathrm{cylinder};\mathbb Z_N)
 \simeq\mathbb Z_N.
 \label{eq:graded-homology}
\end{equation}
and we can proceed with the typical ML analysis to determine the most likely $h$ given knowledge of the error model. For errors of a single type $a$, only the subgroup generated by $q(a)$ can occur. In particular, a grade-one error gives the usual $\mathbb Z_N$ winding problem, while a grade-zero error contributes no graded winding. In the $D_4$ example, the $E$ charge gives grade-one error while the rest of the charges give grade-zero error.

\subsubsection{Examples: $D_4$ and $D_6$.}
For $D_4$, the central element $r^2$ defines the grading
\begin{equation}
 q(A_{xz})=0,
 \qquad
 q(E)=1.
\end{equation}
The grade syndrome is therefore exactly the set of $E$ anyons, and the winding parity of the $E$ error network determines the logical phase on $\operatorname{span}\{\ket{[1]},\ket{[r^2]}\}$.

For $D_6$, using the convention $|D_6|=12$, the central element is $r^3$. There are 6 irreps: $A_{xz}$ with $x,z\in \mathbb{Z}_2$ together with two 2d irreps $E_1$ and $E_2$. The grading is
\begin{equation}
 q(A_{xz})=x, ~~q(E_1)=1, ~~q(E_2)=0.
\end{equation}
Hence an $E_1$ error network again carries a $\mathbb Z_2$ winding
variable, now acting on
$\operatorname{span}\{\ket{[1]},\ket{[r^3]}\}$.
The analog of $\mathcal{S}^{(1)}$ in this case measures whether each side has an even or odd graded charge, i.e.
\begin{equation}
 P_v^{\mathrm{odd}}
 =
 P_v^{E_1}+P_v^{A_{10}}+P_v^{A_{11}}.
\end{equation}
It is not simply a measurement of $E/$no $E$ as in the $D_4$ case. Moreover,
\begin{equation}
 E_1\times E_1
 =
 A_{00}\oplus A_{01}\oplus E_2.
\end{equation}
Even though there is a non-Abelian anyon on the right hand side, all three outcomes are even in terms of grading. 

For recovery, we can actually simply use the odd Abelian anyon strings, because we just need to neutralize the grading. This gives an efficient recovery protocol, since the Abelian anyon movement can be implemented with finite-depth circuits.

\section{Maximum-likelihood decoding for the central qubit}
\label{sec:decoding}
\label{sec:d4-central-numerics}

We now use the three syndrome sets $\mathcal{S}^{(1)},\mathcal{S}^{(2)},$ and $\mathcal{S}^{(3)}$ discussed in the previous section to construct the maximum-likelihood (ML) decoder for the central qubit introduced in Sec.~\ref{sec:cyl-central-subspace}. This amounts to ensuring that the $E$ anyons are cleaned up without inducing a logical $E$ error; the abelian anyons can be cleaned up arbitrarily.

We first review how to use the syndrome probabilities in the ML decoder. We will then show how to numerically probe the threshold by computing syndrome probabilites via tensor network contraction. We then perform a finite-size scaling analysis of the thresholds for ML decoders based on the three syndrome sets. Finally, we turn to a R\'enyi-2 proxy for the intrinsic information theoretic threshold of the decohered state in
Sec.~\ref{sec:renyi2-intrinsic-proxy}.

\subsection{Review of maximum-likelihood decoding}
\label{sec:num-posterior-sampling}

We first review how the ML decoder uses its knowledge of the noise model to obtain an optimal recovery. This will show that at every $p$ and $L$, the ML decoder has higher failure probability for $\mathcal S^{(1)}$ compared to $S^{(2)}$, and $\mathcal S^{(2)}$ compared to $\mathcal{S}^{(3)}$. 

We will focus on recovery of the $E$ anyon with the correct homology; arbitrary recovery of the abelian anyons can be done after $E$ anyon recovery if we are only concerned with the central qubit. Let $\mathcal S$, be the measured syndrome, and choose a deterministic recovery $R_{\mathcal{S}}$ that annihilates all of the $E$ charges. The point of the reference recovery is to remove the $E$ anyon charges, so the $E$ string has a well-defined homology. Changing the reference recovery does not change the outcome of ML recovery, as will be evident below. If $K$ is a noise history compatible with $\mathcal{S}$, then $R_{\mathcal{S}}K$
has no remaining $E$ anyons but may contain a closed $E$ line winding around the cylinder, with holonomy $h_E(K;\mathcal{S})\in\mathbb{Z}_2$. 

Define the positive joint weights
\begin{equation}
 Z_{\mathcal{S}}(h)
 =\Pr_p\!\left(\mathcal{S},h_E=h\right),
 \qquad h=0,1.
 \label{eq:central-joint-weights}
\end{equation}
with $Z_{\mathcal{S}}(0)+Z_{\mathcal{S}}(1)=\mathrm{Pr}_p(\mathcal{S})$, the syndrome probabilities computed in Sec.~\ref{sec:syndrome-decoding-main}. We've just divided the full syndrome probability into homology equivalence classes. The reference recovery serves as a convenient tool for grouping together errors in the same homology equivalence class because if two errors $K,K'$ compose to the trivial logical operator (i.e. they lie in the same homology equivalence class), then $R_{\mathcal{S},\text{ref}}K$ and $R_{\mathcal{S},\text{ref}}K'$ also give the same logical operator. The probability of a particular homology class $h$ is
\begin{equation}
 \Pr_p(h\mid \mathcal{S})
 =\frac{Z_{\mathcal{S}}(h)}
 {Z_{\mathcal{S}}(0)+Z_{\mathcal{S}}(1)},
 \label{eq:num-posterior}
 \end{equation}
 and the maximum likelihood decision is
 \begin{equation}
 \widehat h_E(\mathcal{S})
 =\operatorname*{arg\,max}_{h\in\{0,1\}}Z_{\mathcal{S}}(h).
 \label{eq:central-ml-rule}
\end{equation}
The ML decoder then applies an additional logical operator corresponding to $\widehat h_E(\mathcal{S})$ after the reference recovery (this may be a finite depth circuit, because we can freely insert Abelian anyons that we will clean up arbitrarily later), so the total recovery $\mathcal{R}_{\mathcal{S}}$ is the reference recovery composed with the logical corresponding to $\widehat h_E(\mathcal{S})$. This ensures invariance under different choices of reference recovery; if two reference recoveries differ by a logical $E$ operator, their $\widehat h_E(\mathcal{S})$ would also differ, so the reference recoveries composed with their respective ML computed most likely logical would be the same.

The failure probability for a particular syndrome $\mathcal{S}$ is
\begin{equation}
 \epsilon_{\rm ML}(\mathcal{S})
 =\frac{\min_h Z_{\mathcal{S}}(h)}
 {Z_{\mathcal{S}}(0)+Z_{\mathcal{S}}(1)},
 \label{eq:central-conditional-risk}
 \end{equation}
 Averaging over all of the syndromes gives the total failure probability:
 \begin{equation}
 P_{\rm fail}(p,L,\{\mathcal{S}\})
 =\sum_{\mathcal{S}}\Pr_p(\mathcal{S})\epsilon_{\rm ML}(\mathcal{S})
  =\sum_{\mathcal{S}}\min_h Z_{\mathcal{S}}(h).
 \label{eq:central-average-risk}
\end{equation}
and the threshold is the largest $p$ for which $\lim_{L\to\infty}P_{\mathrm{fail}}(p,L,\mathcal{S})\to 0$.\footnote{Strictly speaking, we refer to the largest $p$ such that all $p\in[0,p_c)$ have vanishing failure probability. For our particular channel $p=1$ gives a toric code state with the central qubit intact. } For the particular case of $\mathcal{S}=\mathcal{S}^{(j)}$ for $j=1,2,3$, we will use the more compact notation $P_{\mathrm{fail}}(p,L,\mathcal{S}^{(j)})=P_{\mathrm{fail}}^{(j)}(p,L)$. From Eq.~\eqref{eq:central-average-risk}, we can see why the decoders obey $p_c^{(1)}\leq p_c^{(2)}\leq p_c^{(3)}$.  If a fine record $\mathcal{S}^{(2)}$ is
mapped to a coarse record $\mathcal{S}^{(1)}$ by ignoring some information, then
\begin{equation}
 \sum_{\mathcal{S}^{(2)}\mapsto \mathcal{S}^{(1)}}\max_h Z_{\mathcal{S}^{(2)}}(h)
 \geq
 \max_h\sum_{\mathcal{S}^{(2)}\mapsto \mathcal{S}^{(1)}}Z_{\mathcal{S}^{(2)}}(h).
 \label{eq:central-refinement-inequality}
\end{equation}
for each $\mathcal{S}^{(1)}$. A similar condition holds for $\mathcal{S}^{(3)}\to\mathcal{S}^{(2)}$. Averaging over the syndromes and using $P_{\rm succ}=1-P_{\rm fail}$ gives 
\begin{equation}
 P_{\rm fail}^{(3)}(p,L)
 \leq P_{\rm fail}^{(2)}(p,L)
 \leq P_{\rm fail}^{(1)}(p,L)
 \label{eq:central-risk-ordering}
\end{equation}
for every $p$ and $L$. It immediately follows that $p_c^{(1)}\leq p_c^{(2)}\leq p_c^{(3)}$. Below, we will show how to probe these failure probabilities numerically.

\subsection{Homology-resolved partition functions via twisted partition functions}
\label{sec:num-homology-weights}
\label{sec:num-twisted-contractions}
In this section, we will show how to compute $Z_{\mathcal{S}}(h)$, which are the homology-resolved syndrome probabilities, using modifications of the full syndrome probabilities computed in Sec.~\ref{sec:syndrome-decoding-main}.

In ML decoding, for every syndrome, we should choose a reference recovery $\mathcal{R}_{\mathcal{S},\text{ref}}$ and divide the sum over errors consistent with $\mathcal{S}$ according to whether an error configuration, combined with the fixed reference recovery, has even or odd winding around the cylinder (i.e. $h_E=0$ or $1$). For numerical calculation, it is actually easier to compute $Z_{\mathcal{S}}(0)+Z_{\mathcal{S}}(1)$ and $Z_{\mathcal{S}}(0)-Z_{\mathcal{S}}(1)$ and then take linear combinations to get $Z_{\mathcal{S}}(h)$. The former is simply the full syndrome probability, which we can compute via a tensor network contraction. The latter is implemented by a modification of the tensor network corresponding to inserting a $\mathbb Z_2$ twist along a seam joining the two open boundaries:\footnote{The choice of seam does not matter, as long as it starts at the top boundary and ends at the bottom boundary.}
each occupied edge crossing the seam, including both the error and the reference recovery, contributes an additional minus sign. $Z_{\mathcal{S}}(0)$ and $Z_{\mathcal{S}}(1)$ can then be obtained by taking the sum and difference of the untwisted and twisted partition functions. 

We now describe the above procedure in more detail. The explicit twisted partition functions end up being slight modifications of the syndrome probabilities in Sec.~\ref{sec:syndrome-decoding-main}. Given the quantum double defined on a $L\times L$ square cylinder, we fix a top-to-bottom dual cut $\mathcal C$ along the seam, and set $c_e=1$ on the $L$ horizontal edges crossed by $\mathcal C$ and $c_e=0$ otherwise. 
\begin{figure}[tbp]
 \centering
 \includegraphics[width=0.98\columnwidth]{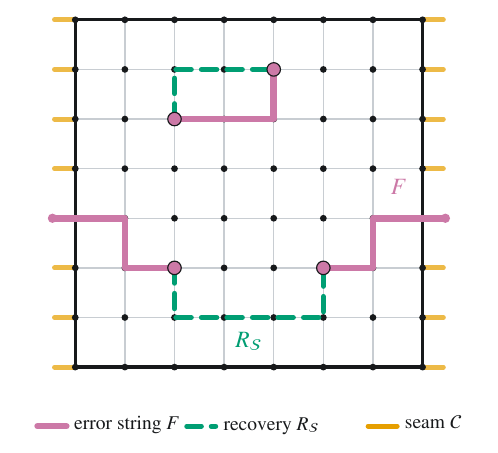}
 \caption{Definition of the residual logical sector. A deterministic
 reference recovery closes the measured $E$ endpoints.  The residual closed
 $E$ line is logical when it crosses the fixed dual cut an odd
 number of times. For the fusion record, the known seam parity of the
 $E$ anyon matching paths is included in the reference recovery. }
 \label{fig:seam}
\end{figure}
Let $F\subseteq\mathcal E$ be the set of occupied edges in the expansion of the syndrome probability (\ref{eq:char-loop-final}) and (\ref{eq:S3-D4-closed-polymer}).  If $\mathcal{R}_{\mathcal{S},\text{ref}}$ is the chosen reference closure, then
\begin{equation}
 h_E(F;\mathcal{S})
 =\bigoplus_{e\in F}c_e
 \underbrace{\bigoplus_{e\in \mathcal{R}_{\mathcal{S},\mathrm{ref}}}c_e}_{b(\mathcal{S})}
 \in\mathbb Z_2.
 \label{eq:num-relative-homology}
\end{equation}
The reference closure is shown schematically in Fig.~\ref{fig:seam}. For the two onsite records we choose for simplicity a referene recovery that does not cross the seam, so $b(\mathcal{S}^{(1)})=b(\mathcal{S}^{(2)})=0$ for any $\mathcal{S}^{(1)},\mathcal{S}^{(2)}$. For the fusion-refined record $\mathcal{S}^{(3)}$, we use the
deterministic MWPM paths $m\in M$. We write
\begin{equation}
 \eta_m=\bigoplus_{e\in m}c_e,
 \qquad b(\mathcal{S}^{(3)})=\bigoplus_{m\in M}\eta_m,
 \label{eq:num-path-seam-parity}
\end{equation}
so $\eta_m$ gives the contribution to the seam parity from a single MWPM path and $b(\mathcal{S}^{(3)})$ includes the contribution from all paths $m\in M$.

Let $\mathcal{A}_{\mathcal{S}^{(j)}}(F)\equiv\Pr_p(\mathcal{S}^{(j)}\mid F)$ be the fixed-$F$ Born factor supplied
by the physical charge and fusion projectors of
Sec.~\ref{sec:syndrome-decoding-main}.  The two weights needed by the decoder
are
\begin{equation}
 Z_{\mathcal{S}^{(j)}}(h)
 =\sum_{\substack{F\subseteq\mathcal E\\
                   \partial F=\gamma\\
                   h_E(F;\mathcal{S}^{(j)})=h}}
 p^{|F|}(1-p)^{N_e-|F|}\mathcal{A}_{\mathcal{S}^{(j)}}(F).
 \label{eq:num-positive-sector-weights}
\end{equation}

The untwisted and twisted partition functions mentioned above are
\begin{align}
 \widetilde Z_{\mathcal{S}^{(j)}}(\kappa)
 =\sum_{h=0}^{1}(-1)^{\kappa h}Z_{\mathcal{S}^{(j)}}(h),
 \qquad \kappa=0,1,
 \label{eq:num-two-point-Fourier}
\end{align}
Here $\widetilde Z_{\mathcal{S}^{(j)}}(0)=\Pr_p(\mathcal{S}^{(j)})$, whereas
$\widetilde Z_{\mathcal{S}^{(j)}}(1)=Z_{\mathcal{S}^{(j)}}(0)-Z_{\mathcal{S}^{(j)}}(1)$. $\epsilon_{\text{ML}}$ can be written in terms of these untwisted/twisted partition functions as
\begin{equation}
 \epsilon_{\rm ML}(\mathcal{S}^{(j)})
 =\frac12\left(1-
 \frac{|\widetilde Z_{\mathcal{S}^{(j)}}(1)|}{\widetilde Z_{\mathcal{S}^{(j)}}(0)}\right).
\end{equation}

The modified partition functions are useful because they both can be computed using tensor network contraction. Substituting Eq.~\eqref{eq:num-positive-sector-weights} into
Eq.~\eqref{eq:num-two-point-Fourier} gives:
\begin{align}
 \widetilde Z_{\mathcal{S}^{(j)}}(\kappa)
 &=\sum_{h=0}^{1}(-1)^{\kappa h}
   \sum_{\substack{F\subseteq\mathcal E\\
                    \partial F=\gamma\\
                    h_E(F;\mathcal{S}^{(j)})=h}}
   p^{|F|}(1-p)^{N_e-|F|}\mathcal{A}_{\mathcal{S}^{(j)}}(F)
 \nonumber\\
 &=\sum_{\substack{F\subseteq\mathcal E\\
                    \partial F=\gamma}}
   p^{|F|}(1-p)^{N_e-|F|}\mathcal{A}_{\mathcal{S}^{(j)}}(F)
   (-1)^{\kappa h_E(F;\mathcal{S}^{(j)})}.
 \label{eq:num-Fourier-fixed-F}
\end{align}
Writing $n_e=\mathbf1[e\in F]$ and using
Eq.~\eqref{eq:num-relative-homology}, the last factor decomposes as
\begin{equation}
 (-1)^{\kappa h_E(F;\mathcal{S}^{(j)})}
 =(-1)^{\kappa b(\mathcal{S}^{(j)})}
  \prod_{e\in\mathcal E}(-1)^{\kappa c_en_e}.
 \label{eq:num-edgewise-homology-character}
\end{equation}

The homology transform therefore inserts a $\mathbb Z_2$ twist
along the seam: for $\kappa=1$, each occupied edge crossing the
seam acquires a minus sign, while all other contributions are
unchanged. The factor $(-1)^{\kappa b(\mathcal{S}^{(j)})}$ is a known overall
sign determined by the reference recovery. Applying this modification
to the syndrome partition functions of
Sec.~\ref{sec:syndrome-decoding-main} gives the following expressions.

For the binary record $\mathcal{S}^{(1)}=\gamma$, using the seam-free reference
choice $b(\gamma)=0$, we obtain
\begin{equation}\label{eq:num-binary-twisted-sum}
\widetilde{Z}_{\mathcal{S}^{(1)}}(\kappa)=\sum_{\substack{F\subseteq\mathcal E\\
                   \partial F=\gamma}}(1-p)^{N_e-|F|}p^{|F|}(-1)^{\kappa\sum_{e\in F}c_e}
\end{equation}

For the complete onsite record $\mathcal{S}^{(2)}=\{\gamma,\mathbf{a}\}$, with
$b(\mathcal{S}^{(2)})=0$, the corresponding expression is
\begin{equation}
 \begin{aligned}
  &\widetilde{Z}_{\mathcal{S}^{(2)}}(\kappa)
 \\&=
 4^{|\gamma|-N_v}
 \sum_{\substack{F\subseteq\mathcal E\\
                  \partial F=\gamma}}
 (1-p)^{N_e-|F|}
 p^{|F|}(-1)^{\kappa\sum_{e\in F}c_e}
\\
 &\quad\times
 4^{|\mathcal{B}_0(F)|}
 \prod_{B\in\mathcal B_0(F)}
 \delta_{X_B,0}
 \delta_{Z_B,0}.
 \end{aligned}
 \label{eq:num-onsite-twisted-contraction}
\end{equation}
Thus the twist modifies only the occupied part of the edge
weight compared to the syndrome probabilities computed in Sec.~\ref{sec:syndrome-decoding-main}; the onsite character factors remain unchanged.

For the fusion-refined record $\mathcal{S}^{(3)}$, we have
\begin{equation}
 \begin{aligned}
&\widetilde{Z}_{\mathcal{S}^{(3)}}(\kappa)
 \\&=4^{|\gamma|/2-N_v}
 \sum_{\substack{F\subseteq\mathcal E\\\partial F=\gamma}}
 (1-p)^{N_e-|F|}p^{|F|}(-1)^{\kappa(\sum_{e\in F}c_e+\sum_{m\in M}\eta_m)}
 \\ &\times 4^{|\mathcal B_{\rm aug}(F,M)|}
 \prod_{B\in\mathcal B_{\rm aug}(F,M)}
 \delta_{X_B^{\rm aug},0}\delta_{Z_B^{\rm aug},0}.
 \end{aligned}
 \label{eq:num-pair-twisted-contraction}
\end{equation}

\begin{figure*}[t]
 \centering
 \includegraphics[width=0.85\textwidth]{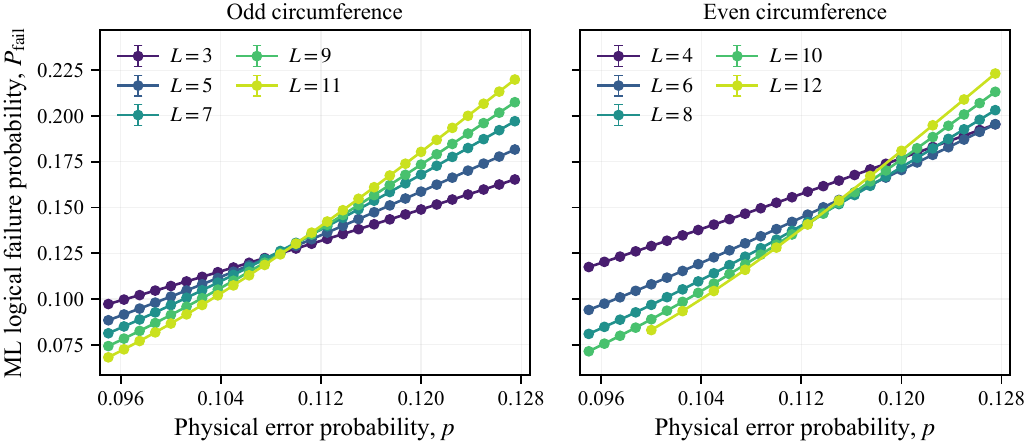}
 \caption{Central-qubit maximum-likelihood failure probability for the
 binary $E$/non-$E$ record $\mathcal S^{(1)}$, shown separately for odd and
 even circumferences. The curves combine exact homology likelihoods with
 reweighted multiple-histogram averaging over $100000$ sampled records. Error bars
 show one-standard-error block-jackknife estimates. The estimated value of $p_c^{(1)}$ is shown in Table~\ref{tab:num-conditional-thresholds}. Further details are described in Appendix~\ref{app:central-numerical-checks}.}
 \label{fig:num-s1-cluster}
\end{figure*}

\subsection{Numerical evaluation of decoding thresholds}
\label{sec:num-evaluation}

We now explain the approach we take to numerically probe the ML thresholds. Additional details are in Appendix~\ref{app:central-numerical-checks}. This task is difficult because we need to not only compute the twisted partition functions, but we also need to average over syndromes, which becomes unmanageable at large system sizes. We will take the following approach:
\begin{itemize}
    \item We Monte Carlo sample a subgraph $F\subseteq\mathcal{E}$. 
    \item For $\mathcal{S}^{(2)}$, we decorate the subgraph with a random allowed Abelian decoration (fusing to vacuum on closed components). This is appropriate because for a fixed $F$, the allowed label assignments have equal probability. For $\mathcal{S}^{(3)}$, we add the MWPM fusion measurement edges to $\partial F$ and jointly sample the onsite Abelian labels and pair-fusion labels subject to neutrality on every augmented component $B\in \mathcal{B}_0$.
    \item We compute the syndrome of the sampled error configuration and compute the twisted partition functions for that syndrome exactly via a tensor network contraction. 
\end{itemize}
In other words, we sample syndromes by sampling errors. Note that in the error sampling, it is crucial that we first sample a subgraph $F$ and then decorate it. Randomly trying to sample a subgraph $F$ and Abelian labels simultaneously for $\mathcal{S}^{(2)}$ would be highly inefficient because most samples would have to be thrown away for not satisfying the hard conditions of (1) Abelian labels must lie on $F$ and (2) Abelian labels on closed components must fuse to vacuum. 

Because we compute the partition functions exactly, we cannot reach very large system sizes. Instead of the naive bond dimension 8 for the order 8 group $D_4$, we manage to reduce the tensor network to bond dimension 5 corresponding to the number of irreps of $D_4.$ In principle, the step of computing the syndrome probability given a syndrome may also be performed via Monte Carlo instead of exact tensor network contraction, but there are some subtleties with this approach (see Sec.~\ref{sec:discussion}). 

\begin{figure*}[t]
 \centering
 \includegraphics[width=0.85\textwidth]{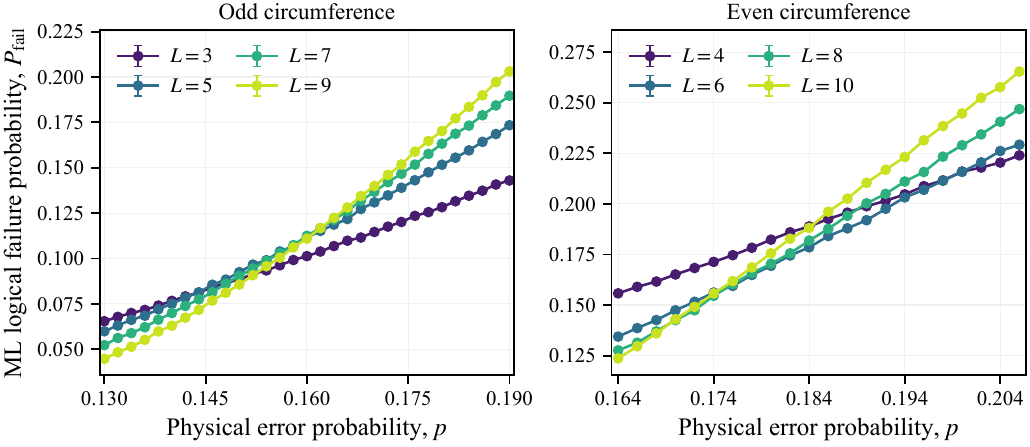}
 \caption{Central-qubit maximum-likelihood failure probability for the
 complete onsite charge record $\mathcal S^{(2)}$, shown separately for odd
 and even circumferences. Conditional failure probabilities are averaged
 over $100000$ directly sampled records, with homology weights evaluated by exact
 five-channel tensor-network contraction. Error bars show one standard
 error. The estimated value of $p_c^{(2)}$ is shown in Table~\ref{tab:num-conditional-thresholds}. Further details are described in Appendix~\ref{app:central-numerical-checks}.}
 \label{fig:num-s2-cluster}
\end{figure*}

\begin{figure*}[t]
 \centering
 \includegraphics[width=0.85\textwidth]{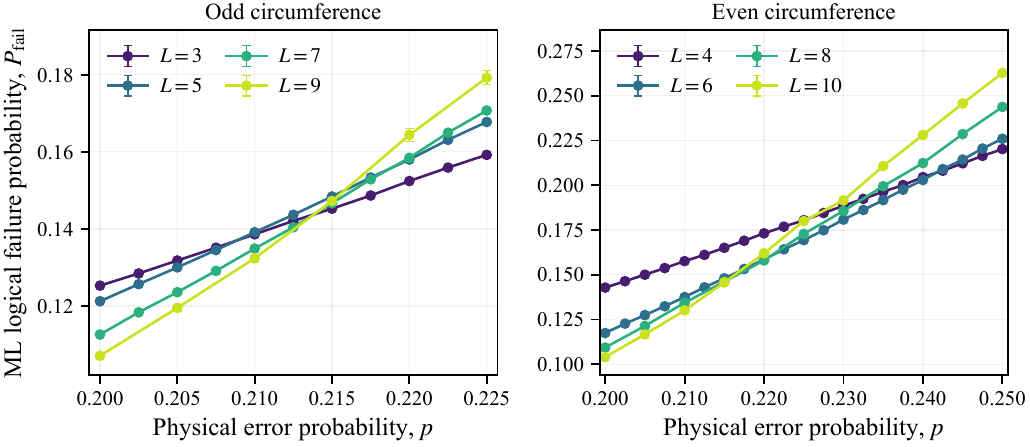}
 \caption{Central-qubit maximum-likelihood failure probability for the
 onsite-plus-pair-fusion record $\mathcal S^{(3)}$, shown separately for odd
 and even circumferences. Conditional failure probabilities are averaged
 over directly $1000000$ sampled records, with homology weights evaluated by exact
 component-frontier contraction. Error bars show one standard error. The estimated value of $p_c^{(3)}$ is shown in Table~\ref{tab:num-conditional-thresholds}. Further details are described in Appendix~\ref{app:central-numerical-checks}.}
 \label{fig:num-s3-cluster}
\end{figure*}

Figures~\ref{fig:num-s1-cluster}--\ref{fig:num-s3-cluster}
show finite-sample estimates of the syndrome-averaged ML failure probability.
The full calculation is not exact because even though the evaluation of syndrome probability is exact, we still sample over the syndromes rather than fully compute the distribution. For $N_s$ sampled records $\mathcal{S}_s$, the plotted estimate has the form
\begin{equation}
 \widehat P_{\rm fail}^{(j)}(p,L)
 =\frac{1}{N_s}\sum_{s=1}^{N_s}\,\epsilon_{\mathrm{ML}}(\mathcal{S}_s)
 \label{eq:num-plotted-sample-estimator}
\end{equation}
for $\mathcal S^{(2)}$ and $\mathcal S^{(3)}$. For $\mathcal S^{(1)}$, we reuse samples across different error rates $p$ using the reweighting procedure described in Appendix~\ref{app:num-binary-estimator}. In each case, $\widehat P_{\rm fail}^{(j)}$ converges to $P_{\rm fail}^{(j)}$ as the number of samples $N_s$ increases.
At finite $N_s$, the plotted values have sampling uncertainty.

For each sampled syndrome, we compute the decoder's conditional failure probability $\epsilon_{\mathrm{ML}}(\mathcal{S})$ using (\ref{eq:central-conditional-risk}) rather than recording whether it succeeds on that particular sampled error. Averaging these probabilities gives the same mean failure rate with less sampling noise (see Appendix~\ref{app:num-failure-observable} for more details).

For $\mathcal S^{(1)}$, the homology-sector weights follow from the
exact two-state Ising contraction in
Eq.~\eqref{eq:num-binary-twisted-sum}.
For efficiency, we reuse samples generated at nearby noise strengths by reweighting
them to each target $p$; Appendix~\ref{app:num-binary-estimator}
specifies the estimator and its sampling uncertainties. 

For $\mathcal S^{(2)}$, we rewrite each edge weight as five products
of functions of its two endpoint variables: one constant term and
four terms from the matrix elements of the $E$ representation in the noise channel (\ref{ChargeDecoherenceChannel}).
This permits the eight-valued group variable to be summed locally
at each vertex, leaving a tensor network whose edge indices take
five values. We contract this network exactly row by row, without a
smaller bond-dimension cutoff. The untwisted and seam-twisted contractions
give the two homology weights through
Eq.~\eqref{eq:num-onsite-twisted-contraction}.
Appendix~\ref{app:num-S2-row-contraction} gives the local tensors,
periodic closure, and numerical precision conventions.

For $\mathcal S^{(3)}$, we instead use a nonnegative expansion over
candidate occupied-edge sets $F'$, with the matching edges always
included. A candidate contributes only when every vertex of the
augmented graph has even degree and every component has zero total
Abelian charge; each surviving component contributes a factor of
four. An exact edge-by-edge sweep performs this sum by tracking
degree parity, connectivity, component charge, and homology,
combining partial configurations with identical remaining data.
Appendix~\ref{app:S3-frontier-contraction} gives the weights and
update rules. 

\begin{table*}[t]
\caption{Primary central-qubit decoding thresholds and nominal
\(95\%\) statistical confidence intervals. The intervals are
conditional on the scaling form and fitting windows shown;
they do not include uncertainty from the finite-size correction
model.}
\label{tab:num-conditional-thresholds}
\begin{ruledtabular}
\begin{tabular}{cccccc}
Record & Circumferences & Probability window
& \(\widehat p_{c}^{(j)}\) & Statistical \(95\%\) interval
& \(\widehat\nu_j\)\\
\hline
\(\mathcal S^{(1)}\)
& \(5\)--\(12\)
& \(0.100\)--\(0.125\)
& \(0.11143\)
& \([0.10934,0.11352]\)
& \(1.548\)\\
\(\mathcal S^{(2)}\)
& \(5\)--\(10\)
& \(0.150\)--\(0.180\)
& \(0.16509\)
& \([0.16408,0.16613]\)
& \(1.600\)\\
\(\mathcal S^{(3)}\)
& \(4\)--\(10\)
& \(0.200\)--\(0.225\)
& \(0.21429\)
& \([0.21276,0.21559]\)
& \(1.373\)\\
\end{tabular}
\end{ruledtabular}
\end{table*}

\begin{figure*}[t]
 \centering
 \includegraphics[width=0.99\textwidth]{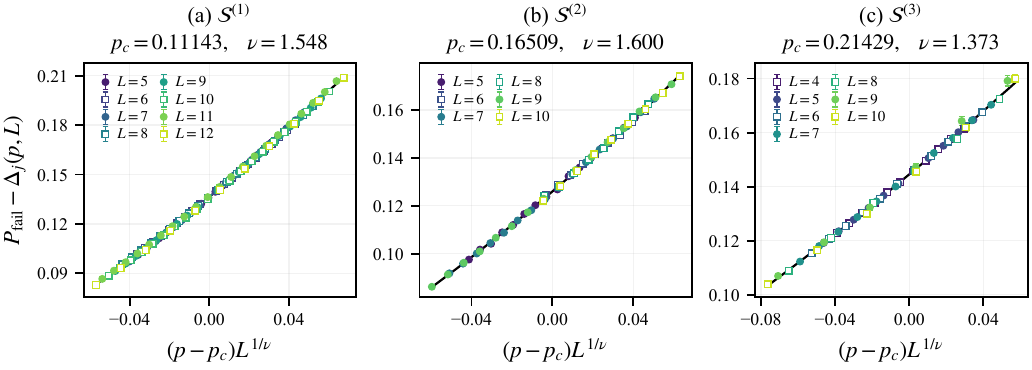}
 \caption{Finite-size scaling of central-qubit maximum-likelihood failure
 for (a) $\mathcal S^{(1)}$, (b) $\mathcal S^{(2)}$, and (c)
 $\mathcal S^{(3)}$. Points show $P_{\mathrm{fail}}-\Delta_j(p,L)$ versus
 $(p-p_c)L^{1/\nu}$, where $\Delta_j$ is the fitted finite-size correction.
 Black curves are the leading scaling functions $F_j$, fitted jointly with
 the corrections to the raw failure probabilities. Error bars show sampling
 standard errors at fixed fit parameters.}
 \label{fig:num-s123-collapses}
\end{figure*}

\subsection{Finite-size scaling and the hierarchy of decoding thresholds}
\label{sec:num-threshold-intervals}
\label{sec:num-scaling-collapses}

Figures~\ref{fig:num-s1-cluster}, \ref{fig:num-s2-cluster},
and~\ref{fig:num-s3-cluster} show the central-qubit decoding
failure probabilities for $\mathcal S^{(1)}$,
$\mathcal S^{(2)}$, and $\mathcal S^{(3)}$, respectively.
Their size dependence identifies three well-separated
transition regions: failure generally decreases with size
below the transition and increases above it, while finite-size
corrections shift the crossings between individual curves.

For a continuous transition with correlation length
$\xi_j\sim|p-p_{c}^{(j)}|^{-\nu_j}$, we use the standard
finite-size scaling ansatz at fixed cylinder
shape~\cite{FisherBarber1972FSS},
\begin{equation}
 \begin{aligned}
 P_{\mathrm{fail}}^{(j)}(p,L)
 &=F_j(x_j)+\Delta_j(p,L),\\
 x_j&=(p-p_{c}^{(j)})L^{1/\nu_j}.
 \end{aligned}
 \label{eq:num-interval-scaling}
\end{equation}
Corrections arise from subleading scaling
fields~\cite{Wegner1972Corrections}. To accommodate the
observed parity dependence, we use the phenomenological form
\begin{equation}
 \Delta_j(p,L)
 =
 L^{-\omega_j}
 \left[B_j(x_j)+(-1)^L C_j(x_j)\right],
 \label{eq:num-interval-corrections}
\end{equation}
with $\omega_j>0$.
Over the restricted fitting windows, we approximate these
smooth functions by low-order polynomials: the primary fits
use a cubic $F_1$, quadratic $F_2,F_3$, constant $B_j,C_j$
for the first two records, and linear $B_3,C_3$.
Sensitivity to these choices and to the fitting windows is
examined in Appendix~\ref{app:central-numerical-checks}.

We determine $p_{c}^{(j)}$ by fitting the raw failure
probabilities, jointly optimizing $p_{c}^{(j)}$, $\nu_j$,
and the scaling-function and correction parameters through
weighted least squares from multiple initial guesses determined by crossings of the raw data.

Odd and even sizes enter the same fit and share
$p_{c}^{(j)}$, $\nu_j$, and $F_j$, but have different corrections,
$L^{-\omega_j}(B_j-C_j)$ and
$L^{-\omega_j}(B_j+C_j)$, respectively.
Their distinct crossing sequences thus constrain a common
thermodynamic threshold.
Figure~\ref{fig:num-s123-collapses} subsequently displays
$P_{\mathrm{fail}}^{(j)}-\Delta_j$ against $x_j$ using these
fitted parameters, so the collapse visualizes the fit rather
than providing an independent threshold estimate.
Appendix~\ref{app:num-scaling-protocol} gives the crossing
diagnostics and uncertainty analysis.

Table~\ref{tab:num-conditional-thresholds} summarizes the fitted
thresholds and nominal $95\%$ statistical confidence intervals.
The primary estimates are
$p_c^{(1)}\simeq0.111$,
$p_c^{(2)}\simeq0.165$, and
$p_c^{(3)}\simeq0.214$, supporting
\begin{equation}
 p_c^{(1)}<p_c^{(2)}<p_c^{(3)}.
 \label{eq:num-interval-hierarchy}
\end{equation}
Resolving onsite Abelian charges therefore improves the
threshold relative to recording only the $E$-anyon locations,
and pair-fusion outcomes provide a further improvement.

The $\mathcal S^{(1)}$ transition gives an independent
benchmark of our numerical approach through its mapping to the random-bond Ising model
along the Nishimori line, for which Merz and Chalker report
$p_{\mathrm{Nish}}=0.1093(2)$ and
$\nu_{\mathrm{Nish}}=1.50(3)$~\cite{merz2022two}. Our fit independently yields $\nu_1=1.548$ with a $95\%$ confidence interval of $[1.496,1.600]$, recovering the established exponent within
statistical uncertainty and providing a consistency check
on the numerical procedure, including its treatment of
finite-size corrections.

\subsection{R\'enyi-2 proxy for the intrinsic threshold}
\label{sec:renyi2-intrinsic-proxy}

The decoding thresholds above depend on which syndrome information is measured. There is also an intrinsic information-theoretic threshold for the decohered mixed state, above which no decoder would possibly be able to decode the state~\cite{fan_diagnostics_2024,sala_stability_2025,sang_stability_2025}. It is difficult for us to access sufficiently large system sizes to probe the intrinsic threshold directly, since this requires exact diagonalization. In this section, we will compute R\'enyi-2 quantities, which serve as a proxy for the von Neumann quantities that describe the intrinsic transition. We simplify these R\'enyi-2 quantities in a way that allows for more efficient numerical calculation via tensor network contraction.

We compare two inputs: a reference maximally entangled with the entire five-dimensional ground-state space, and a reference qubit maximally entangled with the central logical qubit. This is to try to separately probe the intrinsic threshold for the full qudit vs the central qubit; there is a possibility that there is a transition where the memory reduces to the central qubit, and then reduces to nothing. Note that if the central qubit is lost, then the full logical qudit is lost. This is because the central qubit is lost when the $E$ anyons cannot be recovered with the correct homology. If there is a winding $E$ line, then the abelian logicals also become ambiguous because $A_{xz}\times E= E$. Therefore, the rest of the qudit is also lost. On the other hand, the Abelian logicals are allowed to be ambiguous while the $E$ line is guaranteed to be recovered correctly. This is, after all, what we do when we recover the central qubit. Losing the rest of the qudit besides the central qubit can be thought of as the $D_4$ quantum memory transitioning into a $\mathbb{Z}_2$ quantum memory, i.e. a toric code.

For each input, we compute the R\'enyi-2 coherent information between the reference and the physical system after applying noise. This quantity can be evaluated by an exact classical partition function, and serves as a proxy for the usual von Neumann coherent information. Our R\'enyi-2 coherent-information calculations provide finite-size evidence that the full logical qudit and the central qubit are both lost at a common $p_c\simeq0.46$.

\subsubsection{Logical input and the R\'enyi-2 coherent information}
\label{sec:r2proxy-input}

Below we introduce the R\'enyi-2 coherent information as a proxy for diagnosing the intrinsic threshold for the encoded logical information. We will consider the encodings for both the full logical qudit and the central logical qubit, which are defined in the subspace $\mathcal I_5$ and $ \mathcal I_2$, respectively:
\begin{equation}
 \begin{aligned}
 \mathcal I_5&=\{[1],[r^2],[r],[s],[rs]\},\\
 \mathcal I_2&=\{[1],[r^2]\}.
 \end{aligned}
 \label{eq:r2proxy-classes}
\end{equation}

We maximally entangle the full logical qudit or the central logical qubit to a reference qudit/qubit:
\begin{equation}
 |\Phi_d\rangle=\frac{1}{\sqrt d}
    \sum_{C\in\mathcal I_d}|C\rangle_R\otimes|C\rangle_B.
 \label{eq:r2proxy-input}
\end{equation}
where $|C\rangle_B$ is a ground state defined on the physical lattice and $R$ is an external reference. Applying the noise channel  on the physical system $B$ leads to the state
\begin{align}
 \rho_{RB,d}&=(\operatorname{id}_R\otimes\mathcal N_{E,p})
                   (|\Phi_d\rangle\langle\Phi_d|),\notag\\
 \rho_{B,d}&=\operatorname{Tr}_R\rho_{RB,d},
 \qquad
 \label{eq:r2proxy-output}
\end{align}
The R\'enyi-2 coherent information is then defined as the entropy difference
\begin{align}
 &I_{c,d}^{(2)}
 =S_2(\rho_{B,d})-S_2(\rho_{RB,d})
   =\ln\frac{\mathcal P_{RB,d}}{\mathcal P_{B,d}},\notag\\
  &\mathcal P_{X,d}=\operatorname{Tr}\rho_{X,d}^{\,2}.
 \label{eq:r2proxy-definition}
\end{align}
Without any noise,  $\rho_{RB,d}$ is a pure state, so $S_2(\rho_{RB,d}) =0$. Therefore, $I_{c,d}^{(2)} =\ln5$ or $\ln2$, for the full qudit or the central qubit, respectively, when $p=0$.

\subsubsection{Exact result and finite-size evidence}

\begin{figure*}[t]
 \centering
 \includegraphics[width=0.8\textwidth]{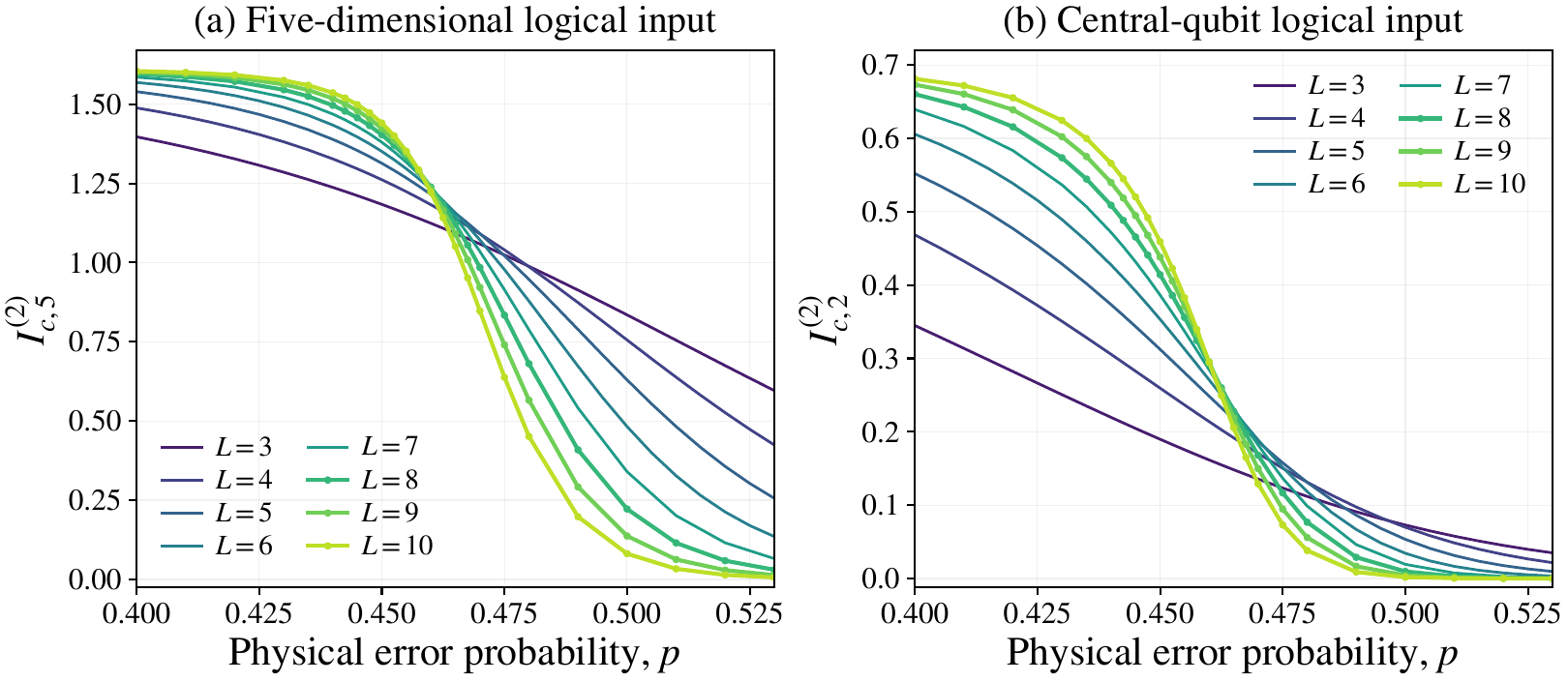}
 \caption{R\'enyi-2 coherent information under independent-edge $E$
 decoherence (Eq.~\ref{eq:r2proxy-definition}), without syndrome measurement or recovery. A reference is maximally entangled with (a) the full five-dimensional logical qudit or (b) the central
 logical qubit spanned by $|[1]\rangle$ and $|[r^2]\rangle$. All curves use exact twisted
 partition-function contractions.}
 \label{fig:r2proxy-raw}
\end{figure*}

\begin{figure*}[t]
 \centering
 \includegraphics[width=0.8\textwidth]{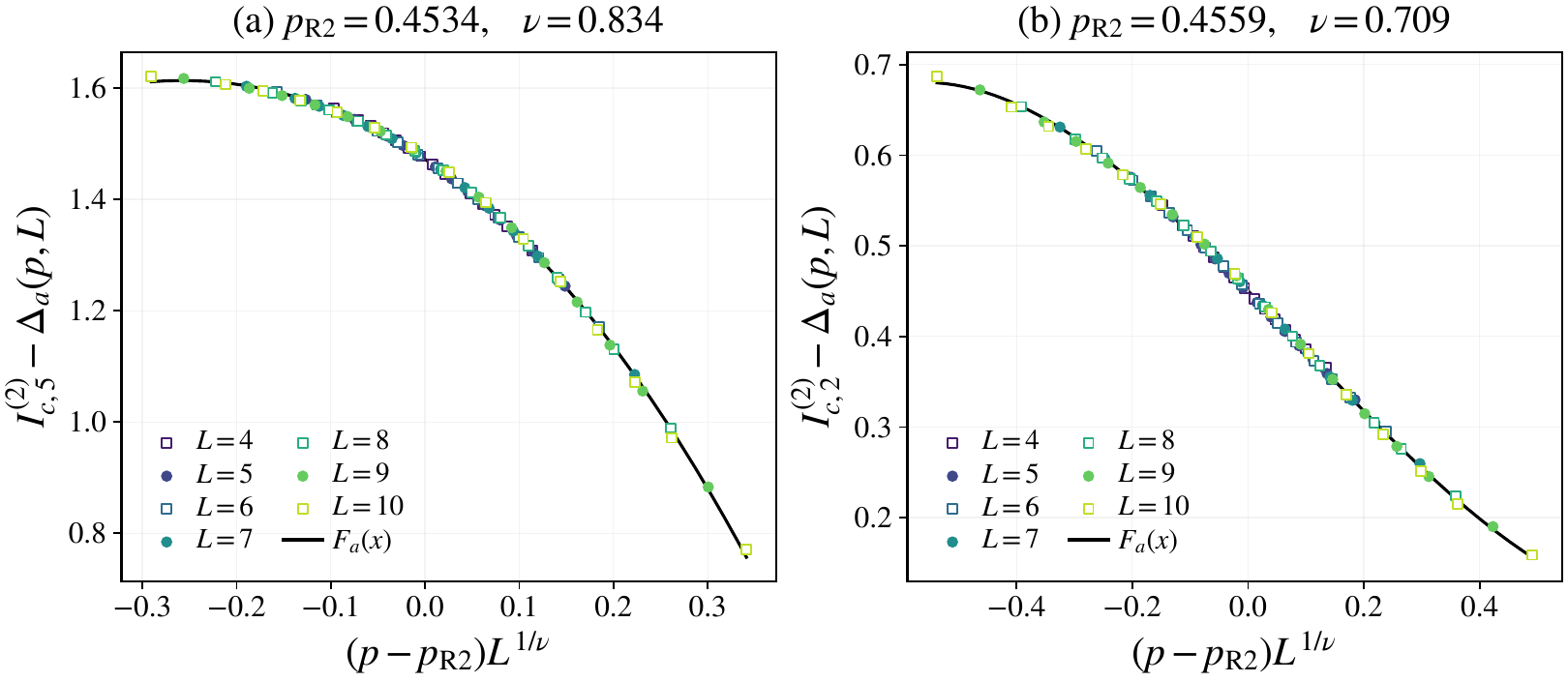}
 \caption{Finite-size scaling for (a) Full five-dimensional logical input and (b) Central-qubit input near the R\'enyi-2 transition, with $0.435\leq p\leq0.475$. The horizontal coordinate is $x=(p-p_{\text{R}2})L^{1/\nu}.$
 Filled circles and open squares distinguish between odd $L$ and even $L$.}
 \label{fig:r2proxy-collapse}
\end{figure*}

We calculate the R\'enyi-2 coherent information without constructing
the full density matrices. Instead, both purities reduce to a small
set of classical partition functions, evaluated by exact
tensor-network contraction.

To see which quantities are needed, define
\begin{equation}
 \begin{aligned}
 X_{CD}&=\mathcal N_{E,p}(|C\rangle\langle D|),\\
 A_{CD}&=\operatorname{Tr}(X_{CD}X_{CD}^{\dagger}).
 \end{aligned}
 \label{eq:r2proxy-Adef}
\end{equation}
For $C=D$, $A_{CC}$ is the purity of the noisy logical basis
state. For $C\ne D$, $A_{CD}$ is the squared Hilbert--Schmidt
norm of the surviving coherence between logical states $C$
and $D$. 

The joint reference--system purity is
$d^{-2}\sum_{C,D}A_{CD}$, whereas the system purity is
$d^{-2}\sum_C A_{CC}$. The latter contains only diagonal terms
because different holonomy classes occupy orthogonal
configuration sectors, which the noise channel does not mix.
Consequently,
\begin{equation}
 I_{c,d}^{(2)}
 =
 \ln\!\left[
 \frac{\displaystyle\sum_{C,D\in\mathcal I_d}A_{CD}}
      {\displaystyle\sum_{C\in\mathcal I_d}A_{CC}}
 \right].
 \label{eq:r2proxy-ratio-from-A}
\end{equation}
Thus the calculation compares the total purity weight,
including logical coherences, with the diagonal contribution.

In Appendices~\ref{sec:r2proxy-purities}
and~\ref{sec:r2proxy-partition}, we rewrite $A_{CD}$ as the
following classical partition function with $D_4$ valued spins $u_v$ on the vertices of the lattice:
\begin{equation}
 \begin{aligned}
 A_{CD}&=A(g_C,g_D)\\
 &=\frac1{8^{N_v}}
 \sum_{\{u_v\in D_4\}}
 \prod_e
 w_p\!\left(
 u_{i_e}g_C^{c_e}u_{f_e}^{-1}g_D^{-c_e}
 \right),
 \end{aligned}
 \label{eq:r2proxy-main-partition}
\end{equation}
where $g_C\in C$ and $g_D\in D$ are fixed conjugacy class representatives, $c_e \in \{0, 1\}$ is an indicator variable for edges that cross a fixed dual seam running from the top to the bottom boundary, and
\begin{equation}
 w_p(x)=|K_{E,p}(x)|^2
 =\left|(1-p)+\frac p2\chi_E(x)\right|^2.
\end{equation}

Evaluating this sum gives the purity of the noisy logical
basis state when $C=D$, and the squared norm of its
surviving coherence with sector $D$ when $C\ne D$.
The variables $u_v$ are mathematical summation variables
introduced for this calculation.

Symmetry organizes the 25 pairs of conjugacy classes $C,D$ into five twist types:
\begin{align}
 A_I&=A(1,1),&
 A_{F_4}&=A(1,r^2),\notag\\
 A_{F_2}&=A(r,r),&
 A_+&=A(1,s),\notag\\
 &&A_-&=A(1,r),
 \label{eq:r2proxy-types}
\end{align}
Summing the appropriate diagonal and off-diagonal entries gives
\begin{equation}
 I_{c,5}^{(2)}
 =
 \ln\!\left(
 1+\frac{2A_{F_4}+10A_++8A_-}{2A_I+3A_{F_2}}
 \right)
 \label{eq:r2proxy-full}
\end{equation}
for the full qudit, and
\begin{equation}
 I_{c,2}^{(2)}
 =
 \ln\!\left(1+\frac{A_{F_4}}{A_I}\right)
 \label{eq:r2proxy-central}
\end{equation}
for the central qubit. The numerical coefficients count the occurrences of each twist type in the $5\times 5$ matrix $A_{CD}$, displayed explicitly in
Appendix~\ref{sec:r2proxy-partition}.

Numerically, we evaluate these partition functions by summing
the spins row by row, retaining the complete boundary
weights between successive rows. The local contraction and
exact symmetry reductions are specified in
Appendix~\ref{sec:r2proxy-numerics}.

The raw data in Fig.~\ref{fig:r2proxy-raw} and the corrected
scaling collapses in Fig.~\ref{fig:r2proxy-collapse} support a
common lower R\'enyi-2 transition near $p\simeq0.46$ for both the central logical qubit ($d=2$) and the full logical qudit ($d=5$). At fixed aspect ratio, we use
\cite{FisherBarber1972FSS,Wegner1972Corrections}
\begin{equation}
 \begin{aligned}
 I_{c,d}^{(2)}(p,L)
 &=F_d(x)+\Delta_d(p,L)+\cdots,\\
 x&=(p-p_{\mathrm{R}2})L^{1/\nu},\\
 \Delta_d(p,L)&=L^{-\omega}G_d(x).
 \end{aligned}
 \label{eq:r2proxy-fss}
\end{equation}
The representative fits use a cubic $F_d$, linear $G_d$,
and $\omega=1$. Separate fits give
$p_{\mathrm{R}2}\simeq0.4534$ for $d=5$ and $0.4559$ for
$d=2$, while a shared-parameter fit gives approximately
$0.4545$. These estimates are consistent with a common
transition on the $p\simeq0.45$--$0.46$ scale. Further details regarding the finite-size scaling are presented in Appendix~\ref{sec:r2proxy-scaling}.

\subsubsection{Physical interpretation}
\label{sec:r2proxy-meaning}

The physical interpretation of the R\'enyi-2 transition concerns the survival of coherence in the logical subspace. In fact,
\begin{equation}
 e^{I_{c,d}^{(2)}}-1
 =\frac{\displaystyle\sum_{C\ne D}\|X_{CD}\|_2^2}
        {\displaystyle\sum_C\|X_{CC}\|_2^2},
 \label{eq:r2proxy-coherence-ratio}
\end{equation}
where $X_{CD}$ is defined in Eq.~\ref{eq:r2proxy-Adef}. The numerator measures the remaining off-diagonal logical coherences and
the denominator measures the diagonal sector weights in the same
purity norm. For the central qubit this compares the coherence between
$|[1]\rangle$ and $|[r^2]\rangle$ with their populations. For the full
qudit it tests coherences among all five holonomy classes. Equivalently,
the twisted partition functions measure the cost of changing the
relative logical holonomy between the two replicas used to compute a
purity. Suppression of these twisted contributions signals loss of
purity-weighted logical coherence.

There is also a direct interpretation in terms of information leaked to the environment. As shown in
Appendix~\ref{app:renyi2-purity-derivation}, $A_{CD}$ is the overlap of the environment states conditioned on $C$ and $D$. At finite size, $I_{c,d}^{(2)}=\ln d$ means those states are identical and the logical state is perfectly recoverable. The opposite endpoint,
$I_{c,d}^{(2)}=0$, means that every off-diagonal block $X_{CD}$ vanishes: the logical basis label survives, but its quantum superpositions have
been completely dephased. The observed common transition scale diagnoses the noisy state's purity-weighted coherence; the von Neumann and optimal-recovery thresholds require a separate analysis.

An important open question is whether the intrinsic loss of the full
logical qudit and that of the central qubit occur at distinct noise
strengths. Since recovery of the full logical space also recovers its
central subspace and failure to recover the central subspace indicates failure to recover the rest of the logical space, the intrinsic recovery thresholds satisfy
\begin{equation}
 p_{c,\mathrm{full}}^{\mathrm{vN}}
 \leq
 p_{c,\mathrm{cent}}^{\mathrm{vN}}.
\end{equation}
A strict inequality would allow an intermediate regime in which the
central qubit remains coherently recoverable, while arbitrary quantum
information in the full five-dimensional logical space does not.
Such a regime would represent partial loss of logical coherence,
rather than complete destruction of the encoded quantum information.

Our R\'enyi-2 numerics instead support a common transition scale near
$p\simeq0.46$, with no resolved intermediate regime. This suggests simultaneous loss of the full qudit and the central qubit. However, the R\'enyi-2 diagnostic weights the
spectrum differently from the von Neumann coherent information, so
the apparent coincidence does not establish equality of the intrinsic
recovery thresholds. Therefore, the existence of an intermediate regime where the central qubit remains recoverable but not the rest of the logical qudit still remains a possibility.

\section{Discussion}
\label{sec:discussion}

In this work we study ML-type decoders for non-Abelian topological order. In contrast to the Abelian topological orders with incoherent Pauli noise,  where the density matrix can be diagonalized in an obvious syndrome basis based on local anyon excitations, the decohered non-Abelian topological orders do not admit such simple optimal syndromes. We show that generally, thresholds can be improved for decoders of non-Abelian topological orders by using finer syndrome sets that include anyon fusion measurements. For the central qubit in the $D_4$ quantum double model on a cylinder, the three syndrome records considered here are progressively more informative and remain logically non-demolition, allowing the threshold for this subspace to reach a remarkably high threshold of over $20\%$ noise rate. Unfortunately, this comes at the cost of disturbing logical information stored in the rest of the logical qudit via processes that are intrinsic to non-Abelian topological orders. In this model, below a certain threshold, fusion measurements also allow for single-shot decoding of the full logical qudit, meaning error correction can be done with only one round of measurements. This is due to the fact that we do not need to wait until after we bring together two non-Abelian anyons to find their fusion outcome -- if we sacrifice some locality, we can measure all of their joint fusion outcomes at the beginning, before movement.

Our work leaves many open questions related to possible improvements on the decoders studied in this work, generalization to other topological orders and noise models, and mixed-state phases of non-Abelian topological orders, which we outline below.

\subsection{Improving the decoders in this work}
In this work, we established single-shot decoding of the full logical qudit for the $D_4$ quantum double with the non-Abelian $E$-charge decoherence below the MWPM and HMWPM threshold. When we include fusion measurements, the threshold of the central qubit can be significantly improved, whereas the logical information stored outside of the central qubit can be disturbed above the MWPM threshold. It would be interesting to study optimal recovery of the full logical qudit in more depth, and to try to prove the success of single-shot decoding at least up to the $S^{(2)}$ threshold (since $S^{(2)}$ would not include such dangerous fusion measurements). More generally, it might be useful to study decoders that preserve a particular subspace to high thresholds even at the expense of the rest of the logical qudit.

\subsection{More general topological orders and noise models}
There are several questions regarding how to generalize decoding to more complex topological orders and noise models. First, in this work, we only studied in detail charge decoherence of the $D_4$ quantum double. In this case, measurement of the fusion tree simplifies because the non-Abelian anyons only fuse to Abelian outcomes. As a result, we only need to measure pairwise fusion; the rest of the fusion tree is completely determined by the pairwise fusion outcomes. For other topological orders that include cyclic anyons, i.e. an anyon produces itself under fusion, we cannot determine the full fusion tree by just pairwise measurements. However, below the information-theoretic threshold, we expect that the anyons generally form clusters of size $\leq\mathcal{O}(\log L)$ with collective vacuum fusion product, so there might be a way to optimally choose the measured fusion tree basis to keep measurements only up to $\mathcal{O}(\log L)$. In Appendix~\ref{app:s3-syndrome-probabilities}, we present a calculation of syndrome probabilities for the syndrome sets $\mathcal S^{(1)},\mathcal S^{(2)},$ and $\mathcal S^{(3)}$ for 2d charge decoherence in the $S_3$ quantum double. In this case, since the charge anyon is cyclic, we are unable to further decompose this full syndrome probability into different classes based on logical operators/homology.

There may also be a balance between size/number of fusion measurements and the gain in threshold. Since we saw that even on-site $E$ fusion measurements in the $D_4$ quantum double improved the threshold of the central qubit, it might be the case that even including only finite-range fusion measurements (i.e. only measure fusion outcomes of anyons in finite-radius clusters) can lead to a nonzero threshold improvement. 

In addition to studying more general topological orders, we can also study more general noise channels. Including flux noise is more subtle in non-Abelian topological orders than in Abelian ones because flux and charge noise do not decouple -- fluxes can fuse to produce charge. However, we expect that the general technique of using fusion measurements to aid with error correction would still improve thresholds. We leave a detailed study of correcting flux and more general dyon noise to future work. 

It would also be interesting to study how robust the threshold improvements are to modifications of the noise channel. For example, in the $D_4$ quantum double with $E$ noise, on-site Abelian charge measurements significantly improve the threshold because these give hard constraints to where the error string must lie. But if there is also Abelian charge noise, the Abelian anyons may not come only from non-Abelian anyon fusion. They therefore do not put as stringent constraints on the non-Abelian error strings. We conjecture that allowing for incoherent Abelian charge noise would bring the thresholds of the $S^{(1)}, S^{(2)}$, and $S^{(3)}$ decoders closer, and at a critical nonzero Abelian noise strength, the thresholds for the three decoders may coincide. These fusion-assisted decoders then may be particularly useful for noise biased toward certain charges~\cite{sala_stability_2025,vadali2026statisticalmechanicssymmetriesnonabelian}.

Finally, in any realistic setting, measurement errors are unavoidable. In ML-type decoders, we assume that the measurements and subsequent recovery are perfect, and the only source of noise comes from finite time exposure to decoherence before the error correction. Measurement error requires keeping measurement records as a function of time, to discern which detected anyons are actual errors in the state vs errors in the measurement. In Abelian topological orders, correction can be implemented at the end using the full space-time syndrome record, but in non-Abelian topological orders, correction must be implemented ``just-in-time" to avoid too much non-Abelian anyon braiding~\cite{brown2020fault,davydova_universal_2025}. Fusion measurements may also assist in the design of optimal just-in-time decoders.

\subsection{Statistical mechanics models and mixed state phases}
Even though we have developed a mapping of syndrome probabilities to classical statistical mechanics models, we have not determined the mixing times of these statistical mechanics models. Because there are hard constraints in the models (i.e. that the Abelian labels must lie on the error strings), one might try to devise efficient Monte Carlo sampling algorithms by staying within this constrained subspace. This is nontrivial, however, because we must also ensure ergodicity within this constrained subspace, meaning that the Monte Carlo moves used must not only keep states within this subspace but also explore all of the subspace. 

Next, while we evaluated the characters and dimensions in the expressions for the syndrome probabilities to obtain a stat mech model mapping for the $D_4$ quantum double, it would be interesting to investigate whether there might be a general further rewriting of the stat mech model in terms of $\mathrm{Tr}(e^{-\beta H})$ where $H$ is some spin Hamiltonian. The ML decoders presented here give stat mech models with a ``Nishimori" condition corresponding to the fact that the decoder knows the precise noise model, and these stat mech models might be interesting in their own right. Such a mapping might also aid the development of numerical algorithms mentioned previously. From the perspective of stat mech models, the increased robustness of the topologically ordered phase with decoders that record finer fusion measurements maps to an increased robustness of the ordered phase. However, as we see explicitly in the $D_4$ syndrome probabilities, the ferromagnetic coupling remains the same for $S^{(1)},S^{(2)},$ and $S^{(3)}$, and is completely determined by the noise strength $p$. Heuristically, we may view the increased robustness of the ordered phase as coming from decreasing entropy rather than increasing energy in the expression for free energy $F=E-TS$, since we are adding more labels and constraints. It would be interesting to make this more precise from the stat mech perspective.

In the context of many-body physics, decodability is often associated with a mixed-state phase~\cite{fan_diagnostics_2024,sang_stability_2025,wang2025fractionalquantumhallstates}. While the noisy state may have some intrinsic information-theoretic threshold below which there is in principle a (possibly impractical but nevertheless $\log L$-local) decoder, a given decoder may see a different mixed state. Even when a decoding problem admits an ML decoder (i.e. it is sufficient to weight classical homology probabilities), it may not see $\mathcal{N}[\rho_0]$ but rather see the $\mathcal{S}$-dependent effective density matrix
\begin{equation}
    \rho_{\mathcal{S}}=\sum_{\mathcal{S},h}\mathrm{Pr}(\mathcal{S},h)|\psi_{\mathcal{S},h}\rangle\langle\psi_{\mathcal{S},h}|
\end{equation}
where $|\psi_{\mathcal{S},h}\rangle$ is a state giving the syndrome $\mathcal{S}$ and in the error homology class $h$. We may then study these effective density matrices from the perspective of mixed state phases of matter, and compute Renyi-2 diagnostics that serve as proxies for the thresholds. These may be easier to compute numerically, because they would correspond to annealed disorder models rather than quenched disorder models. 

\section*{Acknowledgements}
We thank Ehud Altman, Curt von Keyserlingk, Alexey Khudorozhkov, Chiu Fan Bowen Lo, Tibor Rakovszky, Pablo Sala, Ruben Verresen, and Ashvin Vishwanath for helpful discussions. We thank the authors of Ref.~\cite{pablo} for coordinating their submission with ours. RS acknowledges support from the National Science Foundation Graduate Research Fellowship Program. ZS is supported by the U.S. National Science Foundation under Grant No.~DRL I-TEST 2148467. TCL acknowledges support from the RQS Postdoctoral Fellowship through the National Science Foundation QLCI Grant No. OMA-2120757. CZ is supported by the Harvard Society of Fellows and NSERC grants RGPIN-2026-07143 and DGECR-2026-00535. CZ would like to thank the Isaac Newton Institute for Mathematical Sciences, Cambridge, for support and hospitality during the programme Mathematics of Many-Body Entanglement, where work on this paper was undertaken. This work was supported by EPSRC grant EP/Z000580/1. CZ also acknowledges the hospitality of the Aspen Center for Physics, which is supported by National Science Foundation grant PHY-2210452 and the Simons Foundation (1161654, Troyer), and the Kavli Institute for Theoretical Physics, which is supported in part by grant NSF PHY-2309135. ChatGPT 5.5 and above were used to assist with writing and numerics, as well as check analytic calculations.

\bibliography{refs}

\newpage

\onecolumngrid
\appendix
\addtocontents{toc}{\protect\AppendixTOCOnly}

\section{Onsite-syndrome probabilities for a general group}
\label{app:onsite-probability}

This appendix discusses the trace calculation behind
Sec.~\ref{sec:onsiteprob}. Starting from an arbitrary logical density
matrix, we expand the physical noisy state and each onsite projector in
the edge basis, evaluate the matrix element, and obtain the
holonomy-dependent partition function. We then restrict to logical states in the center and derive the onsite syndrome probabilities used in
Secs.~\ref{sec:syndrome-decoding-main} and~\ref{sec:decoding}.

\subsection{Logical input and explicit trace expansion}

Let $\rho_L$ be any normalized density operator on the cylinder ground
space,
\begin{equation}
 \rho_L=\sum_{C,D\in\operatorname{Conj}(G)}
 \rho_{CD}\ket C\!\bra D,
 \qquad \sum_C\rho_{CC}=1,
 \label{eq:general-logical-density-derivation}
\end{equation}
where $\ket C\equiv\ket{[g_C]}$ and $g_C$ is a representative of the
conjugacy class $C$. $\rho_L$ is an arbitrary logical state. Let $\mathcal O_C$
be the set of flat edge configurations appearing in $\ket C$, and set
$N_C=|\mathcal O_C|$. In the edge basis,
\begin{equation}
 \rho_L
 =\sum_{C,D\in\operatorname{Conj}(G)}
 \frac{\rho_{CD}}{\sqrt{N_CN_D}}
 \sum_{\substack{\mathbf g\in\mathcal O_C\\
                  \mathbf k\in\mathcal O_D}}
 \ket{\mathbf g}\!\bra{\mathbf k}.
 \label{eq:general-logical-projector-orbits}
\end{equation}

The noisy density matrix is obtained without any further assumption
on the logical state:
\begin{equation}
 \rho_p(\rho_L)=\sum_{C,D}\frac{\rho_{CD}}{\sqrt{N_CN_D}}
 \sum_{\substack{\mathbf g\in\mathcal O_C\\\mathbf k\in\mathcal O_D}}
 \left[\prod_eK_{\Gamma,p}(g_ek_e^{-1})\right]
 \ket{\mathbf g}\!\bra{\mathbf k}.
 \label{eq:app-onsite-noisy-density}
\end{equation}
To evaluate $\Tr[P_{\mathcal S^{(2)}}\rho_p(\rho_L)]$, we insert the onsite projectors directly into the Born rule.  Writing
$U(\mathbf h)=\prod_vA_v^{h_v}$, Eqs.~\eqref{eq:single-charge-kernel} and
\eqref{eq:central-charge-projector} give
\begin{equation}
 \Pr_p(\mathcal S^{(2)}\mid\rho_L)
 =\frac{1}{|G|^{N_v}}
 \sum_{C,D}
 \frac{\rho_{CD}}{\sqrt{N_CN_D}}
 \sum_{\{h_v\in G\}}
 \sum_{\substack{\mathbf g\in\mathcal O_C\\
                  \mathbf k\in\mathcal O_D}}
 \left[\prod_v d_{\lambda_v}
 \chi_{\lambda_v}(h_v^{-1})\right] \left[\prod_eK_{\Gamma,p}(g_ek_e^{-1})\right]
 \langle\mathbf k|U(\mathbf h)|\mathbf g\rangle .
 \label{eq:general-logical-projector-Born-rule}
\end{equation}
On an oriented edge $e=\langle i_ef_e\rangle$, the vertex action is
\begin{equation}
 g_e\longmapsto g_e^{\mathbf h}
 =h_{i_e}g_eh_{f_e}^{-1}.
 \label{eq:general-logical-projector-vertex-action}
\end{equation}
Consequently,
\begin{equation}
 \langle\mathbf k|U(\mathbf h)|\mathbf g\rangle
 =\delta_{\mathbf k,\mathbf g^{\mathbf h}},
 \qquad
 \langle\mathbf k|U(\mathbf h)|\mathbf g\rangle\neq0
 \Longrightarrow C=D.
 \label{eq:general-logical-projector-matrix-element}
\end{equation}
The delta function sets $\mathbf k=\mathbf g^{\mathbf h}$. Vertex actions
can conjugate the circumferential holonomy but cannot change its conjugacy
class, so the same matrix element also requires $C=D$. This implies that the syndrome probability depends on the logical populations $\rho_{CC}$ instead of the coherences between distinct holonomy classes.

\subsection{Evaluating a fixed logical sector}

It remains to evaluate one diagonal sector $C$.  Fix a top-to-bottom dual
cut $\mathcal{C}$, let $c_e=1$ on an oriented edge crossing the cut and $c_e=0$
otherwise, and choose a representative $g_C\in C$.  A flat configuration in
this sector can be written as
\begin{equation}
 g_e=s_{i_e}g_C^{c_e}s_{f_e}^{-1}.
 \label{eq:general-holonomy-flat-coordinates}
\end{equation}
The intuition here is that we are inserting a twist into the cylinder along the cut $\mathcal{C}$ to ensure that the circumferential holonomy lies in the conjugacy class $C$. We define $r_v=s_v^{-1}h_vs_v$ for future convenience. 

For the surviving term $\mathbf k=\mathbf g^{\mathbf h}$ from Eq.~\eqref{eq:general-logical-projector-matrix-element}, one then finds
\begin{equation}
 \begin{aligned}
 g_ek_e^{-1}
 &=g_eh_{f_e}g_e^{-1}h_{i_e}^{-1}\\
 &=s_{i_e}
 \left(g_C^{c_e}r_{f_e}g_C^{-c_e}r_{i_e}^{-1}\right)
 s_{i_e}^{-1}.
 \end{aligned}
 \label{eq:general-holonomy-relative-edge}
\end{equation}
Both $K_{\Gamma,p}$ and the irreducible characters are class functions, so
\begin{equation}
 \begin{aligned}
 K_{\Gamma,p}(g_ek_e^{-1})
 &=K_{\Gamma,p}
 \left(g_C^{c_e}r_{f_e}g_C^{-c_e}r_{i_e}^{-1}\right),\\
 \chi_{\lambda_v}(h_v^{-1})
 &=\chi_{\lambda_v}(r_v^{-1}).
 \end{aligned}
 \label{eq:general-holonomy-class-functions}
\end{equation}
For each $\mathbf g$, the change of variables
$\{h_v\}\mapsto\{r_v\}$ is bijective.  The resulting sum is independent of
$\mathbf g$; it also depends only on the conjugacy class $C$, since replacing
$g_C$ by a conjugate can be absorbed into a simultaneous conjugation of all
$r_v$.  The orbit sum therefore cancels its normalization.  Defining the
sector partition function by the second line below, the general result is
\begin{equation}
 \begin{aligned}
 \Pr_p(\mathcal S^{(2)}\mid\rho_L)
 &=\sum_{C\in\operatorname{Conj}(G)}
 \rho_{CC}\,\mathcal Z_C(p,\mathcal S^{(2)}),\\[1mm]
 \mathcal Z_C(p,\mathcal S^{(2)})
 &=\frac{1}{|G|^{N_v}}
 \sum_{\{r_v\in G\}}
 \prod_v d_{\lambda_v}\chi_{\lambda_v}(r_v^{-1})
 \prod_{e=\langle i_ef_e\rangle}
 K_{\Gamma,p}
 \left(g_C^{c_e}r_{f_e}g_C^{-c_e}r_{i_e}^{-1}\right).
 \end{aligned}
 \label{eq:general-logical-projector-syndrome-probability-derivation}
\end{equation}

\subsection{Restriction to central holonomy}

We now restrict to the logical subspace spanned by central holonomies: 
\begin{equation}
 \mathcal H_Z
 =
 \operatorname{span}\bigl\{
 \ket{[z]}:z\in Z(G)
 \bigr\}.
\end{equation}
For \(z\in Z(G)\), the conjugation appearing in
Eq.~\eqref{eq:general-logical-projector-syndrome-probability-derivation} is trivial:
\begin{equation}
 z^{c_e}r_{f_e}z^{-c_e}r_{i_e}^{-1}
 =
 r_{f_e}r_{i_e}^{-1}.
\end{equation}
Consequently, every central-holonomy sector has the same syndrome partition function,
\begin{equation}
 \begin{aligned}
 \mathcal Z_{\mathrm{cen}}
 (p,\mathcal S^{(2)})
 =
 \frac{1}{|G|^{N_v}}
 \sum_{\{r_v\in G\}}
 \prod_v
 d_{\lambda_v}\chi_{\lambda_v}(r_v^{-1})
 \prod_{e=\langle i_ef_e\rangle}
 K_{\Gamma,p}
 \bigl(r_{f_e}r_{i_e}^{-1}\bigr).
 \end{aligned}
 \label{eq:central-holonomy-syndrome-probability-derivation}
\end{equation}
If \(\rho_L\) is any normalized logical density operator supported on
\(\mathcal H_Z\), Eq.~\eqref{eq:general-logical-projector-syndrome-probability-derivation}
therefore reduces to
\begin{equation}
 \Pr_p(\mathcal S^{(2)}\mid\rho_L)
 =
 \sum_{z\in Z(G)}
 \rho_{zz}\,
 \mathcal Z_{\mathrm{cen}}
 =
 \mathcal Z_{\mathrm{cen}},
\end{equation}
where we used \(\sum_{z\in Z(G)}\rho_{zz}=1\). Thus the onsite syndrome
distribution is independent of both the populations and coherences of the
logical state within the central-holonomy subspace.

\subsection{Projection onto central winding sectors}
\label{app:onsite-winding}

This subsection justifies the local edge modification in
Sec.~\ref{sec:decoding}. We first expand the central-sector likelihood in occupied edges,
\begin{equation}
 \Pr_p(y)=\sum_{F\subseteq\mathcal E}w_p(F)\,\mathcal A_y(F),
 \qquad w_p(F)=p^{|F|}(1-p)^{N_e-|F|},
 \label{eq:app-onsite-occupied-weight}
\end{equation}
where $y=\mathcal S^{(2)}$ and
\begin{equation}
 \mathcal A_y(F)=\frac{1}{|G|^{N_v}}
 \sum_{\{r_v\}}\left[\prod_vd_{\lambda_v}
 \chi_{\lambda_v}(r_v^{-1})\right]
 \prod_{e\in F}\frac{\chi_\Gamma(r_{f_e}r_{i_e}^{-1})}{d_\Gamma}.
 \label{eq:app-onsite-fixed-support}
\end{equation}
This is the conditional Born probability for the fixed set of noisy
edges, so the final coefficient $\mathcal A_y(F)$ is nonnegative even
when its character expansion has complex terms.

For an oriented charge channel, assign to an occupied edge the central
character $\omega_\Gamma\in\widehat{Z(G)}$. Reversing its orientation
replaces this character by its inverse. Local fusion conserves these
characters. Hence, once the measured charges have been removed along
a fixed reference correction, the remaining logical action is a
character $a\in\widehat{Z(G)}$. Let
\begin{equation}
 n_{\mathcal C}(F)=\sum_{e\in F} c_e,
 \qquad a_F(z)=\omega_\Gamma(z)^{n_{\mathcal C}(F)}.
\end{equation}
Choose a compatible reference with character $a_0$; the residual
character is $a_Fa_0^{-1}$. Define its sector weights by
\begin{equation}
 Z_y(a)=\sum_{F:\,a_Fa_0^{-1}=a}w_p(F)\mathcal A_y(F).
 \label{eq:app-general-winding-weights}
\end{equation}
The finite Fourier transform and its inverse are
\begin{equation}
 \widetilde Z_y(z)=\sum_{a\in\widehat{Z(G)}}a(z)Z_y(a),
 \qquad
 Z_y(a)=\frac{1}{|Z(G)|}\sum_{z\in Z(G)}a(z)^*\widetilde Z_y(z).
 \label{eq:app-general-winding-fourier}
\end{equation}
Substituting the occupied-edge expansion and distributing its phase
over the edges gives
\begin{equation}
 \widetilde Z_y(z)=\frac{a_0(z)^{-1}}{|G|^{N_v}}
 \sum_{\{r_v\}}\prod_vd_{\lambda_v}\chi_{\lambda_v}(r_v^{-1})
 \prod_{e=\langle i_ef_e\rangle}
 \left[(1-p)+\frac{p}{d_\Gamma}\omega_\Gamma(z)^{c_e}
 \chi_\Gamma(r_{f_e}r_{i_e}^{-1})\right].
 \label{eq:app-general-winding-local-product}
\end{equation}
The transformed weights need not be
probabilities, but the inverse transform returns the nonnegative
sector weights. If the noise accesses only a subgroup of
$\widehat{Z(G)}$, the other sector weights vanish. For $D_4$ with
$\Gamma=E$, $Z(G)=\{1,r^2\}$ and $\omega_E(r^2)=-1$; these equations
reduce to the two-point transform and seam signs of Sec.~\ref{sec:decoding}.

\section{Fusion-syndrome probabilities for disjoint ribbons}
\label{app:fusion-probability}

This appendix derives the joint likelihood stated in
Sec.~\ref{sec:fusion-syndrome-probabilities}, following the same
edge-basis trace calculation as Appendix~\ref{app:onsite-probability}.
We resolve the diagonal spine projectors, evaluate the combined ribbon
and onsite action, and perform the remaining group sums explicitly.
The calculation permits arbitrary charges on the ribbon interiors;
its assumptions are a central logical input and disjoint microscopic
ribbon supports. This derivation provides the framework for performing the ML decoding for the central logical qubit discussed in Sec.~\ref{sec:d4-central-numerics}.

\subsection{Input state, record, and projector trace}

Let the initial state be an arbitrary normalized density operator on
\begin{equation}
 \mathcal H_Z
 =\operatorname{span}\{\ket{[z]}:z\in Z(G)\},
 \label{eq:S3-central-logical-subspace}
\end{equation}
written as
\begin{equation}
 \rho_L^{\mathrm{cen}}
 =\sum_{z,z'\in Z(G)}\rho_{zz'}\ket{[z]}\!\bra{[z']},
 \qquad \sum_z\rho_{zz}=1.
 \label{eq:S3-arbitrary-central-state-derivation}
\end{equation}
As in the calculation for the onsite syndrome distribution, let \(\mathcal O_z\) be the orbit of flat
edge configurations with holonomy \(z\), and let \(N_z=|\mathcal O_z|\).
Then the initial state can be expressed as 
\begin{equation}
 \rho_L^{\mathrm{cen}}
 =\sum_{z,z'}\frac{\rho_{zz'}}{\sqrt{N_zN_{z'}}}
 \sum_{\substack{\mathbf g\in\mathcal O_z\\
                  \mathbf k\in\mathcal O_{z'}}}
 \ket{\mathbf g}\!\bra{\mathbf k}.
 \label{eq:S3-central-state-edge-basis}
\end{equation}

For a fixed onsite record
\(\mathcal S^{(2)}=(\gamma,\mathbf a)\), the protocol chooses a
deterministic pairing and path family,
\begin{equation}
 M=\operatorname{pair}(\gamma)
 =\{m=(i_m,j_m):m=1,\ldots,N_M\}.
 \label{eq:S3-general-pair-set}
\end{equation}
We assume the microscopic ribbon supports, including their spine
projectors, are disjoint. For each pair $m \in M$, let $p_{i,j}$ be its prescribed oriented path from $i_m$ to $j_m$. Define
\begin{equation}
 C_m=V(m),\qquad
 \mathcal P=\bigsqcup_{m\in M}C_m.
 \label{eq:S3-ribbon-vertex-sets-derivation}
\end{equation}
The refined record is
\begin{equation}
 \mathcal S^{(3)}=(\gamma,\mathbf a,\mathbf q),
 \qquad \mathbf q=\{q_m:m\in M\},
 \label{eq:S3-record-projector-outcomes}
\end{equation}
where \(q_m\) is the total electric charge of the complete region
\(C_m\).  Because the onsite projectors and fusion projectors commute, its Born
probability is
\begin{equation}
 \Pr_p(\mathcal S^{(3)}\mid\rho_L^{\mathrm{cen}})\quad=
 \operatorname{Tr}\!\left[
 \rho_p(\rho_L^{\mathrm{cen}})
 \left(\prod_vP_v^{\lambda_v}\right)
 \left(\prod_{m\in M}P_{m}^{q_m}\right)
 \right].
 \label{eq:S3-projector-trace-derivation}
\end{equation}
The paths are fixed by the observed onsite record and the deterministic protocol, so there is no additional probability factor for choosing them.

\subsection{Expanding the trace}

Insert both character-projector expansions,
\begin{equation}
 P_v^{\lambda_v}
 =\frac{d_{\lambda_v}}{|G|}
   \sum_{\widetilde h_v\in G}
   \chi_{\lambda_v}(\widetilde h_v^{-1})A_v^{\widetilde h_v},\qquad P_{m}^{q_m}
 =\frac{d_{q_m}}{|G|}
   \sum_{\widetilde k_m\in G}
   \chi_{q_m}(\widetilde k_m^{-1})
   \mathcal F_{m}(\widetilde k_m).
\end{equation}
Write the operator in one term of this expansion as
\begin{equation}
 \mathcal O(\widetilde{\mathbf h},\widetilde{\mathbf k})
 =
 \left(\prod_vA_v^{\widetilde h_v}\right)
 \left(\prod_{m\in M}\mathcal F_{m}(\widetilde k_m)\right).
 \label{eq:S3-operator-term}
\end{equation}
The channel multiplies \(\ket{\mathbf g}\!\bra{\mathbf k}\) by
\(\prod_eK_{\Gamma,p}(g_ek_e^{-1})\), while $\operatorname{Tr}\!\left[
 \ket{\mathbf g}\!\bra{\mathbf k}\,\mathcal O
 \right]
 =\bra{\mathbf k}\mathcal O\ket{\mathbf g}$, so the fully expanded trace is
\begin{equation}
 \begin{aligned}
 \Pr_p(\mathcal S^{(3)}\mid\rho_L^{\mathrm{cen}})
 &=\frac{1}{|G|^{N_v}}
  \left(\prod_{m\in M}\frac{d_{q_m}}{|G|}\right)
 \sum_{z,z'}\frac{\rho_{zz'}}{\sqrt{N_zN_{z'}}}
 \sum_{\{\widetilde h_v\}}\sum_{\{\widetilde k_m\}}\\
 &\sum_{\substack{\mathbf g\in\mathcal O_z\\
                  \mathbf k\in\mathcal O_{z'}}}
 \left[\prod_vd_{\lambda_v}
       \chi_{\lambda_v}(\widetilde h_v^{-1})\right]
 \left[\prod_{m\in M}\chi_{q_m}(\widetilde k_m^{-1})\right]
 \left[\prod_eK_{\Gamma,p}(g_ek_e^{-1})\right]
 \bra{\mathbf k}
 \mathcal O(\widetilde{\mathbf h},\widetilde{\mathbf k})
 \ket{\mathbf g}.
 \end{aligned}
 \label{eq:S3-expanded-Born-rule-main}
\end{equation}
All group-variable sums run over \(G\).
This is the direct fusion analogue of
Eq.~\eqref{eq:general-logical-projector-Born-rule}; the extra character
factors and ribbon operators come from the fusion measurement.

\subsection{Resolving the microscopic ribbon action}

We next evaluate the matrix element, just as in the onsite calculation.
For a configuration \(\mathbf g\), the diagonal spine projectors
select its actual path labels:
\begin{equation}
 \mathsf T_{m}(\mathbf x)\ket{\mathbf g}
 =\delta_{\mathbf x,\mathbf x_{m}(\mathbf g)}
  \ket{\mathbf g}.
 \label{eq:S3-spine-projector-resolution}
\end{equation}
Hence the sum over spine labels in each ribbon leaves exactly one term.
Let \(X_{m,v}(\mathbf g)\) be the ordered product along the spine
from \(i_m\) to \(v\), and define
\begin{equation}
 b_v=
 \begin{cases}
 X_{m,v}^{-1}\widetilde k_mX_{m,v},&v\in C_m,\\
 1,&v\notin\mathcal P,
 \end{cases}
 \qquad
 a_v=\widetilde h_v b_v.
 \label{eq:S3-resolved-vertex-action}
\end{equation}
Using the resolved-sector ribbon action established above, the combined
operator takes \(\mathbf g\) to a single configuration \(\mathbf g'\):
\begin{equation}
 g'_e=a_{i_e}g_ea_{f_e}^{-1},
 \qquad e=\langle i_ef_e\rangle.
 \label{eq:S3-final-edge-configuration}
\end{equation}
The order \(a_v=\widetilde h_vb_v\) follows because the ribbons act
first and the onsite operators act second. Consequently,
\begin{equation}
 \bra{\mathbf k}\mathcal O\ket{\mathbf g}
 =\delta_{\mathbf k,\mathbf g'},\qquad 
 \bra{\mathbf k}\mathcal O\ket{\mathbf g}\ne0
 \ \Longrightarrow\ z'=z.
 \label{eq:S3-central-sector-diagonal}
\end{equation}
The first identity performs the \(\mathbf k\) sum.
The second follows because a vertex action can only conjugate the
circumferential holonomy; it cannot change a central element \(z\).
Thus, exactly as for the onsite record, logical coherences drop out of
the scalar probability:
\begin{equation}
 \Pr_p(\mathcal S^{(3)}\mid\rho_L^{\mathrm{cen}})
 =\sum_{z\in Z(G)}\rho_{zz}\,
   \mathcal Z_z^{(3)}(p,\mathcal S^{(3)}),
 \label{eq:S3-central-sector-mixture}
\end{equation}
where the remaining sum in one sector is explicitly
\begin{equation}
 \mathcal Z_z^{(3)}
 ={}\frac{1}{|G|^{N_v}N_z}
 \left(\prod_{m\in M}\frac{d_{q_m}}{|G|}\right)
 \sum_{\mathbf g\in\mathcal O_z}
 \sum_{\{\widetilde h_v\}}\sum_{\{\widetilde k_m\}}
 \left[\prod_vd_{\lambda_v}
       \chi_{\lambda_v}(\widetilde h_v^{-1})\right]
 \left[\prod_{m\in M}\chi_{q_m}(\widetilde k_m^{-1})\right]\prod_eK_{\Gamma,p}\!\left(g_e(g'_e)^{-1}\right).
 \label{eq:S3-collapsed-edge-basis-sum}
\end{equation}

\subsection{Removing the flat-configuration sum}

It remains to remove the dependence on the particular flat configuration
\(\mathbf g\), again following the derivation in the onsite syndrome distribution. Choose the same
top-to-bottom dual cut and write
\begin{equation}
 g_e=s_{i_e}z^{c_e}s_{f_e}^{-1},
 \qquad z\in Z(G),
 \label{eq:S3-flat-coordinates-main}
\end{equation}
where \(c_e=1\) on edges crossing the cut in its chosen orientation
and \(c_e=0\) elsewhere. Use the coordinates of this \emph{initial}
configuration for both sets of group variables:
\begin{equation}
 h_v=s_v^{-1}\widetilde h_vs_v,
 \qquad
 k_m=s_{i_m}^{-1}\widetilde k_ms_{i_m}.
 \label{eq:S3-root-coordinate-label}
\end{equation}
If the spine segment from \(i_m\) to \(v\) crosses the cut a signed
number \(\nu_{m,v}\) of times, then
\begin{equation}
 X_{m,v}=s_{i_m}z^{\nu_{m,v}}s_v^{-1}.
 \label{eq:S3-central-path-holonomy}
\end{equation}
Its transported action is therefore
\begin{equation}
 \begin{aligned}
 X_{m,v}^{-1}\widetilde k_mX_{m,v}
 &=s_v\bigl(z^{-\nu_{m,v}}k_mz^{\nu_{m,v}}\bigr)s_v^{-1}\\
 &=s_vk_ms_v^{-1},
 \end{aligned}
 \label{eq:S3-central-transport-cancellation}
\end{equation}
where the centrality removes the winding-dependent conjugation.

Define
\begin{equation}
 Q_v(\mathbf k)=
 \begin{cases}
 k_m,&v\in C_m,\\
 1,&v\notin\mathcal P,
 \end{cases}
 \label{eq:S3-Qv-disjoint}
\end{equation}
and combine the two successive actions into
\begin{equation}
 c_v=h_vQ_v,\qquad a_v=s_vc_vs_v^{-1}.
 \label{eq:S3-combined-local-transformation}
\end{equation}
Here \(c_v\) is simply the total group action at \(v\); no second choice of local coordinates is needed. The initial and final edge labels give
\begin{equation}
 \begin{aligned}
 g'_e&=s_{i_e}c_{i_e}z^{c_e}c_{f_e}^{-1}s_{f_e}^{-1},\\
 g_e(g'_e)^{-1}
 &=s_{i_e}\bigl(z^{c_e}c_{f_e}z^{-c_e}c_{i_e}^{-1}\bigr)s_{i_e}^{-1}=s_{i_e}(c_{f_e}c_{i_e}^{-1})s_{i_e}^{-1}.
 \end{aligned}
 \label{eq:S3-relative-edge-element}
\end{equation}
Since the kernel and characters are class functions,
\begin{equation} K_{\Gamma,p}\!\left(g_e(g'_e)^{-1}\right)
 =K_{\Gamma,p}(c_{f_e}c_{i_e}^{-1}),\qquad
 \chi_{q_m}(\widetilde k_m^{-1})
 =\chi_{q_m}(k_m^{-1}).
 \label{eq:S3-edge-kernel-main}
\end{equation}
For the onsite source, \(h_v=c_vQ_v^{-1}\) gives
\begin{equation}
 \chi_{\lambda_v}(\widetilde h_v^{-1})=\chi_{\lambda_v}(h_v^{-1})=\chi_{\lambda_v}(Q_vc_v^{-1})=\chi_{\lambda_v}(c_v^{-1}Q_v),
 \label{eq:S3-source-change-main}
\end{equation}
where the cyclicity of the trace is used in the last equality.

For each fixed \(\mathbf g\), these changes of summation variables are
bijective. The transformed summand is independent of \(\mathbf g\),
so \(\sum_{\mathbf g\in\mathcal O_z}1=N_z\) cancels the orbit
normalization in Eq.~\eqref{eq:S3-collapsed-edge-basis-sum}.
It is also independent of \(z\), giving the explicit vertex model
\begin{equation}
 \mathcal Z_z^{(3)}
 =\frac{1}{|G|^{N_v}}
 \left(\prod_{m\in M}\frac{d_{q_m}}{|G|}\right)
 \sum_{\{c_v\}}\sum_{\{k_m\}}
 \left[\prod_vd_{\lambda_v}
       \chi_{\lambda_v}(c_v^{-1}Q_v)\right]
 \left[\prod_{m\in M}\chi_{q_m}(k_m^{-1})\right]\prod_eK_{\Gamma,p}(c_{f_e}c_{i_e}^{-1})
 \equiv\mathcal Z_{\mathrm{cen}}^{(3)}.
 \label{eq:S3-uncontracted-vertex-model}
\end{equation}
Using \(\sum_z\rho_{zz}=1\) then yields
\begin{equation}
 \Pr_p(\mathcal S^{(3)}\mid\rho_L^{\mathrm{cen}})
 =\mathcal Z_{\mathrm{cen}}^{(3)}.
 \label{eq:S3-central-state-independence}
\end{equation}

\subsection{Final result as a partition function}

The final expression can be organized exactly like the onsite partition
function. Each \(k_m\) appears only in the source factors of \(C_m\);
performing this finite group sum defines
\begin{equation}
 \mathcal V_q^{C_m}\!
 \left(\{c_v\}_{v\in C_m}\right)
 ={}\frac{d_q}{|G|}
 \sum_{k\in G}\chi_q(k^{-1})
 \prod_{v\in C_m}
 d_{\lambda_v}\chi_{\lambda_v}(c_v^{-1}k).
 \label{eq:S3-general-regional-factor-derivation}
\end{equation}
Consequently,
\begin{equation}
 \Pr_p(\mathcal S^{(3)}\mid\rho_L^{\mathrm{cen}})
 =\frac{1}{|G|^{N_v}}\sum_{\{c_v\}}
 \prod_eK_{\Gamma,p}(c_{f_e}c_{i_e}^{-1})\prod_{v\notin\mathcal P}
 d_{\lambda_v}\chi_{\lambda_v}(c_v^{-1})\prod_{m\in M}
 \mathcal V_{q_m}^{C_m}\!
 \left(\{c_v\}_{v\in C_m}\right).
 \label{eq:S3-partition-main-derivation}
\end{equation}
Relative to the onsite model, the edge kernels are unchanged.
The fusion measurement replaces the product of onsite source factors
within each measured region by one joint character sum
\(\mathcal V_{q_m}^{C_m}\). The complete likelihood is independent of the initial logical state
within \(\mathcal H_Z\), and in particular within the \(D_4\) central qubit.

The outcome \(q_m\) refers to the entire region, including its
intermediate vertices. It is the endpoint-pair channel when these
vertices are vacuum. For the \(D_4\) protocol, their known
one-dimensional charges can instead be removed by a deterministic
relabeling, as shown in Sec.~\ref{sec:fusion-D4-components}. General non-Abelian
interior charges cannot be removed using their onsite labels alone.

Finally, the regular-character identity
\begin{equation}
 \sum_{q\in\operatorname{Irr}(G)}d_q\chi_q(k^{-1})
 =|G|\delta_{k,1}
 \label{eq:regular-character-identity}
\end{equation}
gives
\begin{equation}
 \sum_q\mathcal V_q^{C_m}\!
 \left(\{c_v\}_{v\in C_m}\right)
 =
 \prod_{v\in C_m}d_{\lambda_v}\chi_{\lambda_v}(c_v^{-1}).
 \label{eq:S3-regional-factor-marginal}
\end{equation}
Thus forgetting the fusion outcomes restores the onsite probability:
\begin{equation}
 \sum_{\mathbf q}
 \Pr_p(\gamma,\mathbf a,\mathbf q\mid\rho_L^{\mathrm{cen}})
 =
 \Pr_p(\gamma,\mathbf a\mid\rho_L^{\mathrm{cen}}).
 \label{eq:fusion-syndrome-marginalization-derivation}
\end{equation}
The specialization to $D_4$ and its component interpretation are given
in Sec.~\ref{sec:fusion-D4-components}.

\section{Path-dependent fusion and logical holonomy}
\label{app:path-dependent-fusion}

This appendix proves the general-group statements summarized in
Sec.~\ref{sec:fusion-logical-backaction}. We relate the two lattice
fusion projectors through their relative path holonomy, construct a test pair with the reference-path fusion outcome known, and
evaluate its winding-path vacuum probability. The final argument shows
why every noncentral holonomy is detectable by some charge, while the central logical subspace cannot be distinguished by any fusion outcomes.

\subsection{The two microscopic path projectors}

We compare the lattice fusion projectors associated with two paths
joining the same charges. Their path dependence follows directly from
the ordered products of edge labels entering the microscopic ribbon
operators. We assume that the state is flux free and that the
intermediate vertices of both ribbons carry vacuum charge.
This ensures that both projectors measure the same endpoint pair,
rather than different collections of enclosed charges.

Orient both paths \(p_0,p_1\) from \(i\) to \(j\), and suppose that
\(p_1\overline p_0\) winds once around the cylinder.
For an edge configuration \(\mathbf g\), define
\begin{equation}
 X_\mu(\mathbf g)
 =\prod_{e\in p_\mu}g_e^{\epsilon_{\mu e}},
 \qquad \mu=0,1,
\end{equation}
where the product follows the path and
\(\epsilon_{\mu e}=\pm1\) specifies whether the path follows or
opposes the stored edge orientation. The relative loop has holonomy
\begin{equation}
 H(\mathbf g)=X_1(\mathbf g)X_0(\mathbf g)^{-1}.
 \label{eq:path-relative-lattice-holonomy-derivation}
\end{equation}
In logical sector \([H]\), the element \(H(\mathbf g)\) belongs to
\([H]\), although it needs not equal the same representative
for every configuration.

For either path, the lattice fusion projector is
\begin{equation}
 P_{p_\mu}^{\lambda}
 =\frac{d_\lambda}{|G|}
 \sum_{h\in G}\chi_\lambda(h^{-1})
 \mathcal F_{p_\mu}(h).
 \label{eq:path-fusion-character-projector}
\end{equation}
The diagonal spine projectors select the actual path labels of
\(\mathbf g\). In this resolved term, the corresponding vertex
actions at the endpoints are \(h\) at \(i\) and
\begin{equation}
 Y_\mu(h;\mathbf g)
 =X_\mu(\mathbf g)^{-1}hX_\mu(\mathbf g)
\end{equation}
at \(j\) according to Fig.~\ref{fig:pair-ribbon-spine}. Consequently, changing the path replaces the transported
group element by
\begin{equation}
 Y_1(h;\mathbf g)
 =X_0^{-1}H(\mathbf g)^{-1}hH(\mathbf g)X_0.
 \label{eq:path-fusion-projector-conjugation-derivation}
\end{equation}
This is the microscopic origin of the holonomy dependence.

We first consider central holonomy. If \(H(\mathbf g)=z\in Z(G)\),
then \(H(\mathbf g)^{-1}zH(\mathbf g)=z\), so both paths give
the same transported action at the second endpoint. After resolving
the spine projectors, any remaining difference consists of vertex
actions at the intermediate vertices. These leave the state unchanged
because those vertices carry vacuum. Consequently,
\begin{equation}
 \left.P_{p_1}^{\lambda}\right|_{[z]}
 =
 \left.P_{p_0}^{\lambda}\right|_{[z]}.
 \label{eq:central-path-independence-derivation}
\end{equation}
This equality applies on the flux-free, vacuum-interior subspace
specified above.

For noncentral holonomy, the transported action can change.
To establish that this change has an observable consequence, we use
a simple test: prepare two charges that fuse to vacuum
along \(p_0\), and measure whether they also fuse to vacuum along
\(p_1\). We now construct this test state and calculate its outcome
directly from the lattice projectors.

\subsection{Preparing the reference pair}
Let \(\Gamma\) be an irrep and take distinct endpoints \(i\ne j\).
The character-contracted open charge ribbon acts diagonally:
\begin{equation}
 \mathcal Z_\Gamma(p_0)\ket{\mathbf g}
 =
 \chi_\Gamma(X_0(\mathbf g))\ket{\mathbf g}.
\end{equation}
This is the trace of the matrix-valued charge ribbon along the
open path \(p_0\). Acting on a ground state creates a
\(\Gamma,\overline\Gamma\) pair at its endpoints. The intermediate
vertices remain vacuum because their vertex actions leave the
ordered path product unchanged.

Let \(\mathcal O_H\) be the flat configurations appearing in
\(\ket{[H]}\), and let \(N_H=|\mathcal O_H|\). Define
\begin{equation}
 \begin{aligned}
 \ket{\Psi_{0,H}}
 &:=\mathcal Z_\Gamma(p_0)\ket{[H]}\\
 &=\frac1{\sqrt{N_H}}
   \sum_{\mathbf g\in\mathcal O_H}
   \chi_\Gamma(X_0(\mathbf g))\ket{\mathbf g}.
 \end{aligned}
 \label{eq:path-lattice-pair-preparation}
\end{equation}
These states are normalized. A vertex action at \(i\) multiplies
\(X_0\) by an arbitrary group element and hence permutes
\(\mathcal O_H\), so the above sum is equal weighted for every \(X_0\in G\). Character orthogonality therefore gives
\begin{equation}
 \langle\Psi_{0,H}|\Psi_{0,H}\rangle
 =\frac1{|G|}\sum_{x\in G}|\chi_\Gamma(x)|^2=1.
\end{equation}

By definition, the state \(\ket{\Psi_{0,H}}\) has anyons at $i$ and $j$ that fuse to vacuum along $p_0$. Indeed, \(\mathcal F_{p_0}(h)\) preserves
the spine labels, and hence \(X_0\), while permuting the flat
configurations in \(\mathcal O_H\). It therefore leaves this
state unchanged for every \(h\). Averaging over \(h\) gives
\begin{equation}
 P_{p_0}^{\mathbf1}\ket{\Psi_{0,H}}
 =\ket{\Psi_{0,H}}.
 \label{eq:path-reference-vacuum-preparation}
\end{equation}
Thus \(\ket{\Psi_{0,H}}\) is a normalized pair state with a
certain vacuum outcome along \(p_0\):
\begin{equation}
 \Pr_{[H]}(\mathbf1\mid p_0)=1.
\end{equation}

\subsection{Measuring along the winding path}
For this state, the vacuum probability along \(p_1\) is
\begin{equation}
 \Pr_{[H]}(\mathbf1\mid p_1)
 =
 \bigl\|P_{p_1}^{\mathbf1}\ket{\Psi_{0,H}}\bigr\|^2.
\end{equation}
We therefore need only calculate the vacuum component selected
by \(P_{p_1}^{\mathbf1}\).

For a fixed \(h\), the two endpoint actions of
\(\mathcal F_{p_1}(h)\) transform the reference path product as
\begin{equation}
 \begin{aligned}
 X_0
 &\longmapsto hX_0X_1^{-1}h^{-1}X_1\\
 &=hH(\mathbf g)^{-1}h^{-1}X_1.
 \end{aligned}
\end{equation}
To obtain the coefficient of a given final configuration
\(\mathbf g\), we evaluate the original coefficient on the
configuration that maps to it. This uses the inverse ribbon
\(\mathcal F_{p_1}(h^{-1})\), so the reference path product
in that coefficient is
\[
 h^{-1}H(\mathbf g)^{-1}hX_1(\mathbf g).
\]
The vacuum projector averages these coefficients over \(h\):
\begin{equation}
 \begin{aligned}
 \bra{\mathbf g}P_{p_1}^{\mathbf1}\ket{\Psi_{0,H}}
 =\frac1{|G|\sqrt{N_H}}\sum_{h\in G}\chi_\Gamma\!\left(
 h^{-1}H(\mathbf g)^{-1}hX_1(\mathbf g)
 \right).
 \end{aligned}
 \label{eq:path-vacuum-lattice-average}
\end{equation}
This group sum is evaluated by the character identity of an irrep $\Gamma$:
\begin{equation}
 \frac1{|G|}\sum_{h\in G}\chi_\Gamma(h^{-1}AhB)
 =
 \frac{\chi_\Gamma(A)\chi_\Gamma(B)}{d_\Gamma}.
\end{equation}
Here \(A=H(\mathbf g)^{-1}\) and \(B=X_1(\mathbf g)\).
The first character equals \(\chi_\Gamma(H^{-1})\), independently
of \(\mathbf g\), because all \(H(\mathbf g)\) belong to the same
conjugacy class. The second character is precisely the coefficient
defining \(\ket{\Psi_{1,H}}\). Therefore,
\begin{equation}
 P_{p_1}^{\mathbf1}\ket{\Psi_{0,H}}
 =
 \frac{\chi_\Gamma(H^{-1})}{d_\Gamma}
 \ket{\Psi_{1,H}}.
 \label{eq:path-projected-pair-state}
\end{equation}
In words, the vacuum component selected along \(p_1\) is the
vacuum-pair state associated with that path, multiplied by a
holonomy-dependent amplitude. Since \(\ket{\Psi_{1,H}}\) is
normalized, the probability is the squared magnitude of this
amplitude:
\begin{equation}
 \Pr_{[H]}(\mathbf1\mid p_1)
 =\frac{|\chi_\Gamma(H)|^2}{d_\Gamma^2}.
 \label{eq:path-dependent-vacuum-probability-derivation}
\end{equation}

\subsection{Detecting noncentral logical holonomy}
For a central element \(z\), Schur's lemma makes
\(\Gamma(z)\) a scalar unitary, so
\(|\chi_\Gamma(z)|=d_\Gamma\). The vacuum outcome is therefore
certain along both paths.

Conversely, every noncentral \(H\) has an irrep \(\Gamma\)
for which \(\Gamma(H)\) is not scalar. For a non-scalar unitary matrix, the eigenvalue phases are not
all equal, so \(|\chi_\Gamma(H)|<d_\Gamma\). Hence
\begin{equation}
 \begin{gathered}
 H\notin Z(G)\ \Longrightarrow\ \exists\,\Gamma:\,
 \Pr_{[H]}(\mathbf1\mid p_1)<1
 =\Pr_{[H]}(\mathbf1\mid p_0).
 \end{gathered}
 \label{eq:noncentral-holonomy-detectable}
\end{equation}
A suitable choice of $\Gamma$-$\bar{\Gamma}$ pair fusion measurement therefore has a certain vacuum outcome along
the reference path but a nonzero probability of a nonvacuum
outcome along the winding path. This difference reveals
information about noncentral logical holonomy.

For an arbitrary logical input \(\rho_L\), prepare the same
reference pair and measure its fusion along \(p_1\). Its vacuum probability is consequently
\(\operatorname{Tr}_L(\mathsf E_{\mathbf1}^{(\Gamma)}\rho_L)\),
where
\begin{equation}
 \mathsf E_{\mathbf1}^{(\Gamma)}
 =
 \sum_{C\in\operatorname{Conj}(G)}
 \frac{|\chi_\Gamma(C)|^2}{d_\Gamma^2}
 \ket{C}\!\bra{C}.
 \label{eq:logical-vacuum-effect-general}
\end{equation}
The coefficient is one on every central class and strictly
smaller than one on any noncentral class detected by \(\Gamma\).
The detecting charge type may depend on the class, and multiple charges may need to be used to distinguish all logical sectors from each other.

\section{Microscopic construction of the boundary-to-boundary flux ribbon}
\label{app:flux-ribbon-proof}

This appendix constructs the flux ribbon used in
Sec.~\ref{sec:cyl-charge-lines} and proves that its conjugacy-class average
preserves every stabilizer of the cylinder Hamiltonian.

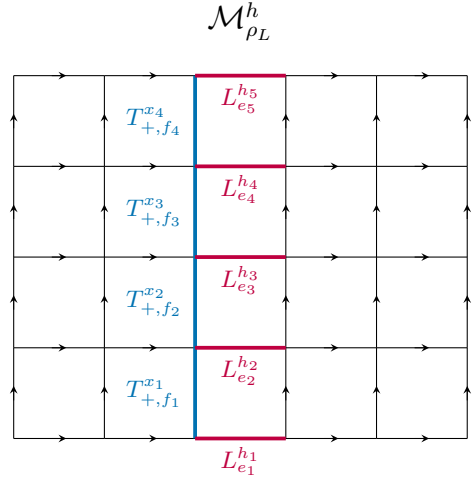
\begin{figure}[t]
\centering
\begin{tikzpicture}[x=1.2cm,y=1.2cm]

  \foreach \y in {0,-1,-2,-3,-4}{
    \foreach \x in {0,1,2,3,4}{
      \draw (\x,\y)--({\x+1},\y);
    }
  }
  \foreach \x in {0,1,2,3,4,5}{
    \foreach \y in {-4,-3,-2,-1}{
      \draw (\x,\y)--(\x,{\y+1});
    }
  }

  \foreach \y in {0,-1,-2,-3,-4}{
    \foreach \x in {0,1,3,4}{
      \draw[-stealth,line width=0.3pt]
        ({\x+0.42},\y)--({\x+0.58},\y);
    }
  }

  \foreach \x in {0,1,3,4,5}{
    \foreach \y in {-4,-3,-2,-1}{
      \draw[-stealth,line width=0.3pt]
        (\x,{\y+0.42})--(\x,{\y+0.58});
    }
  }

  \node[font=\large] at (2.5,0.60)
    {$\mathcal M_{\rho_L}^{h}$};

  \foreach \j/\y in {
    1/-4,
    2/-3,
    3/-2,
    4/-1
  }{
    \draw[
      line width=0.5mm,
      myBlue
    ] (2,\y)--(2,{\y+1});

    \node[
      myBlue,
      fill=white,
      inner sep=0.5pt
    ] at (1.55,{\y+0.5})
      {$T_{+,f_{\j}}^{x_{\j}}$};
  }

  \foreach \j/\y in {
    1/-4,
    2/-3,
    3/-2,
    4/-1,
    5/0
  }{
    \draw[
      line width=0.5mm,
      purple
    ] (2,\y)--(3,\y);

    \node[
      purple,
      fill=white,
      inner sep=0.5pt
    ] at (2.5,{\y-0.25})
      {$L_{e_{\j}}^{h_{\j}}$};
  }

\end{tikzpicture}

\caption{
Microscopic action of the boundary-to-boundary flux ribbon
$\mathcal M_{\rho_L}^{h}$ on a local portion of the oriented lattice.
The blue operators $T_{+,f_j}^{x_j}$ resolve the edge labels along one
side of the ribbon, while the purple operators $L_{e_j}^{h_j}$ apply
the corresponding transported shifts to the crossed horizontal edges.
Horizontal lattice edges are oriented to the right and vertical edges
are oriented upward.
}
\label{fig:explicit-flux-ribbon}
\end{figure}

\subsection{Ribbon operator}

Orient the ribbon $\rho_L$ from the bottom flux-condensing boundary to the
top boundary.  Let
\[
e_1,e_2,\ldots,e_n
\]
be the horizontal edges crossed by the ribbon, ordered from bottom
to top.  All crossed edges are oriented to the right.

Let $f_j$ be the vertical edge connecting the left endpoint of
$e_j$ to the left endpoint of $e_{j+1}$.  Thus the $f_j$ edges form one
side of the one-plaquette-wide strip containing the ribbon.  They are not
crossed by the ribbon.  Instead, their group labels determine how the
inserted flux is parallel transported between consecutive crossings.

The vertical edges are oriented upward.  If $x_j$ is the group label on
$f_j$, define
\begin{equation}
 X_0=1,
 \qquad
 X_j=x_1x_2\cdots x_j.
 \label{eq:app-ribbon-X}
\end{equation}
The flux label applied to the $j$th crossed edge is
\begin{equation}
 h_j=X_{j-1}^{-1}hX_{j-1}.
 \label{eq:app-ribbon-hj}
\end{equation}
Consequently,
\begin{equation}
 h_{j+1}=x_j^{-1}h_jx_j,
 \qquad
 h_jx_j=x_jh_{j+1}.
 \label{eq:app-ribbon-recursion}
\end{equation}

Using
\[
T_{+,f_j}^{x_j}\ket{g}_{f_j}
=\delta_{x_j,g}\ket{g}_{f_j},
\]
the framed $h$-flux ribbon in
Fig.~\ref{fig:explicit-flux-ribbon} is
\begin{equation}
 \mathcal M_{\rho_L}^{h}
 =
 \sum_{x_1,\ldots,x_{n-1}\in G}\left(
 \prod_{j=1}^{n}
 L_{e_j}^{\,X_{j-1}^{-1}hX_{j-1}}
 \right)
 \left(
 \prod_{j=1}^{n-1}
 T_{+,f_j}^{x_j}
 \right).
 \label{eq:app-explicit-flux-ribbon}
\end{equation}
The projectors act first.  They resolve the side-edge labels used to
determine the transported shifts, but the ribbon does not change the
$f_j$ edges themselves.

If $y_j$ is the initial group label on the crossed edge $e_j$, then
Eq.~\eqref{eq:app-explicit-flux-ribbon} acts as
\begin{equation}
 y_j\longmapsto
 h_jy_j
 =
 X_{j-1}^{-1}hX_{j-1}y_j.
 \label{eq:app-flux-ribbon-action}
\end{equation}
The first few crossed-edge labels transform according to
\begin{align}
 y_1&\longmapsto h y_1,\nonumber\\
 y_2&\longmapsto x_1^{-1}hx_1\,y_2,\nonumber\\
 y_3&\longmapsto
 (x_1x_2)^{-1}h(x_1x_2)\,y_3.
\end{align}

\subsection{Plaquette stabilizers}

A plaquette not intersected by the ribbon is unchanged.  Consider the
plaquette between the consecutive crossed edges $e_j$ and $e_{j+1}$.
Let their initial labels be $y_j$ and $y_{j+1}$, and let $c_j$ be the
label of the vertical edge on the opposite side of the plaquette.  Its
oriented flux is
\begin{equation}
 \Phi_p
 =
 y_jc_jy_{j+1}^{-1}x_j^{-1}.
 \label{eq:app-plaquette-flux}
\end{equation}

The ribbon acts on the bottom and top edges as
\begin{equation}
 y_j\longmapsto h_jy_j,
 \qquad
 y_{j+1}\longmapsto h_{j+1}y_{j+1}.
\end{equation}
The transformed plaquette flux is therefore
\begin{equation} 
\Phi_p'
 =
 (h_jy_j)c_j
 (h_{j+1}y_{j+1})^{-1}x_j^{-1}=
 h_jy_jc_jy_{j+1}^{-1}
 h_{j+1}^{-1}x_j^{-1}.
 \label{eq:app-plaquette-step-one}
\end{equation}
From Eq.~\eqref{eq:app-ribbon-recursion},
\begin{equation}
 h_{j+1}^{-1}
 =
 x_j^{-1}h_j^{-1}x_j.
\end{equation}
Substituting this relation gives
\begin{equation}
 \Phi_p'
 =
 h_jy_jc_jy_{j+1}^{-1}
 x_j^{-1}h_j^{-1}=
 h_j\Phi_ph_j^{-1}.
 \label{eq:app-plaquette-conjugation}
\end{equation}
Thus the ribbon only conjugates the plaquette flux.  In particular,
\begin{equation}
 \Phi_p=1
 \quad\Longleftrightarrow\quad
 \Phi_p'=1.
\end{equation}
Because $B_p^e$ projects onto the identity flux, this proves
\begin{equation}
 [\mathcal M_{\rho_L}^{h},B_p^e]=0
 \qquad
 \text{for every bulk plaquette }p.
 \label{eq:app-ribbon-B-commutation}
\end{equation}

At the two endpoints, the ribbon can change the flux of a boundary
plaquette.  Those plaquette terms are absent on a flux-condensing
boundary, so no boundary Hamiltonian term is violated.

\subsection{Vertex stabilizers}

Let $u_j$ be the left endpoint of the crossed edge $e_j$, so that
$f_j$ connects $u_j$ to $u_{j+1}$.  Let $q_j\in G$ be the group element
used in a vertex action at $u_j$.  Since $f_j$ is oriented from $u_j$ to
$u_{j+1}$, its label transforms as
\begin{equation}
 x_j\longmapsto q_jx_jq_{j+1}^{-1}.
 \label{eq:app-side-edge-transformation}
\end{equation}
The side-path holonomy consequently transforms as
\begin{equation}
 X_j\longmapsto q_1X_jq_{j+1}^{-1}.
 \label{eq:app-X-transformation}
\end{equation}

If the initial ribbon label is transformed according to
\begin{equation}
 h\longmapsto q_1hq_1^{-1},
\end{equation}
then the transported label at the $j$th crossing transforms as
\begin{equation}
 h_j
 =X_{j-1}^{-1}hX_{j-1}\longmapsto
 q_jh_jq_j^{-1}.
 \label{eq:app-hj-transformation}
\end{equation}
This is exactly the transformation of the corresponding left shift,
because
\begin{equation}
 L^{q_j}L^{h_j}L^{q_j^{-1}}
 =
 L^{q_jh_jq_j^{-1}}.
 \label{eq:app-L-covariance}
\end{equation}
At the right endpoint of a crossed edge, the vertex action is a right
multiplication and therefore commutes with $L^{h_j}$.

The sums over $x_j$ in
Eq.~\eqref{eq:app-explicit-flux-ribbon} ensure that the side-edge
projectors are relabelled consistently.  Hence, for an arbitrary collection
of vertex actions
\begin{equation}
 \mathcal A(\mathbf q)=\prod_v A_v^{q_v},
\end{equation}
the framed ribbon satisfies
\begin{equation}
 \mathcal A(\mathbf q)
 \mathcal M_{\rho_L}^{h}
 \mathcal A(\mathbf q)^{-1}
 =
 \mathcal M_{\rho_L}^{q_1hq_1^{-1}}.
 \label{eq:app-ribbon-gauge-covariance}
\end{equation}
All vertex actions away from the basepoint $u_1$ therefore commute with
the framed ribbon.  A vertex action at $u_1$ conjugates its initial flux
label.

The physical ribbon is obtained by averaging over this conjugation:
\begin{equation}
 \mathcal M_{\rho_L,[h]}
 =
 \frac{1}{|[h]|}
 \sum_{k\in[h]}
 \mathcal M_{\rho_L}^{k}.
 \label{eq:app-class-ribbon}
\end{equation}
Indeed,
\begin{equation}
 A_{u_1}^{q}
 \mathcal M_{\rho_L,[h]}
 (A_{u_1}^{q})^{-1}
 =
 \frac{1}{|[h]|}
 \sum_{k\in[h]}
 \mathcal M_{\rho_L}^{qkq^{-1}}=
 \mathcal M_{\rho_L,[h]},
 \label{eq:app-class-ribbon-invariance}
\end{equation}
because conjugation permutes the elements of $[h]$.  It follows that
\begin{equation}
 [\mathcal M_{\rho_L,[h]},A_v^q]=0
 \qquad
 \text{for every }v\text{ and }q\in G.
 \label{eq:app-ribbon-Ah-commutation}
\end{equation}
Since
\[
A_v=\frac{1}{|G|}\sum_{q\in G}A_v^q,
\]
we also have
\begin{equation}
 [\mathcal M_{\rho_L,[h]},A_v]=0.
 \label{eq:app-ribbon-A-commutation}
\end{equation}

Combining Eqs.~\eqref{eq:app-ribbon-B-commutation} and
\eqref{eq:app-ribbon-A-commutation}, the physical ribbon preserves every
stabilizer in the cylinder Hamiltonian:
\begin{equation}
 \mathcal M_{\rho_L,[h]}:
 \cH_0\longrightarrow\cH_0.
 \label{eq:app-ribbon-preserves-groundspace}
\end{equation}

\subsection{Connection to the logical operator}

The preceding subsections supply the microscopic construction and the
stabilizer-preservation proof deferred from Sec.~\ref{sec:cyl-charge-lines}. The action of the class-averaged ribbons on the cylinder
ground space was already derived in
Eq.~\eqref{eq:cyl-framed-shift-action}, so we record here only the microscopic simplification used for the central
qubit. 

For the central element $r^2\in D_4$, every transported label in
Eq.~\eqref{eq:app-ribbon-hj} remains $r^2$. The side-edge projectors in
Eq.~\eqref{eq:app-explicit-flux-ribbon} therefore sum to the identity, and
\begin{equation}
 \mathcal M_{\rho_L}^{r^2}
 =\prod_{e\cap\rho_L\neq\varnothing}L_e^{r^2}.
 \label{eq:app-central-ribbon-product}
\end{equation}
It exchanges $\ket{[1]}$ and $\ket{[r^2]}$ and hence realizes logical Pauli $X$ on the central qubit.

\section{Auxiliary positive-current representation}
\label{app-current-rep}

This appendix provides an algebraic check on the charge channel in
Sec.~\ref{sec:onsiteprob} and clarifies the distinction between auxiliary
noise labels and the physical-record sampler of
Sec.~\ref{sec:num-evaluation}. It derives a positive random-unitary unraveling of the
$D_4$ $E$-charge channel. It is useful as an algebraic check and as a latent
representation of the channel, but its current labels are neither measured
syndrome values nor inputs to any decoder used in the numerics.

The Schur kernel admits an exact random-unitary
decomposition,
\begin{equation}
    \mathcal{N}[\rho]=\sum_ip_iU_i\rho U_i^\dagger
\end{equation}
where $p_i\ge0$ and the $U_i$ are unitary. This is not a rule for reading the
complete physical onsite charge from the divergence of a sampled branch.
Write a group element as
\begin{equation}\label{d4elements}
 g=r^{2t+u}s^q,\qquad t,u,q\in\mathbb Z_2,
\end{equation}
where $r$ is a rotation and $s$ is a reflection.  The two-dimensional irrep
can be chosen as
\begin{equation}\label{Eirrep}
    \Gamma_E(g)=\begin{pmatrix} i& 0\\ 0 & -i\end{pmatrix}^{2t+u}\begin{pmatrix}0 & 1\\ 1 & 0 \end{pmatrix}^q
\end{equation}

Since every $Z_{ab}^E$ is diagonal in the group basis, it is enough to calculate the action of the channel (\ref{ChargeDecoherenceChannel}) on $|g\rangle\langle h|$ as in (\ref{eq:single-charge-kernel}). 

From (\ref{Eirrep}) we get
\begin{equation}
    \chi_E(r^{2t+u}s^q)=\begin{cases} 2(-1)^t\quad u=0,q=0\\ 0\qquad\quad \text{ otherwise}\end{cases}
\end{equation}
so the character is only nonzero on $1$ and $r^2$. Writing $g=r^{2t+u}s^{q},h=r^{2t'+u'}s^{q'}$, we have
\begin{equation}
    \frac{1}{2}\chi_E(gh^{-1})=(-1)^{t+ t'}\delta_{u,u'}\delta_{q,q'}
\end{equation}

We will now show that we can obtain the same action in a positive current representation. We define diagonal unitaries
\begin{equation}
 \begin{aligned}
 U_{fxz}\ket{r^{2t+u}s^q}
 &=(-1)^{ft+xu+zq}\ket{r^{2t+u}s^q}, \quad (f,x,z)\in\mathbb Z_2^3,
 \end{aligned}
 \label{eq:main-current-unitaries}
\end{equation}
with
\begin{equation}
    U_{fxz}^2=1.
\end{equation}
For fixed $x,z$, we have
\begin{equation}
    U_{1xz}|g\rangle\langle h|U_{1xz}^\dagger =(-1)^{t+ t'+x(u+ u')+z(q+ q')}|g\rangle\langle h|
\end{equation}
Averaging uniformly over $x,z$ recovers
\begin{equation}
    \frac{1}{4}\sum_{x,z\in\mathbb{Z}_2}U_{1xz}|g\rangle\langle h|U_{1xz}^\dagger=(-1)^{t+ t'}\delta_{u,u'}\delta_{q,q'}|g\rangle\langle h|
\end{equation}
It follows that the $D_4$ $E$-charge channel (\ref{ChargeDecoherenceChannel}) obeys the exact identity
\begin{equation}
 \mathcal N_e(\rho)
 =(1-p)\rho+\frac p4\sum_{x,z\in\mathbb Z_2}
 U_{1xz,e}\rho U_{1xz,e}^{\dagger}.
 \label{eq:main-random-unitary-channel}
\end{equation}
This rewriting of the decoherence channel is convenient because it means that we can associate to each edge an auxiliary current
\begin{equation}
 a_e=(f_e,x_e,z_e)\in A\equiv\mathbb Z_2^3
\end{equation}
with probability
\begin{equation}
 w_p(a)=
 \begin{cases}
 1-p,&a=(0,0,0),\\[1mm]
 p/4,&a=(1,x,z),\quad x,z\in\mathbb Z_2,\\[1mm]
 0,&\text{otherwise}.
 \end{cases}
 \label{eq:main-current-weight}
\end{equation}
Because the phase operators satisfy
\begin{equation}
 U_aU_{a'}=U_{a\oplus a'},
 \label{eq:main-current-composition}
\end{equation}
the full noisy state is exactly
\begin{equation}
 \rho_p=\sum_{\mathbf a}
 \left[\prod_e w_p(a_e)\right]
 U_{\mathbf a}\rho_0U_{\mathbf a}^{\dagger},
 \qquad U_{\mathbf a}=\prod_eU_{a_e,e}.
 \label{eq:main-full-current-ensemble}
\end{equation}
Thus each term in the sum is a definite unitary error branch, with a positive
probability.

\section{Numerical sampling, exact contractions, and analysis protocol}
\label{app:central-numerical-checks}

This appendix provides the details discussed in Sec.~\ref{sec:decoding}. We normalize the outer physical-record samplers, spell out the two different
exact inner contractions, and record the estimators and
finite-size analysis protocol.

\subsection{Generation of the physical records}
\label{app:physical-record-sampling}

This subsection specifies the outer record sampler used in
Sec.~\ref{sec:num-evaluation}. We denote its latent occupied set by
$F^{\rm gen}$ to distinguish it from the candidate sets summed over by the
conditional-likelihood algorithms. Only the resulting measurement record is
retained from $F^{\rm gen}$.

Each edge is independently included in $F^{\rm gen}$ with probability $p$. Thus,
\begin{equation}
 \Pr_p(F^{\rm gen})
 =p^{|F^{\rm gen}|}(1-p)^{N_e-|F^{\rm gen}|},
 \label{app:num-F-prior}
\end{equation}
and $\gamma=\partial F^{\rm gen}$ gives the complete binary record
$\mathcal{S}^{(1)}$.

For $\mathcal S^{(2)}$, every vertex in $\gamma$ carries the $E$ charges, whereas every
$v\notin\gamma$ carries an Abelian label
$\boldsymbol a_v=(x_v,z_v)\in\mathbb Z_2^2$. Let
$\mathcal C_0(F^{\rm gen})$ be the set of connected components of
$(V,F^{\rm gen})$ containing no vertex in $\gamma$, including isolated
vertices. The normalized conditional law obtained from
Eq.~\eqref{eq:char-loop-final} is
\begin{equation}
 \Pr(\mathcal S^{(2)}\mid F^{\rm gen})
 =4^{|\gamma|-N_v+|\mathcal C_0(F^{\rm gen})|}\prod_{B\in\mathcal C_0(F^{\rm gen})}
 \mathbf1\!\left[
  \bigoplus_{v\in B}\boldsymbol a_v=0
 \right].
 \label{app:num-S2-fixed-F-law}
\end{equation}
A component meeting $\gamma$ has no Abelian neutrality constraint. If it
contains $n_B$ ordinary vertices, it admits $4^{n_B}$ assignments. An
endpoint-free component containing $n_B$ vertices admits $4^{n_B-1}$
assignments, because its last label is fixed by neutrality. Hence the total
number of allowed records is
\begin{equation}
 4^{N_v-|\gamma|-|\mathcal C_0(F^{\rm gen})|},
\end{equation}
the reciprocal of the prefactor in
Eq.~\eqref{app:num-S2-fixed-F-law}. Operationally, on a component meeting
$\gamma$ we draw every ordinary label independently and uniformly. On an
endpoint-free component we draw all but one label uniformly and fix the last
by charge neutrality; an isolated vertex is therefore assigned the vacuum label. 

For $\mathcal S^{(3)}$, construct the deterministic matching
$M=\operatorname{pair}(\gamma)$ and the augmented multigraph
\begin{equation}
 \mathcal G_{\rm aug}=(V,F^{\rm gen}\sqcup M),
\end{equation}
where $\sqcup$ retains a physical edge and a matching edge with the same
endpoints as distinct edges. In addition to the onsite variables, introduce
one endpoint-corrected fusion label
$\bar{\boldsymbol a}_m=(\bar x_m,\bar z_m)$ for every $m\in M$. For a
component $B$ of $\mathcal G_{\rm aug}$, define
\begin{equation}
 \boldsymbol Q_B
 =\bigoplus_{v\in B\setminus\gamma}\boldsymbol a_v\bigoplus_{m\in M_B}\bar{\boldsymbol a}_m.
 \label{app:num-augmented-component-charge}
\end{equation}
The normalized fixed-$F^{\rm gen}$ law is
\begin{equation}
 \Pr(\mathcal S^{(3)}\mid F^{\rm gen})
 =4^{|M|-N_v+|\mathcal C(\mathcal G_{\rm aug})|}\prod_{B\in\mathcal C(\mathcal G_{\rm aug})}
 \mathbf1[\boldsymbol Q_B=0].
 \label{app:num-S3-fixed-F-law}
\end{equation}
In each component we draw all but one of its onsite and matching-edge labels
independently and uniformly, then fix the remaining label so that
$\boldsymbol Q_B=0$. Since $|M|=|\gamma|/2$, there are
\begin{equation}
 4^{N_v-|M|-|\mathcal C(\mathcal G_{\rm aug})|}
\end{equation}
allowed joint assignments, again proving normalization.

If $I_m$ denotes the internal vertices of the physical ribbon, the outcome is reconstructed from the sampled endpoint-pair channel by
\begin{equation}
 \boldsymbol a_m^{\rm reg}
 =\bar{\boldsymbol a}_m
  \bigoplus_{v\in I_m}\boldsymbol a_v.
 \label{app:num-restore-regional-outcome}
\end{equation}
Thus the sampler produces the complete physical fusion record, including
all onsite charges in the ribbon interior.

The matching uses cylindrical Manhattan distance with deterministic tie
breaking and path routing. When lattice routes touch or cross, fixed local
detours in the thickened lattice make their microscopic ribbon supports
disjoint without changing their endpoints or seam parities
\begin{equation}
 \eta_m=\bigoplus_{e\in m}c_e.
\end{equation}
After the record is generated, $F^{\rm gen}$ and its homology are discarded.
The decoder receives only the physical record and the known matching
convention.

\subsection{Binary multiple-histogram estimator}
\label{app:num-binary-estimator}

This subsection explains how we reuse sampled error configurations
to estimate the $\mathcal S^{(1)}$ failure probability at different
noise strengths, and how we obtain its statistical uncertainties.

At noise strength $p$, an occupied-edge support
$F\subseteq\mathcal E$ has probability
\begin{equation}
 P_p(F)=p^{|F|}(1-p)^{N_e-|F|}
\end{equation}
and produces the binary syndrome $\gamma=\partial F$.
For each target $p$, the exact contraction in
Eq.~\eqref{eq:num-binary-twisted-sum} determines the conditional
ML failure probability
\begin{equation}
 q_p(\gamma)=\min_{h\in\{0,1\}}\Pr_p(h\mid\gamma).
\end{equation}
The desired average is therefore
\begin{equation}
 P_{\rm fail}^{(1)}(p)
 =\sum_F P_p(F)\,q_p(\partial F).
\end{equation}

\subsubsection{The plotted estimator}
For each lattice size, we generate equal numbers of configurations
at nearby anchor probabilities $p_1,\ldots,p_{N_a}$.
Their pooled proposal distribution is
\begin{equation}
 Q(F)=\frac{1}{N_a}\sum_{a=1}^{N_a}P_{p_a}(F).
 \label{app:num-S1-proposal}
\end{equation}
The same pool is reused at every target $p$, which need not
coincide with an anchor. To correct for sampling from $Q$ rather
than $P_p$, configuration $F_i$ receives the importance weight
\begin{equation}
 w_i(p)=\frac{P_p(F_i)}{Q(F_i)}.
 \label{app:num-S1-importance-weight}
\end{equation}
This gives
$\mathbb E_{F\sim Q}[w_p(F)q_p(\partial F)]
=P_{\rm fail}^{(1)}(p)$.
Consequently, for $N_s$ pooled samples, ordinary importance
sampling provides the unbiased estimator
\begin{equation}
 \widehat P_{\rm IS}^{(1)}(p)
 =\frac{1}{N_s}\sum_{i=1}^{N_s}
 w_i(p)q_p(\partial F_i).
 \label{eq:num-S1-reweighted-risk}
\end{equation}
The conditional quantity $q_p$ is evaluated at the target $p,$ not at the anchor probability. 

The reported curves instead use the self-normalized average
\begin{equation}
 \widehat P_{\rm fail}^{(1)}(p)
 =
 \frac{\sum_iw_i(p)q_p(\partial F_i)}
      {\sum_iw_i(p)}.
 \label{app:num-S1-reweighted-risk}
\end{equation}
This normalization keeps the estimate within $[0,1/2]$ and
reproduces constant observables exactly. This statistical choice can reduce sampling
fluctuations and can provide a variance improvement over ordinary
importance sampling.

\subsubsection{Jackknife uncertainties}
We randomly divide the pooled configurations into $B=80$
approximately equal blocks. For each block $b$, we omit that
block and recompute the entire self-normalized ratio, including
its denominator. Writing the resulting estimate as
$\widehat P_{(-b)}(p)$ and its average over blocks as
$\overline P_{\rm JK}(p)$, the estimated standard error is
\begin{equation}
 \mathrm{SE}_{\rm JK}(p)
 =
 \left[
 \frac{B-1}{B}\sum_{b=1}^{B}
 \bigl(
 \widehat P_{(-b)}(p)-\overline P_{\rm JK}(p)
 \bigr)^2
 \right]^{1/2}.
\end{equation}
The same blocks are omitted at every target $p$, preserving
the correlations between estimates obtained from the same samples.
This pooled-block procedure gives approximate sampling
uncertainties and does not exploit the fixed sample counts
at each anchor.
\subsection{Exact bond-dimension-five contraction for the onsite record}
\label{app:num-S2-row-contraction}

This subsection derives the tensor network used for the
$\mathcal S^{(2)}$ likelihoods in
Sec.~\ref{sec:num-evaluation} and specifies its row-by-row contraction. The construction first sums the group variable locally at each vertex, then contracts the remaining auxiliary
edge indices. 

\subsubsection{Local factorization and group sums.}
For a fixed onsite record $\mathcal{S}^{(2)}=\{\lambda_v\}$, the twisted
bond weight is
\begin{equation}
 B_e^{(\kappa)}(g,h)
 =(1-p)+(-1)^{\kappa c_e}\frac p2\chi_E(hg^{-1}),
\end{equation}
where $c_e$ is the seam-crossing parity of edge $e$.
Using the real orthogonal representation in
Eq.~\eqref{eq:D4-real-E-representation},
\begin{equation} \chi_E(hg^{-1})=\operatorname{Tr}\!\left[E(h)E(g)^{\mathsf T}\right]=\sum_{a,b=1,2}E(g)_{ab}E(h)_{ab}.
\end{equation}
The occupied contribution therefore separates into four
products of endpoint functions. Together with the constant
unoccupied contribution, this gives five terms.

Define
\begin{equation}
 \begin{gathered}
 \phi_0(g)=1,\qquad \phi_{ab}(g)=E(g)_{ab},\\
 \mathcal I=\{0,(1,1),(1,2),(2,1),(2,2)\}.
 \end{gathered}
\end{equation}
Then
\begin{equation}
 B_e^{(\kappa)}(g,h)
 =\sum_{\mu\in\mathcal I}
 \beta_e^{(\kappa)}(\mu)\phi_\mu(g)\phi_\mu(h),
 \label{eq:num-rank-five-factorization}
\end{equation}
with
\begin{equation}
 \beta_e^{(\kappa)}(\mu)=
 \begin{cases}
  1-p,&\mu=0,\\
  (-1)^{\kappa c_e}p/2,&\mu=(a,b).
 \end{cases}
\end{equation}
The auxiliary index $\mu$ specifies a term in this sum, not a
physical degree of freedom. Each edge carries one such index,
shared by its two endpoints.

For fixed edge indices, all factors depending on $g_v$ can now
be collected and summed independently at vertex $v$:
\begin{equation}
 T_v^{\lambda_v}
 \!\left(\{\mu_e\}_{e\ni v}\right)
 ={}\frac{d_{\lambda_v}}8
 \sum_{g_v\in D_4}\chi_{\lambda_v}(g_v^{-1})\prod_{e\ni v}\phi_{\mu_e}(g_v).
 \label{eq:num-S2-local-tensor-main}
\end{equation}
The transformed partition function becomes
\begin{equation}
 \widetilde Z_{\mathcal{S}^{(2)}}^{(2)}(\kappa)
 =\sum_{\{\mu_e\in\mathcal I\}}
 \left[\prod_e\beta_e^{(\kappa)}(\mu_e)\right]
 \left[\prod_v
 T_v^{\lambda_v}\!\left(\{\mu_e\}_{e\ni v}\right)\right].
 \label{eq:num-S2-virtual-network-main}
\end{equation}
Every edge coefficient appears once. A bulk tensor has four
indices, while a tensor on an open boundary has three.
Each index takes five values, giving bond dimension five.

\subsubsection{Row contraction and periodic closure.}
Use coordinates $(x,y)$ with $x=1,\ldots,L$ around the periodic
direction and $y=1,\ldots,L$ between the open boundaries.
Let $h_{x,y}$ denote the horizontal index from $(x,y)$ to
$(x+1,y)$ and $v_{x,y}$ the vertical index between rows $y$
and $y+1$. The local tensor has arguments
\begin{equation}
 \bigl(h_{x-1,y},h_{x,y},v_{x,y-1},v_{x,y}\bigr),
\end{equation}
with missing vertical arguments omitted at the open boundaries.
Periodicity requires $h_{0,y}=h_{L,y}$.

Let $\boldsymbol d$ and $\boldsymbol u$ be the incoming and
outgoing vertical-index strings of a row. Define
\begin{equation}
 \mathsf R_y^{(\kappa)}(\boldsymbol d,\boldsymbol u)
 =\sum_{\boldsymbol h\in\mathcal I^L}
 \prod_{x=1}^{L}
 T_{x,y}^{\lambda_{x,y}}
 (h_{x-1,y},h_{x,y},d_x,u_x)\prod_{x=1}^{L}
 \beta_{e^h_{x,y}}^{(\kappa)}(h_{x,y})\prod_{x=1}^{L}
 \beta_{e^v_{x,y}}^{(\kappa)}(u_x).
 \label{app:num-S2-row-tensor}
\end{equation}
The first row has no incoming string, and the last has no
outgoing string or outgoing vertical coefficients. Each vertical
coefficient is included when its edge first appears as an
outgoing edge, so no edge weight is counted twice.

The boundary array is updated according to
\begin{align}
 \Psi_1^{(\kappa)}(\boldsymbol u)
 &=\mathsf R_1^{(\kappa)}(\varnothing,\boldsymbol u),\\
 \Psi_y^{(\kappa)}(\boldsymbol u)
 &=\sum_{\boldsymbol d\in\mathcal I^L}
 \Psi_{y-1}^{(\kappa)}(\boldsymbol d)
 \mathsf R_y^{(\kappa)}(\boldsymbol d,\boldsymbol u),
 \qquad 2\leq y<L,\\
 \widetilde Z_{\mathcal{S}^{(2)}}^{(2)}(\kappa)
 &=\sum_{\boldsymbol d\in\mathcal I^L}
 \Psi_{L-1}^{(\kappa)}(\boldsymbol d)
 \mathsf R_L^{(\kappa)}(\boldsymbol d,\varnothing).
 \label{app:num-S2-row-recursion}
\end{align}
Thus $\Psi_y$ retains the full dependence on the vertical
indices connecting the processed rows to the remaining lattice.

The implementation does not form a dense row tensor.
For each possible closing index $h_{L,y}$, it contracts the row
from left to right and retains only terms whose final horizontal
index equals that initial index. Summing over the closing index
implements the periodic boundary condition.

With the seam chosen across the periodic closing bond in each
row, the twisted coefficients there are
\begin{equation}
 \beta_{e^h_{L,y}}^{(1)}(0)=1-p,
 \qquad
 \beta_{e^h_{L,y}}^{(1)}(a,b)=-\frac p2.
\end{equation}
Away from the seam, the four occupied channels have coefficient
$+p/2$. The two contractions therefore use the same local tensors
and differ only in these seam coefficients.

\subsection{Component-frontier contraction for the fusion record}
\label{app:S3-frontier-contraction}

This subsection gives the nonnegative component expansion and
exact edge-summation algorithm used for the
$\mathcal S^{(3)}$ likelihoods in
Sec.~\ref{sec:num-evaluation}. Unlike the outer record sampler,
the inner calculation does not draw candidate error sets:
it sums their contributions by combining partial configurations
that have identical effects on the unprocessed graph.

\subsubsection{Component weights.}
Fix the measured record $\mathcal{S}^{(3)}$ and its prescribed matching $M$.
Write $F'\subseteq\mathcal E$ for a candidate physical edge set,
distinguishing it from the support $F^{\rm gen}$ used to
generate the record. The augmented multigraph is
\begin{equation}
 H'=H(F',M)=(V,F'\sqcup M).
\end{equation}
Matching edges are always present and are retained as distinct
edges even when a physical edge has the same endpoints.
They carry no Bernoulli occupation weight.

Let $\mathcal C(H')$ be the connected components, including
isolated vertices, and let $M_C$ denote the matching edges
in component $B$. Its total Abelian charge is
\begin{equation}
 \boldsymbol Q_B
 =
 \bigoplus_{v\in B\setminus\gamma}(x_v,z_v)
 \bigoplus_{m\in M_B}(\bar x_m,\bar z_m),
\end{equation}
where the barred labels are the endpoint-corrected fusion
outcomes defined in
Appendix~\ref{app:physical-record-sampling}.
All sums of charge labels are in $\mathbb Z_2^2$.

After the local group sums, each component contributes four
if its total charge vanishes and zero otherwise. This is the
character-orthogonality identity
\begin{equation}
 \sum_{\boldsymbol t\in\mathbb Z_2^2}
 (-1)^{\boldsymbol t\cdot\boldsymbol Q_B}
 =
 4\,\mathbf1[\boldsymbol Q_B=(0,0)].
\end{equation}
The factor of four counts the values of the two binary group
coordinates shared by that component. Including the syndrome
constraint gives
\begin{equation}
 \mathcal N_y(F')
 =4^{|\mathcal C(H')|}
 \prod_{v\in V}
 \mathbf1[\deg_{H'}(v)\text{ is even}]
 \prod_{B\in\mathcal C(H')}
 \mathbf1[\boldsymbol Q_B=(0,0)].
\end{equation}
The even-degree condition applies to every vertex; the augmented
graph need not be connected. Since the matching contributes one
incident edge at each vertex of $\gamma$ and none elsewhere,
\begin{equation}
 \deg_{H'}(v)
 =\deg_{F'}(v)+\mathbf1[v\in\gamma].
\end{equation}
Consequently, this condition is exactly
$\partial F'=\gamma$.

The homology-sector weights are
\begin{equation}
 \begin{aligned}
 Z_{\mathcal S^{(3)}}(h)
 =4^{|M|-N_v}
 \sum_{F'\subseteq\mathcal E}
 (1-p)^{N_e-|F'|}p^{|F'|}\mathcal N_y(F')\mathbf1\!\left[
 \bigoplus_{e\in F'}c_e \bigoplus_{m\in M}\eta_m=h
 \right].
 \end{aligned}
 \label{eq:num-component-frontier-weight}
\end{equation}
Here $c_e$ is the physical edge's seam-crossing parity, while
$\eta_m$ is the known seam-crossing parity of the reference
path associated with matching edge $m$.
Every contribution is nonnegative: incompatible configurations
have zero weight, and compatible configurations carry their
Bernoulli weight multiplied by the component factors.

\subsubsection{Frontier states and local updates}
For $p<1$, factor out $(1-p)^{N_e}$ and define
$t_p=p/(1-p)$. Physical edges then have absent and occupied
weights $1$ and $t_p$, respectively; matching edges are forced
with weight one. The common factor is restored after contraction.

Choose a fixed ordering of the physical and matching edges.
After any prefix has been processed, the \emph{frontier}
contains the vertices incident on both processed and
unprocessed edges. A state stores:
\begin{enumerate}
 \item the degree parity of every frontier vertex;
 \item their connectivity through processed occupied or
 forced edges;
 \item the accumulated $\mathbb Z_2^2$ charge of each
 component meeting the frontier; and
 \item the accumulated homology bit $h$.
\end{enumerate}
Initially the frontier is empty, its weight is one, and $h=0$.
Before an edge is processed, any previously unseen endpoint
is introduced as a singleton component with degree parity zero
and charge
\begin{equation}
 \boldsymbol q_v=
 \begin{cases}
 (x_v,z_v),&v\notin\gamma,\\
 (0,0),&v\in\gamma.
 \end{cases}
\end{equation}
Each vertex charge is inserted only once.

The edge updates are:
\begin{enumerate}
 \item An absent physical edge contributes weight one and
 leaves the parity, connectivity, charge, and homology unchanged.

 \item An occupied physical edge contributes $t_p$, toggles
 both endpoint degree parities, and joins their components.
 If those components were distinct, their charges are combined
 by XOR; if they were already connected, their charge is
 unchanged. The homology becomes $h\oplus c_e$.

 \item A forced matching edge makes the same parity and
 connectivity update, then inserts its charge
 $(\bar x_m,\bar z_m)$ into the resulting component and changes
 the homology to $h\oplus\eta_m$. Its charge is inserted exactly
 once, including when the endpoints were already connected.
 There is no factor of $t_p$.
\end{enumerate}

After processing a vertex's last incident edge, its degree
parity must be even; otherwise that branch is discarded.
The vertex can then be forgotten, with its accumulated charge
remaining attached to its component. When the final frontier
vertex of a component is forgotten, that component closes.
It is retained only if its total charge vanishes, in which
case its weight is multiplied by four.
Isolated vertices obey the same closure rule and therefore
must carry the vacuum charge.

At every step, states with identical remaining parity,
connectivity, charge, and homology data are combined by adding
their weights. This data contains everything needed to determine
the contribution of any future edge choices.

\subsubsection{Final weights and exactness.}
The resulting table has a simple invariant: each entry is the
sum over all partial occupied sets producing its stored data,
weighted by $t_p$ for every occupied physical edge and by four
for every component already closed. The local updates preserve
this invariant.

Once all edges and vertices have been processed, the frontier
is empty. If $W_{\rm final}(h)$ denotes its two remaining weights,
then
\begin{equation}
 Z_{\mathcal S^{(3)}}(h)
 =
 4^{|M|-N_v}(1-p)^{N_e}W_{\rm final}(h),
 \qquad h=0,1.
\end{equation}
Equivalently, one may use physical-edge branch weights
$1-p$ and $p$ directly, in which case the final prefactor is
only $4^{|M|-N_v}$.

The algorithm thus sums all candidate supports without
explicitly listing them individually. Its cost is governed
by the maximum frontier width and the number of reachable
frontier states; long matching edges can increase that width.
\subsection{Estimators, error bars, and validation}
\label{app:num-estimators}

This subsection explains how the exactly contracted record risks are
averaged and reports independent checks of both inner algorithms. For an
exactly contracted record define
\begin{equation}
 q(\mathcal{S})=\min_h\frac{Z_{\mathcal{S}}(h)}{Z_{\mathcal{S}}(0)+Z_{\mathcal{S}}(1)}.
 \label{app:num-record-risk}
\end{equation}
For directly sampled independent records, the estimator and its standard
error are
\begin{equation}
 \widehat P_{\rm fail}^{(i)}=\frac{1}{N_s}\sum_{i=1}^{N_s}q(\mathcal{S}^{(i)}),
 \qquad
 \operatorname{SE}(\widehat P_{\rm fail})=\frac{s_q}{\sqrt{N_s}}.
 \label{app:num-direct-risk-estimator}
\end{equation}
This is used for $\mathcal S^{(2)}$ and for the 
$L=3,4,5,6$ $\mathcal S^{(3)}$ data. The larger-size $\mathcal S^{(3)}$
production data use antithetic pairs. For each independent set of uniforms $U_e$, define
$F_e^+=\mathbf1[U_e<p]$ and $F_e^-=\mathbf1[U_e>1-p]$. Both supports have
the correct Bernoulli-$p$ marginal. The independent sampling unit is
\begin{equation}
 Q_a=\frac12\bigl[q(\mathcal{S}_a^+)+q(\mathcal{S}_a^-)\bigr],
 \label{app:num-antithetic-unit}
\end{equation}
and the reported error is $s_Q/\sqrt{N_{\rm pair}}$, not the standard error
obtained by treating the two members of a pair as independent.

\subsection{Averaging the conditional decoding failure probability}
\label{app:num-failure-observable}

This subsection explains the observable plotted in
Figs.~\ref{fig:num-s1-cluster}--\ref{fig:num-s3-cluster}.
The average ML failure probability can be estimated either by
counting failed decoding attempts or by averaging the conditional
failure probability of each measured syndrome. We use the latter
quantity, which averages over the unresolved homology for each
syndrome.

Fix a record type $j\in\{1,2,3\}$, and denote the measured
syndrome by $\mathcal S$. Its conditional failure probability is
\begin{equation}
 q_j(\mathcal S)
 \equiv\epsilon_{\rm ML}(\mathcal S)
 =
 \frac{\min_{h\in\{0,1\}}Z_{\mathcal S}(h)}
      {Z_{\mathcal S}(0)+Z_{\mathcal S}(1)}.
 \label{eq:num-plotted-conditional-risk}
\end{equation}
For example, if the two posterior sector probabilities are
$0.98$ and $0.02$, the decoder selects the first sector and
this syndrome contributes $0.02$ to the average, rather than
a zero or one determined by the particular sampled error.

To establish the equivalence, let $H_j$ denote the hidden
residual homology and let
$\widehat h_E(\mathcal S)=\operatorname*{arg\,max}_h
Z_{\mathcal S}(h)$, with a fixed tie-breaking convention.
The failure indicator for an individual realization is
\begin{equation}
 X_j=\mathbf1[\widehat h_E(\mathcal S)\neq H_j].
\end{equation}
Under the physical joint distribution of errors and syndromes,
\begin{equation}
 \mathbb E[X_j\mid\mathcal S]=q_j(\mathcal S).
\end{equation}
Consequently,
\begin{equation}
 \mathbb E[X_j]
 =\mathbb E_{\mathcal S\sim\Pr_p}[q_j(\mathcal S)]
 =P_{\rm fail}^{(j)}(p,L).
\end{equation}
Replacing the binary indicator by its conditional expectation
is a Rao--Blackwellization: it preserves the physical mean
while removing the fluctuations associated with the unresolved
homology. Explicitly,
\begin{equation}
 \begin{aligned}
 \operatorname{Var}(X_j)
 &=\operatorname{Var}\!\left(q_j(\mathcal S)\right)\\
 &\quad+\mathbb E\!\left[
 q_j(\mathcal S)\bigl(1-q_j(\mathcal S)\bigr)\right]\\
 &\geq\operatorname{Var}\!\left(q_j(\mathcal S)\right).
 \end{aligned}
\end{equation}

This variance comparison applies to direct sampling from the
physical distribution. The importance reweighting used for
$\mathcal S^{(1)}$ is a separate outer estimation step, described
in Appendix~\ref{app:num-binary-estimator}; in particular,
Rao--Blackwellization does not remove the finite-sample bias
of its self-normalized ratio estimator.

\subsection{Crossings and finite-size scaling}
\label{app:num-scaling-protocol}

This subsection specifies the data, uncertainty estimates, and correction
models behind Sec.~\ref{sec:num-threshold-intervals}. It also explains why
the opposite crossing drifts seen for some parity sequences do not by
themselves bound the infinite-size threshold. All conditional likelihoods
in these figures are contracted exactly; sampling enters only the outer
average over physical records.

\subsubsection{Datasets and crossing estimates}
The binary calculation uses 339,000 sampled supports across $p$;
for $L=3,\ldots,12$, the counts are
$(60000,60000,45000,45000,36000,30000,24000,18000,12000,9000)$.
The onsite calculation uses 100,000 independent records per point, on
$p=0.130,0.132,\ldots,0.190$ for odd $L=3,5,7,9$ and
$p=0.164,0.166,\ldots,0.206$ for even $L=4,6,8,10$.
For fusion, each $L=3,4,5,6,7$ point uses one million records on $p=0.200,0.2025,\ldots,0.250$.

Same-parity crossings are estimated by linear interpolation for
$\mathcal S^{(1)}$ and local quadratic fits to curve differences for
$\mathcal S^{(2)},\mathcal S^{(3)}$. They are diagnostics rather than
the threshold estimator: changing the fitting window or polynomial order
can shift a crossing beyond its sampling error.

To see the origin of their drift, set $\epsilon=(-1)^L$ and
$A_{j,\epsilon}=B_j(0)+\epsilon C_j(0)$. Expanding
Eqs.~\eqref{eq:num-interval-scaling} and
\eqref{eq:num-interval-corrections} for two same-parity sizes $L,L'$,
with $s=L'/L>1$, gives to leading order
\begin{equation}
 p_{\times,j}(L,L')-p_{c,j}
 =\frac{A_{j,\epsilon}(1-s^{-\omega_j})}
 {F_j'(0)(s^{1/\nu_j}-1)}
 L^{-(\omega_j+1/\nu_j)}.
 \label{eq:num-parity-crossing-drift}
\end{equation}
This assumes $F_j'(0)\neq0$ and that the retained correction dominates.
For an increasing failure curve, a negative correction amplitude produces
crossings approaching from below and a positive amplitude crossings approaching from above. 

\subsubsection{Fits and statistical uncertainty.}
We fit $p_c,\nu,\omega$ simultaneously, solve for the polynomial
coefficients by weighted linear least squares at each nonlinear parameter
choice, and use multiple initial conditions. The size ranges and windows
are in Table~\ref{tab:num-conditional-thresholds}; only sampled points in
those windows enter. The leading polynomial degrees are $(3,2,2)$,
the correction degrees are $(0,0,1)$, and the fitted effective correction
exponents are $(5.37,3.76,5.73)$. The conditional standard errors are
\begin{equation}
 \begin{aligned}
 p_c^{(1)}&=0.11143\pm0.00107,\\
 p_c^{(2)}&=0.16509\pm0.00052,\\
 p_c^{(3)}&=0.21429\pm0.00076.
 \end{aligned}
 \label{eq:num-conditional-fit-pc}
\end{equation}

\section{Syndrome probabilities for the \(S_3\) quantum double}
\label{app:s3-syndrome-probabilities}

This appendix applies the general likelihoods of
Secs.~\ref{sec:onsiteprob} and~\ref{sec:fusion-syndrome-probabilities}
to $S_3$ quantum double, following their sequence of derivation and
physical interpretation. It derives the three syndrome distributions
for decoherence by the two-dimensional electric charge of \(S_3\). We follow the same sequence
as for \(D_4\): insert the physical charge projectors, evaluate the syndrome probabilities using properties of the group, and then interpret the resulting positive statistical model. 
We note that \(S_3\) denotes the group, whereas \(\mathcal S^{(1)}\), \(\mathcal S^{(2)}\), and \(\mathcal S^{(3)}\) denote the measurement records.

\subsection{Setup and character data}

We first specify the charge labels and the initial state used throughout
the calculation. Write
\begin{equation}
 \begin{aligned}
 S_3&=\langle r,s\mid r^3=s^2=1,\ srs=r^{-1}\rangle,\\
 g&=r^ts^q,\qquad t\in\mathbb Z_3,\quad q\in\mathbb Z_2,
 \qquad \omega=e^{2\pi i/3}.
 \end{aligned}
\end{equation}
The irreducible charges are the vacuum \(A\), the sign charge \(B\), and
the two-dimensional charge \(C\). Their characters are
\begin{equation}
 \begin{aligned}
 \chi_A(r^ts^q)&=1,\\
 \chi_B(r^ts^q)&=(-1)^q,\\
 \chi_C(r^ts^q)&=\delta_{q,0}(\omega^t+\omega^{-t}),
 \qquad d_A=d_B=1,\quad d_C=2.
 \end{aligned}
 \label{eq:app-s3-characters}
\end{equation}
The corresponding fusion rules are
\begin{equation}
 B\otimes B=A,\qquad B\otimes C=C,\qquad
 C\otimes C=A\oplus B\oplus C.
 \label{eq:app-s3-fusion-rules}
\end{equation}

We use the same independent edge-decoherence channel as in the main text,
with the noisy charge now \(\Gamma=C\). Its contribution to the
character contraction on each edge is
\begin{equation}
 K_{C,p}(g)=(1-p)+\frac p2\chi_C(g).
 \label{eq:app-s3-kernel}
\end{equation}
On the cylinder, we take the initial state to be
\(\rho_0=\ket{[1]}\!\bra{[1]}\). This specifies the trivial-holonomy sector
in which the untwisted formulas below apply. Since \(Z(S_3)=\{1\}\), the
central-holonomy subspace is one-dimensional; these formulas alone do not
establish a protected central qubit or state-independent fusion statistics
on the full cylinder ground space.

For a charge outcome \(\lambda\in\{A,B,C\}\), the onsite projector is
\begin{equation}
 P_v^\lambda
 =\frac{d_\lambda}{6}
 \sum_{h\in S_3}\chi_\lambda(h^{-1})A_v^h.
 \label{eq:app-s3-onsite-projector}
\end{equation}
As in the main-text derivation, inserting these projectors into the Born
trace gives one group variable per vertex and the kernel
\(K_{C,p}(h_{f_e}h_{i_e}^{-1})\) on an oriented edge
\(e=\langle i_ef_e\rangle\).
We denote the numbers of vertices and edges by \(N_v\) and \(N_e\).

\subsection{The coarse record \(\mathcal S^{(1)}\)}

The first record specifies only the set \(\gamma\) of vertices carrying
\(C\) charge. We derive an oriented \(\mathbb Z_3\) model for this
distribution by summing over the unresolved \(A/B\) outcomes.

\paragraph{Derivation.}
The projector is \(P_v^C\) at \(v\in\gamma\) and \(P_v^A+P_v^B\) elsewhere.
Consequently,
\begin{equation}
 \Pr_p(\mathcal S^{(1)}=\gamma)
 =\frac1{6^{N_v}}
 \sum_{\{h_v\in S_3\}}
 \prod_{v\in\gamma}2\chi_C(h_v^{-1})
 \prod_{v\notin\gamma}
 [\chi_A(h_v^{-1})+\chi_B(h_v^{-1})]
 \prod_e K_{C,p}(h_{f_e}h_{i_e}^{-1}).
 \label{eq:app-s3-s1-character}
\end{equation}
Writing \(h_v=r^{t_v}s^{q_v}\), both kinds of vertex factor force
\(q_v=0\): the unresolved factor is \(2\delta_{q_v,0}\), while the
\(C\) factor is \(2\delta_{q_v,0}(\omega^{t_v}+\omega^{-t_v})\).
The \(q_v\) sums therefore give
\begin{equation}
 \Pr_p(\gamma)
 =\frac1{3^{N_v}}
 \sum_{\{t_v\in\mathbb Z_3\}}
 \prod_{v\in\gamma}(\omega^{t_v}+\omega^{-t_v})
 \prod_e\left[
 (1-p)+\frac p2\sum_{k=\pm1}
 \omega^{k(t_{f_e}-t_{i_e})}\right].
 \label{eq:app-s3-s1-three-state}
\end{equation}

Expand each edge factor by assigning \(k_e\in\{0,+1,-1\}\), with weight
\begin{equation}
 \widetilde w_p(k)=
 \begin{cases}
  1-p,&k=0,\\
  p/2,&k=\pm1.
 \end{cases}
 \label{eq:app-s3-coarse-edge-weight}
\end{equation}
Let \(\epsilon_{ve}=+1\) for an edge entering \(v\) and \(-1\) for an
edge leaving \(v\), and define
\begin{equation}
 K_v=(\partial k)_v
 =\sum_{e\ni v}\epsilon_{ve}k_e\pmod3.
 \label{eq:app-s3-mod3-divergence}
\end{equation}
Collecting all edge phases at a vertex leaves \(\omega^{K_vt_v}\).
The two local sums are
\begin{equation}
 \begin{aligned}
 \frac13\sum_{t\in\mathbb Z_3}\omega^{Kt}
 &=\delta_{K,0},\\
 \frac13\sum_{t\in\mathbb Z_3}
 (\omega^t+\omega^{-t})\omega^{Kt}
 &=\delta_{K,1}+\delta_{K,-1}
 =1-\delta_{K,0}.
 \end{aligned}
 \label{eq:app-s3-coarse-vertex-sums}
\end{equation}
Thus the measured set is precisely the set of nonzero divergences in this
auxiliary expansion. Define
\begin{equation}
 \begin{aligned}
 \mathcal K_\gamma
 &=\{\mathbf k:K_v\ne0\ \text{if and only if}\ v\in\gamma\},\\
 F(\mathbf k)&=\{e:k_e\ne0\}.
 \end{aligned}
 \label{eq:app-s3-current-support}
\end{equation}
The probability becomes
\begin{equation}
 \Pr_p(\mathcal S^{(1)}=\gamma)
 =
 \sum_{\mathbf k\in\mathcal K_\gamma}
 (1-p)^{N_e-|F(\mathbf k)|}
 \left(\frac p2\right)^{|F(\mathbf k)|}.
 \label{eq:app-s3-s1-edge}
\end{equation}
Normalization follows directly: every \(\mathbf k\) determines one
\(\gamma\), and \(\sum_k\widetilde w_p(k)=1\).

\paragraph{Physical interpretation.}
The resulting model is a gas of oriented occupied edges. An unoccupied
edge has weight \(1-p\), and either orientation of an occupied edge has
weight \(p/2\). The coarse measurement specifies where the mod-three flow
has a nonzero source or sink, without resolving its sign.

This differs from the binary support constraint for \(D_4\): the set
\(F(\mathbf k)\) alone does not determine the \(C\)-charge syndrome,
because the orientations also matter. Three incoming unit flows, for
example, sum to zero modulo three, so branching is allowed without a
\(C\) charge at the junction. A connected component can also contain an
odd number of \(C\) charges, provided its signed divergences sum to zero.

\subsection{The complete onsite record \(\mathcal S^{(2)}\)}

The second record also distinguishes \(A\) from \(B\) at every non-\(C\)
vertex. We first derive the additional binary constraints and then sum the
binary edge labels to obtain a component model directly analogous to the
\(D_4\) onsite model.

\paragraph{Derivation.}
For \(\boldsymbol\lambda=\{\lambda_v\}\), inserting every onsite
projector gives
\begin{equation}
 \Pr_p(\mathcal S^{(2)}=\boldsymbol\lambda)
 =
 \frac1{6^{N_v}}
 \sum_{\{h_v\in S_3\}}
 \prod_v d_{\lambda_v}\chi_{\lambda_v}(h_v^{-1})
 \prod_e K_{C,p}(h_{f_e}h_{i_e}^{-1}).
 \label{eq:app-s3-s2-character}
\end{equation}
Now retain both coordinates of \(h_v\). For
\(a=(k,z)\in\mathbb Z_3\oplus\mathbb Z_2\), introduce
\begin{equation}
 u_{kz}(r^ts^q)=\omega^{kt}(-1)^{zq},
 \qquad
 \frac16\sum_{g\in S_3}u_a(g)u_b(g)^*=\delta_{a,b}.
 \label{eq:app-s3-fourier-basis}
\end{equation}
These form a basis of functions on the group elements. For \(g=r^ts^q\) and \(h=r^{t'}s^{q'}\),
\begin{equation}
 \begin{aligned}
 \frac12\chi_C(gh^{-1})
 &=\delta_{q,q'}\cos\!\left[\frac{2\pi(t-t')}{3}\right]\\
 &=\frac14\sum_{k=\pm1}\sum_{z\in\mathbb Z_2}
 u_{kz}(g)u_{kz}(h)^*.
 \end{aligned}
 \label{eq:app-s3-edge-character-resolution}
\end{equation}
Hence each edge has the expansion
\begin{equation}
 K_{C,p}(h_{f_e}h_{i_e}^{-1})
 =
 \sum_{\ell_e}w_p(\ell_e)
 u_{\ell_e}(h_{f_e})u_{\ell_e}(h_{i_e})^*,
 \qquad \ell_e=(k_e,z_e),
 \label{eq:app-s3-edge-expansion}
\end{equation}
with
\begin{equation}
 w_p(k,z)=
 \begin{cases}
  1-p,&(k,z)=(0,0),\\
  p/4,&k=\pm1,\ z\in\mathbb Z_2,\\
  0,&(k,z)=(0,1).
 \end{cases}
 \label{eq:app-s3-edge-weight}
\end{equation}
An unoccupied edge therefore has \(z_e=0\).

Collect the edge phases at each vertex into
\begin{equation}
 b_v=(K_v,J_v)
 =(\partial\ell)_v
 =\sum_{e\ni v}\epsilon_{ve}\ell_e
 \in\mathbb Z_3\oplus\mathbb Z_2.
 \label{eq:app-s3-incidence}
\end{equation}
The remaining local group sum is
\begin{equation}
 \mathcal V_\lambda(K,J)
 =
 \frac{d_\lambda}{6}
 \sum_{t\in\mathbb Z_3}\sum_{q\in\mathbb Z_2}
 \chi_\lambda((r^ts^q)^{-1})
 \omega^{Kt}(-1)^{Jq}.
 \label{eq:app-s3-vertex-filter-definition}
\end{equation}
For \(A\) and \(B\), the two sums impose \(K=0\) and respectively
\(J=0\) or \(J=1\). For \(C\), the character fixes \(q=0\), so \(J\)
drops out; the remaining sum is the second identity in
Eq.~\eqref{eq:app-s3-coarse-vertex-sums}. Therefore
\begin{equation}
 \begin{aligned}
 \mathcal V_A(K,J)&=\delta_{K,0}\delta_{J,0},\\
 \mathcal V_B(K,J)&=\delta_{K,0}\delta_{J,1},\\
 \mathcal V_C(K,J)&=1-\delta_{K,0}.
 \end{aligned}
 \label{eq:app-s3-explicit-filters}
\end{equation}
Performing all vertex sums yields the positive model
\begin{equation}
 \Pr_p(\mathcal S^{(2)}=\boldsymbol\lambda)
 =
 \sum_{\boldsymbol\ell}
 \prod_e w_p(\ell_e)
 \prod_v\mathcal V_{\lambda_v}(b_v).
 \label{eq:app-s3-s2-edge}
\end{equation}

To expose its component structure, let
\(\gamma=\{v:\lambda_v=C\}\) and
\(\mathcal B=\{v:\lambda_v=B\}\).
Fix \(\mathbf k\in\mathcal K_\gamma\) and write \(F=F(\mathbf k)\).
At every \(v\notin\gamma\), the binary labels must satisfy
\(J_v=\mathbf1[v\in\mathcal B]\); at \(v\in\gamma\), \(J_v\) is
unconstrained. On a connected component \(D\) of the spanning graph
\((V,F)\), write \(n_D=|V(D)|\), \(m_D=|E(D)|\), and
\(g_D=|D\cap\gamma|\). The binary incidence matrix has rank \(n_D-1\).
If \(g_D=0\), its only compatibility condition is even total \(B\) parity.
If \(g_D>0\), the remaining \(n_D-g_D\) rows are independent.
The number of allowed binary assignments is consequently
\begin{equation}
 n_z(D)=
 \begin{cases}
  2^{m_D-n_D+1}\,
  \mathbf1[|D\cap\mathcal B|\ \text{is even}],&g_D=0,\\
  2^{m_D-n_D+g_D},&g_D>0.
 \end{cases}
 \label{eq:app-s3-binary-component-count}
\end{equation}

Let \(\mathcal C_0(F)\) be the components containing no \(C\) vertex,
including isolated vertices. Multiplying
Eq.~\eqref{eq:app-s3-binary-component-count} over components and combining
with the occupied-edge weight \((p/4)^{|F|}\) gives
\begin{equation}
 \begin{aligned}
 \Pr_p(\mathcal S^{(2)}=(\gamma,\mathcal B))
 =2^{|\gamma|-N_v}
 \sum_{\mathbf k\in\mathcal K_\gamma}
 (1-p)^{N_e-|F|}
 \left(\frac p2\right)^{|F|}
 2^{|\mathcal C_0(F)|}
 \prod_{D\in\mathcal C_0(F)}
 \mathbf1[|D\cap\mathcal B|\ \text{is even}].
 \end{aligned}
 \label{eq:app-s3-s2-component}
\end{equation}
For fixed \(\mathbf k\), there are
\(2^{N_v-|\gamma|-|\mathcal C_0(F)|}\) allowed \(A/B\) assignments.
Summing them cancels the component prefactor and recovers
Eq.~\eqref{eq:app-s3-s1-edge} exactly.

\paragraph{Physical interpretation.}
The complete onsite record adds a binary decoration to the oriented
mod-three model. A vertex with zero mod-three divergence is vacuum when
its binary divergence is even and carries \(B\) when it is odd.
A \(C\) vertex does not constrain this binary divergence.

After the binary labels are summed, the content is particularly simple.
Every component without a \(C\) charge must contain an even number of
\(B\) charges, so its total Abelian charge is vacuum. A component containing
\(C\) charges has no such restriction on its \(B\) content, consistent
with \(B\otimes C=C\). In particular, an isolated vertex must be \(A\),
so a measured \(B\) charge certifies that the vertex touches the occupied
support. These extra neutrality conditions eliminate configurations that
remain compatible with the coarse record. Relative to the \(D_4\) model,
there is one binary decoration and one Abelian parity constraint per
\(C\)-free component, producing factors of two rather than four.

\subsection{The pair-fusion record \(\mathcal S^{(3)}\)}

The third record supplements the complete onsite charges with
fusion outcomes of prescribed pairs of \(C\) anyons. We first
write the corresponding character contraction, then use the
same edge expansion as for \(\mathcal S^{(2)}\) to derive the
additional pair constraints.

\paragraph{Derivation.}
For a fixed onsite record
\(\boldsymbol\lambda=\{\lambda_v\}\), choose a deterministic
set \(M\) of disjoint \(C\)-anyon pairs. Each pair
\(m=(i_m,j_m)\) has a prescribed ribbon with vertex set
\(R_m=\{i_m,j_m\}\cup I_m\), where the intermediate vertices
\(I_m\) carry only known \(A/B\) charges. We assume disjoint
microscopic ribbon supports; any unpaired \(C\) anyons
retain only their onsite measurement.

As in the \(D_4\) calculation of
Sec.~\ref{sec:fusion-D4-components}, distinguish the measured
total ribbon charge \(\Lambda_m\) from the endpoint-pair
outcome \(q_m\in\{A,B,C\}\). They are related by
\begin{equation}
 \pi_m=\sum_{v\in I_m}\mathbf1[\lambda_v=B]\pmod2,
 \qquad
 \Lambda_m=B^{\otimes\pi_m}\otimes q_m.
 \label{eq:app-s3-interior-B-parity}
\end{equation}
Thus an odd interior \(B\) parity exchanges \(A\) and \(B\),
while leaving \(C\) unchanged. We label the refined record
by the corrected outcomes,
\(\mathcal S^{(3)}=(\boldsymbol\lambda,\mathbf q)\).

Using the general ribbon projectors
\eqref{app:pair-fusion-projector} with \(G=S_3\), the joint
probability is
\begin{equation}
 \Pr_p(\mathcal S^{(3)})
 =
 \operatorname{Tr}\!\left[
 \mathcal N_{C,p}(\rho_0)
 \left(\prod_vP_v^{\lambda_v}\right)
 \left(\prod_{m\in M}P_m^{\Lambda_m}\right)
 \right].
 \label{eq:app-s3-fusion-Born-trace}
\end{equation}
The ribbon construction and trace evaluation are unchanged
from Sec.~\ref{sec:fusion-syndrome-probabilities} and
Appendix~\ref{app:fusion-probability}; only the group and
its characters are specialized to \(S_3\).

In the regional source
\eqref{eq:S3-general-regional-factor}, each intermediate
one-dimensional character factorizes as
\(\chi_{\lambda_v}(c_v^{-1}k)
=\chi_{\lambda_v}(c_v^{-1})\chi_{\lambda_v}(k)\).
Their product contributes \(\chi_B(k)^{\pi_m}\), which
converts \(\chi_{\Lambda_m}(k^{-1})\) into
\(\chi_{q_m}(k^{-1})\). The remaining endpoint factor is
\begin{equation}
 V_q(c_i,c_j)
 =
 \frac{d_q}{6}\sum_{k\in S_3}
 \chi_q(k^{-1})
 \bigl[2\chi_C(c_i^{-1}k)\bigr]
 \bigl[2\chi_C(c_j^{-1}k)\bigr].
 \label{eq:app-s3-pair-character-factor}
\end{equation}
Writing \(V(M)=\bigcup_{m\in M}\{i_m,j_m\}\), the resulting
character contraction is
\begin{equation}
 \begin{aligned}
 \Pr_p(\boldsymbol\lambda,\mathbf q)
 ={}&\frac1{6^{N_v}}\sum_{\{c_v\in S_3\}}
 \prod_e K_{C,p}(c_{f_e}c_{i_e}^{-1})\\
 &\times\prod_{v\notin V(M)}
 d_{\lambda_v}\chi_{\lambda_v}(c_v^{-1})
 \prod_{m\in M}V_{q_m}(c_{i_m},c_{j_m}).
 \end{aligned}
 \label{eq:app-s3-s3-character}
\end{equation}
Thus, relative to the onsite expression
\eqref{eq:app-s3-s2-character}, the physical-edge kernels
are unchanged, while each measured pair replaces its two
onsite source factors by \(V_{q_m}\).
The intermediate \(A/B\) vertices retain their ordinary
onsite factors.

We now use the Fourier basis
\eqref{eq:app-s3-fourier-basis} and edge expansion
\eqref{eq:app-s3-edge-expansion}, with the same weights
\(w_p(\ell_e)\) as in \(\mathcal S^{(2)}\).
Collecting the edge phases at each vertex gives
\(u_{b_v}(c_v)\), where
\begin{equation}
 b_v=(K_v,J_v)
 =\sum_{e\ni v}\epsilon_{ve}\ell_e
 \in\mathbb Z_3\oplus\mathbb Z_2.
\end{equation}
At a vertex outside \(V(M)\), the group sum produces the
onsite filter \(\mathcal V_{\lambda_v}(b_v)\) of
Eq.~\eqref{eq:app-s3-explicit-filters}.

For a measured endpoint, Fourier orthogonality gives the
shifted version of the same local sum:
\begin{equation}
 \frac{2}{6}\sum_{c\in S_3}
 \chi_C(c^{-1}k)u_b(c)
 =
 \mathcal V_C(b)u_b(k).
 \label{eq:app-s3-shifted-source-sum}
\end{equation}
Applying this identity to both endpoints in \(V_q\)
leaves
\begin{equation}
 \begin{aligned}
 &\frac1{6^2}\sum_{c_i,c_j\in S_3}
 V_q(c_i,c_j)u_{b_i}(c_i)u_{b_j}(c_j)\\
 &\qquad=
 \mathcal V_C(b_i)\mathcal V_C(b_j)
 \Phi_q(b_i,b_j),
 \end{aligned}
 \label{eq:app-s3-pair-contraction}
\end{equation}
where
\begin{equation}
 \begin{aligned}
 \Phi_q(b_i,b_j)
 &=\frac{d_q}{6}\sum_{k\in S_3}
 \chi_q(k^{-1})u_{b_i}(k)u_{b_j}(k)\\
 &=\mathcal V_q(b_i+b_j).
 \end{aligned}
 \label{eq:app-s3-regional-sum}
\end{equation}
The last equality follows because the two Fourier phases
add their mode labels. Substituting the explicit onsite
filters therefore gives
\begin{equation}
 \begin{aligned}
 \Phi_A(b_i,b_j)
 &=\delta_{K_i+K_j,0}\delta_{J_i+J_j,0},\\
 \Phi_B(b_i,b_j)
 &=\delta_{K_i+K_j,0}\delta_{J_i+J_j,1},\\
 \Phi_C(b_i,b_j)
 &=1-\delta_{K_i+K_j,0},
 \end{aligned}
 \label{eq:app-s3-pair-factors}
\end{equation}
with sums taken modulo three and two in the respective
coordinates.

Performing all vertex sums yields the positive model
\begin{equation}
 \begin{aligned}
 \Pr_p(\mathcal S^{(3)}
       =(\boldsymbol\lambda,\mathbf q))
 =\sum_{\boldsymbol\ell}
 &\left[\prod_e w_p(\ell_e)\right]
 \left[\prod_v\mathcal V_{\lambda_v}(b_v)\right]\\
 &\times\prod_{m\in M}
 \Phi_{q_m}(b_{i_m},b_{j_m}).
 \end{aligned}
 \label{eq:app-s3-s3-edge}
\end{equation}
This is the same decorated-edge ensemble as
Eq.~\eqref{eq:app-s3-s2-edge}, with one additional
constraint for each measured pair. Since
\(\Phi_A+\Phi_B+\Phi_C=1\), summing over the pair outcomes
recovers the onsite distribution:
\begin{equation}
 \sum_{\mathbf q}
 \Pr_p(\boldsymbol\lambda,\mathbf q)
 =
 \Pr_p(\boldsymbol\lambda).
 \label{eq:app-s3-fusion-marginal}
\end{equation}

\paragraph{Physical interpretation.}
The onsite record identifies each \(C\) anyon with a
nonzero mod-three divergence, \(K_i=\pm1\), but does not
resolve its sign or binary coordinate \(J_i\).
A pair-fusion measurement constrains how these unresolved
labels combine. Opposite mod-three labels can fuse to
\(A\) or \(B\), distinguished by the parity of \(J_i+J_j\);
equal mod-three labels fuse to \(C\).
The factors in Eq.~\eqref{eq:app-s3-pair-factors} therefore
distinguish configurations with identical onsite charges
but different fusion outcomes for the selected pair.

Known intermediate \(A/B\) charges only determine the
relabeling used to extract the endpoint-pair outcome.
After that correction, each fusion factor depends solely
on the two endpoints. 

Unlike \(E\otimes E\) for \(D_4\), a pair of \(C\) anyons
can fuse to a non-Abelian charge. These measurements
resolve the fusion channel of each selected pair, but
need not resolve all remaining collective fusion
information. 

\section{R\'enyi-2 purity calculation and exact contraction}
\label{app:renyi2-purity-derivation}

This appendix presents the derivation of the R\'enyi-2 diagnostic used in
Sec.~\ref{sec:renyi2-intrinsic-proxy}. We first expand the physical
density-matrix traces, then reduce them to twisted eight-state partition
functions and specify their exact local contraction. We also prove the
environment interpretation of the two endpoint values. The final
subsection records the finite-size fits underlying
Fig.~\ref{fig:r2proxy-collapse}.

\subsection{Expanding the physical purity traces}
\label{sec:r2proxy-purities}

This subsection derives the quantities $A_{CD}$ and explains
how their sums give the two purities used in
Sec.~\ref{sec:renyi2-intrinsic-proxy}. We first evaluate the
channel on logical matrix elements, before rewriting the
result as a classical partition function.

Let $\mathcal O_C$ denote the flat edge configurations with
holonomy class $C$. The normalized logical state is
\begin{equation}
 |C\rangle
 =\frac1{\sqrt{N_C}}
 \sum_{\mathbf g\in\mathcal O_C}|\mathbf g\rangle,
 \qquad
 N_C=|\mathcal O_C|=|C|\,8^{N_v-1}.
 \label{eq:r2proxy-code-state}
\end{equation}
The counting factor is derived in
Appendix~\ref{sec:r2proxy-partition}.

Using Eq.~\eqref{eq:onsite-noisy-state-explicit}, expanding the logical matrix element gives
\begin{equation}
 X_{CD}
 =\frac1{\sqrt{N_CN_D}}
 \sum_{\substack{\mathbf g\in\mathcal O_C\\
                 \mathbf k\in\mathcal O_D}}
 \left[\prod_eK_{E,p}(g_ek_e^{-1})\right]
 |\mathbf g\rangle\langle\mathbf k|.
 \label{eq:r2proxy-Xexpand}
\end{equation}

The physical matrix units
$|\mathbf g\rangle\langle\mathbf k|$ are orthonormal under
the Hilbert--Schmidt inner product. Taking the squared norm
of this expansion thus retains only equal configuration pairs:
\begin{equation}
 A_{CD}=\operatorname{Tr}(X_{CD}X_{CD}^{\dagger})=\frac1{N_CN_D}
 \sum_{\substack{\mathbf g\in\mathcal O_C\\
                 \mathbf k\in\mathcal O_D}}
 \prod_e w_p(g_ek_e^{-1}),
 \label{eq:r2proxy-orbitsum}
\end{equation}
where $w_p(x)=|K_{E,p}(x)|^2.$ In other words, $A_{CD}$ is the uniform average of the squared
coherence-retention factor $K_{E,p}(x)$ over pairs of configurations from
the two logical sectors. 

We next assemble the two output density matrices:
\begin{equation}
 \begin{aligned}
 \rho_{RB,d}
 &=\frac1d\sum_{C,D\in\mathcal I_d}
 |C\rangle\langle D|_R\otimes X_{CD},\\
 \rho_{B,d}
 &=\frac1d\sum_{C\in\mathcal I_d}X_{CC}.
 \end{aligned}
 \label{eq:r2proxy-blocks}
\end{equation}
For the joint purity, tracing the reference indices pairs
$X_{CD}$ with $X_{DC}=X_{CD}^{\dagger}$:
\begin{equation}
 \operatorname{Tr}\rho_{RB,d}^{\,2}
 =
 \frac1{d^2}\sum_{C,D}
 \operatorname{Tr}(X_{CD}X_{DC})
 =
 \frac1{d^2}\sum_{C,D}A_{CD}.
\end{equation}
For the system purity, the initial double sum is
$d^{-2}\sum_{C,D}\operatorname{Tr}(X_{CC}X_{DD})$.
However, $X_{CC}$ is supported entirely on the configuration
subspace spanned by $\mathcal O_C$. These subspaces are
orthogonal for distinct classes, so
$X_{CC}X_{DD}=0$ whenever $C\ne D$. Therefore
\begin{equation}
 \mathcal P_{B,d}
 =\frac1{d^2}\sum_C A_{CC},
 \qquad
 \mathcal P_{RB,d}
 =\frac1{d^2}\sum_{C,D}A_{CD}.
 \label{eq:r2proxy-purities}
\end{equation}
Their logarithmic ratio gives
Eq.~\eqref{eq:r2proxy-ratio-from-A}.

This distinction also clarifies the physical meaning of
$A_{CD}$. The diagonal entry $A_{CC}$ is a state purity,
whereas an off-diagonal entry is the squared norm of
$\mathcal N_{E,p}(|C\rangle\langle D|)$.
The equivalent environment-overlap interpretation is derived
in Eq.~\eqref{eq:r2proxy-environment}.

\subsection{Reduction to a twisted classical partition function}
\label{sec:r2proxy-partition}

This subsection converts the noisy logical matrix elements
in Eq.~\eqref{eq:r2proxy-Xexpand} into an eight-state classical
partition function. We then interpret its spins and seam
twists, and assemble the partition functions needed for the
two coherent-information calculations.

\subsubsection{From edge configurations to vertex spins}

Taking the squared Hilbert--Schmidt norm of
Eq.~\eqref{eq:r2proxy-Xexpand} gives the normalized
two-configuration sum in Eq.~\eqref{eq:r2proxy-orbitsum}.
To evaluate this sum, we parameterize each flat edge
configuration using vertex variables and a fixed seam twist.

Choose a dual-lattice seam joining the open cylinder
boundaries, and orient the circumference edges consistently.
Let $c_e=1$ on edges crossing the seam and $c_e=0$ elsewhere.
For a representative $g_C\in C$, write
\begin{equation}
 g_e(\mathbf s;g_C)
 =s_{i_e}g_C^{c_e}s_{f_e}^{-1},
 \qquad s_v\in D_4,
 \label{eq:r2proxy-potentials}
\end{equation}
where $i_e$ and $f_e$ are the initial and final vertices of
edge $e$. Away from the seam, the edge label is determined
by neighboring vertex variables; the seam inserts the
required holonomy.

The vertex variables are unrestricted, so this
parameterization is redundant. Each physical configuration
occurs $|Z_{D_4}(g_C)|$ times, where $Z_{D_4}(g_C)$ is the
centralizer: simultaneous right multiplication
$s_v\mapsto s_vz$ leaves all edge labels unchanged when
$zg_C=g_Cz$. Consequently,
$N_C=8^{N_v}/|Z_{D_4}(g_C)|=|C|8^{N_v-1}$, and
\begin{equation}
 \frac1{N_C}\sum_{\mathbf g\in\mathcal O_C}f(\mathbf g)
 =
 8^{-N_v}\sum_{\mathbf s}
 f\bigl(\mathbf g(\mathbf s;g_C)\bigr).
 \label{eq:r2proxy-average}
\end{equation}

Apply this identity to both configurations in
Eq.~\eqref{eq:r2proxy-orbitsum}, using vertex variables
$\mathbf s,\mathbf t$ and representatives $g_C,g_D$.
With $u_v=t_v^{-1}s_v$, conjugation invariance of $w_p$ gives
\begin{equation}
 w_p\!\left(
 g_e(\mathbf s;g_C)g_e(\mathbf t;g_D)^{-1}
 \right)=
 w_p\!\left(
 u_{i_e}g_C^{c_e}u_{f_e}^{-1}g_D^{-c_e}
 \right).
 \label{eq:r2proxy-relative}
\end{equation}
The summand is therefore independent of $\mathbf t$.
Its sum contributes $8^{N_v}$, cancelling one of the two
normalization factors and leaving
\begin{equation}
 A_{CD}=A(g_C,g_D)
 =8^{-N_v}\sum_{\{u_v\in D_4\}}
 \prod_{e:c_e=0}w_p(u_{i_e}u_{f_e}^{-1})
 \prod_{e:c_e=1}
 w_p(u_{i_e}g_Cu_{f_e}^{-1}g_D^{-1}).
 \label{eq:r2proxy-partition}
\end{equation}
This is a normalized classical partition function with one
eight-valued spin per vertex. Changing representatives by
$g_C\mapsto ag_Ca^{-1}$ and $g_D\mapsto bg_Db^{-1}$ is
absorbed by $u_v\mapsto bu_va^{-1}$, so its value depends
only on $C$ and $D$.

\subsubsection{Physical interpretation}

The classical spins describe the relative configuration of
the two terms entering the purity trace.
Specifically, $u_v=t_v^{-1}s_v$ compares their vertex
variables.
The bulk interactions compare these relative variables at
neighboring vertices, while the seam imposes the boundary
condition associated with the two logical holonomies.

For the $D_4$ channel, the local weights are
\begin{equation}
 w_p(x)=
 \begin{cases}
  1,&x=1,\\
  (1-2p)^2,&x=r^2,\\
  (1-p)^2,&x\notin\{1,r^2\}.
 \end{cases}
 \label{eq:r2proxy-boltzmann}
\end{equation}
Thus neighboring spins are distinguished according to whether
they are equal, differ by the central element $r^2$, or
belong to different central pairs.
At $p=0$, all assignments have equal weight and every
normalized partition function equals one.
Increasing $p$ toward $1/2$ penalizes local mismatches and
favors agreement of the relative spins.

The seam tests how these local preferences accommodate the
chosen logical sectors. Writing the bulk interaction as
$W(u,v)=w_p(uv^{-1})$, its seam counterpart is
\begin{equation}
 W\bigl(u,T_{CD}(v)\bigr),
 \qquad
 T_{CD}(v)=g_Dv g_C^{-1}.
\end{equation}
Both logical labels matter, not merely their product
$g_Cg_D^{-1}$. Even equal noncentral representatives produce
a conjugation twist: $A(1,1)$ is untwisted, whereas
$A(r,r)$ need not be.

The physical meaning of the partition function follows
from its equality to $A_{CD}$. For $C=D$, it is the purity
of the noisy logical basis state. For $C\ne D$, it measures
the surviving $C$--$D$ coherence in the squared
Hilbert--Schmidt norm.
Since the individual purities can decrease with system size,
the coherent-information diagnostic compares off-diagonal
partition functions with the diagonal ones.

For example, the central-qubit ratio defines a dimensionless
twist free-energy cost,
\begin{equation}
 \Delta\mathcal F
 =-\ln\frac{A_{F_4}}{A_I},
 \qquad
 I_{c,2}^{(2)}
 =\ln\!\left(1+e^{-\Delta\mathcal F}\right).
\end{equation}
A vanishing twist cost gives the maximal value $\ln2$,
whereas a diverging cost suppresses the central-qubit
coherence relative to its diagonal purity.

\subsubsection{Twist types and coherent information}

The symmetry of the bulk interactions reduces the number
of partition functions needed for the logical-sector sums.
The weights are invariant under permutations of the four
central pairs $\{u,r^2u\}$ and exchanges within each pair.
Classifying the seam permutations under these relabelings
gives the five types in Eq.~\eqref{eq:r2proxy-types}.
The labels $F_2$ and $F_4$ denote exchanges within two and
four central pairs, respectively.

In the logical-class order
$([1],[r^2],[r],[s],[rs])$, the complete matrix is
\begin{equation}
 (A_{CD})=
 \begin{pmatrix}
 A_I&A_{F_4}&A_-&A_+&A_+\\
 A_{F_4}&A_I&A_-&A_+&A_+\\
 A_-&A_-&A_{F_2}&A_-&A_-\\
 A_+&A_+&A_-&A_{F_2}&A_+\\
 A_+&A_+&A_-&A_+&A_{F_2}
 \end{pmatrix}.
 \label{eq:r2proxy-type-matrix}
\end{equation}
The diagonal contains two $A_I$ and three $A_{F_2}$
entries. The off-diagonal entries contain two $A_{F_4}$,
ten $A_+$, and eight $A_-$ terms.
These coefficients count logical-sector pairs, not
additional conjugacy-class multiplicities.

Substituting into Eq.~\eqref{eq:r2proxy-purities} gives
\begin{equation}
 \begin{aligned}
 \mathcal P_{B,5}
 &=\frac{2A_I+3A_{F_2}}{25},\\
 \mathcal P_{RB,5}
 &=\frac{
 2A_I+3A_{F_2}+2A_{F_4}+10A_++8A_-
 }{25}.
 \end{aligned}
 \label{eq:r2proxy-full-purities}
\end{equation}
The upper-left $2\times2$ block gives the central-qubit
purities:
\begin{equation}
 \mathcal P_{B,2}=\frac{A_I}{2},
 \qquad
 \mathcal P_{RB,2}=\frac{A_I+A_{F_4}}2.
 \label{eq:r2proxy-central-purities}
\end{equation}
Taking their logarithmic ratios yields
Eqs.~\eqref{eq:r2proxy-full}
and~\eqref{eq:r2proxy-central}.
Appendix~\ref{sec:r2proxy-numerics} specifies the exact
row contraction used to evaluate these partition functions.

\subsection{Exact contraction by local row operations}
\label{sec:r2proxy-numerics}

The partition function is evaluated exactly by successively summing one row of
spins at a time. For completeness, we explicitly derive the transfer contraction of
Eq.~\eqref{eq:r2proxy-partition}. A row state is
$\mathbf u=(u_1,\ldots,u_L)\in D_4^L$, and its horizontal weight is
\begin{equation}
 R_{H,H'}(\mathbf u)=
 \left[\prod_{x=1}^{L-1}w_p(u_xu_{x+1}^{-1})\right]
 w_p(u_LHu_1^{-1}H'^{-1}).
 \label{eq:r2proxy-row}
\end{equation}
Set $W_{uv}=w_p(uv^{-1})$ and start with
$v_1(\mathbf u)=8^{-L}R_{H,H'}(\mathbf u)$. Adding a row gives
\begin{align}
 v_{y+1}(\mathbf u)&=R_{H,H'}(\mathbf u)
 \sum_{\mathbf v}\left[\prod_{x=1}^L\frac{W_{u_xv_x}}8\right]
 v_y(\mathbf v),\notag\\
 A(H,H')&=\sum_{\mathbf u}v_L(\mathbf u).
 \label{eq:r2proxy-transfer}
\end{align}
Thus each vertical layer is a product of $L$ local $8\times8$ operations. With
$a=(1-2p)^2$ and $b=(1-p)^2$, the local matrix can also be written
\begin{equation}
 W=bJ+(1-b)I+(a-b)P_{r^2},
 \label{eq:r2proxy-local-matrix}
\end{equation}
where $J$ is the all-ones matrix and $P_{r^2}$ exchanges central partners.
The implementation applies the nonnegative entries of $W$ directly.

\subsection{Environment interpretation and exact endpoint statements}

We construct an explicit environment representation of the
channel to show that $A_{CD}$ measures the overlap of the
environment states associated with logical inputs $C$ and $D$.
This also establishes the bounds on $I_{c,d}^{(2)}$ used in
Sec.~\ref{sec:r2proxy-meaning}.

For each edge, introduce orthonormal environment states
$|0\rangle$ and $|a,b\rangle$, with $a,b=1,2$, and define
\begin{equation}
 |\varepsilon_g\rangle
 =
 \sqrt{1-p}\,|0\rangle
 +
 \sqrt{\frac p2}
 \sum_{a,b=1}^{2}E(g)_{ab}|a,b\rangle.
\end{equation}
These states are normalized because
$\operatorname{Tr}[E(g)^\dagger E(g)]=2$, and their overlaps are
\begin{equation}
 \begin{aligned}
 \langle\varepsilon_k|\varepsilon_g\rangle
 &=(1-p)+\frac p2
   \operatorname{Tr}[E(k)^\dagger E(g)]\\
 &=(1-p)+\frac p2\chi_E(gk^{-1})
 =K_{E,p}(gk^{-1}).
 \end{aligned}
\end{equation}
For independent edge environments, set
\begin{equation}
 |e_{\mathbf g}\rangle
 =\bigotimes_e|\varepsilon_{g_e}\rangle,
 \qquad
 V|\mathbf g\rangle
 =|\mathbf g\rangle|e_{\mathbf g}\rangle.
\end{equation}
The orthonormality of the configuration states and normalization
of the environment states make $V$ an isometry, with
\begin{equation}
 \langle e_{\mathbf k}|e_{\mathbf g}\rangle
 =\prod_eK_{E,p}(g_ek_e^{-1}).
 \label{eq:r2proxy-dilation}
\end{equation}
Tracing out the environment then reproduces the noise channel:
\begin{equation}
\operatorname{Tr}_{\mathrm{env}}
 \!\left[V|\mathbf g\rangle\langle\mathbf k|V^\dagger\right]=
 \langle e_{\mathbf k}|e_{\mathbf g}\rangle
 |\mathbf g\rangle\langle\mathbf k|
 =\mathcal N_p(|\mathbf g\rangle\langle\mathbf k|).
\end{equation}
Thus the environment acquires a configuration-dependent state,
whose overlap determines the coherence remaining between
configurations.

For the logical state
$|C\rangle=N_C^{-1/2}\sum_{\mathbf g\in\mathcal O_C}
|\mathbf g\rangle$, tracing out the physical system removes
terms with unequal configuration labels:
\begin{equation}
 \sigma_C
 =\operatorname{Tr}_B[V|C\rangle\langle C|V^\dagger]
 =\frac1{N_C}\sum_{\mathbf g\in\mathcal O_C}
 |e_{\mathbf g}\rangle\langle e_{\mathbf g}|.
\end{equation}
Consequently,
\begin{equation}
 \begin{aligned}
 \operatorname{Tr}(\sigma_C\sigma_D)
 &=
 \frac1{N_CN_D}
 \sum_{\substack{\mathbf g\in\mathcal O_C\\
                 \mathbf k\in\mathcal O_D}}
 |\langle e_{\mathbf k}|e_{\mathbf g}\rangle|^2\\
 &=
 \frac1{N_CN_D}
 \sum_{\substack{\mathbf g\in\mathcal O_C\\
                 \mathbf k\in\mathcal O_D}}
 \prod_e|K_{E,p}(g_ek_e^{-1})|^2\\
 &=A_{CD},
 \end{aligned}
 \label{eq:r2proxy-environment}
\end{equation}
where the last equality is Eq.~\eqref{eq:r2proxy-orbitsum}.
The environment overlap is therefore exactly the quantity
computed by the purity partition function.

Substituting into
Eq.~\eqref{eq:r2proxy-coherence-ratio} gives
\begin{equation}
 I_{c,d}^{(2)}
 =
 \ln\!\left[
 \frac{\operatorname{Tr}
       \left(\sum_C\sigma_C\right)^2}
      {\sum_C\operatorname{Tr}\sigma_C^2}
 \right],
\end{equation}
with sums over the $d$ logical basis states.
The numerator is the denominator plus nonnegative cross terms,
so $I_{c,d}^{(2)}\geq0$. Expanding the squared
Hilbert--Schmidt distances yields
\begin{equation}
d\sum_C\operatorname{Tr}\sigma_C^2
 -
 \operatorname{Tr}\left(\sum_C\sigma_C\right)^2=
 \sum_{C<D}\|\sigma_C-\sigma_D\|_2^2
 \geq0,
 \label{eq:r2proxy-upperbound}
\end{equation}
where $\|X\|_2^2=\operatorname{Tr}(X^\dagger X)$.
Hence the ratio is at most $d$, establishing
$0\leq I_{c,d}^{(2)}\leq\ln d$.

\subsection{Finite-size analysis of the R\'enyi-2 transition}
\label{sec:r2proxy-scaling}

This subsection specifies the fits displayed in
Fig.~\ref{fig:r2proxy-collapse} and summarizes their systematic
sensitivity. The exact contractions have no sampling variance, so fit
residuals quantify finite-size-model error instead of statistical uncertainty.

We use Eq.~\eqref{eq:r2proxy-fss} with a cubic leading function, a linear
correction function, and the representative choice $\omega=1$. Each
input uses the same 15 probabilities in $0.435\leq p\leq0.475$ and
$L=4,\ldots,10$, with equal weights for the 105 exact points. The
separate least-squares fits give
\begin{align}
 p_{{\mathrm{R}2},\mathrm{full}}&\simeq0.4534,
 &\nu_{\mathrm{full,eff}}&\simeq0.834,\notag\\
 p_{{\mathrm{R}2},\mathrm{cent}}&\simeq0.4559,
 &\nu_{\mathrm{cent,eff}}&\simeq0.709.
 \label{eq:r2proxy-representative-fits}
\end{align}
Their root-mean-square residuals are $0.00356$ and $0.00226.$
A joint fit with common $p_{\mathrm{R}2}$ and $\nu$ gives $0.4545$ and
$0.799$, respectively, with residual $0.00327$.

\end{document}